\documentclass[twocolumn, twocolappendix, astrosymb]{aastex631} 

\usepackage{amsmath}	

\newcommand{\txn}{\textnormal}
\newcommand{\us}{\,} 

\newcommand{\Ang}{\ensuremath{\us \txn{\AA}}}

\newcommand{\pkpc}{\ensuremath{\us \txn{pkpc}}} 
\newcommand{\pcmsq}{\ensuremath{\us \txn{cm}^{-2}}}
\newcommand{\kmps}{\ensuremath{\us \txn{km}\, \txn{s}^{-1}}}

\newcommand{\Msun}{\ensuremath{\us \txn{M}_{\sun}}}
\newcommand{\Lstar}{\ensuremath{\us \txn{L}_{*}}}
\newcommand{\Mstellar}{\ensuremath{\us \txn{M}_{\star}}}
\newcommand{\Mvir}{\ensuremath{\us \txn{M}_{\txn{vir}}}}
\newcommand{\Rvir}{\ensuremath{\us \txn{R}_{\txn{vir}}}}

\newcommand{\dex}{\ensuremath{\us \txn{dex}}}

\shorttitle{Simulated and observed CGM \ion{Ne}{8} absorption}
\shortauthors{N.\ A.\ Wijers, C.-A.\ Faucher-Gigu\`ere, J.\ Stern, et al.}

\begin{document}

\title{\ion{Ne}{8} in the warm-hot circumgalactic medium of FIRE simulations and in observations}

\correspondingauthor{Nastasha A.\ Wijers}
\email{nastasha.wijers@northwestern.edu}

\author[0000-0001-6374-7185]{Nastasha A.\ Wijers}
\affiliation{CIERA and Department of Physics and Astronomy, Northwestern University, 1800 Sherman Ave, Evanston, IL 60201, USA}

\author[0000-0002-4900-6628]{Claude-Andr{\'e} Faucher-Gigu{\`e}re}
\affiliation{CIERA and Department of Physics and Astronomy, Northwestern University, 1800 Sherman Ave, Evanston, IL 60201, USA}

\author[0000-0002-7541-9565]{Jonathan Stern}
\affiliation{School of Physics \& Astronomy, Tel Aviv University, Tel Aviv 69978, Israel}

\author[0000-0002-8408-1834]{Lindsey Byrne}
\affiliation{CIERA and Department of Physics and Astronomy, Northwestern University, 1800 Sherman Ave, Evanston, IL 60201, USA}

\author[0000-0003-2341-1534]{Imran Sultan}
\affiliation{CIERA and Department of Physics and Astronomy, Northwestern University, 1800 Sherman Ave, Evanston, IL 60201, USA}



\begin{abstract}
The properties of warm-hot gas around $\sim \Lstar$ galaxies can be studied with absorption lines from highly ionized metals. 
We predict \ion{Ne}{8} column densities from cosmological zoom-in simulations of halos with masses in $\sim 10^{12}$ and $\sim 10^{13} \Msun$ from the FIRE project. 
\ion{Ne}{8} traces the volume-filling, virial-temperature gas in $\sim 10^{12} \Msun$ halos. 
In $\sim 10^{13} \Msun$ halos the \ion{Ne}{8} gas is clumpier, and biased towards the cooler part of the warm-hot phase. 
We compare the simulations to observations from the CASBaH and CUBS surveys.
We show that when inferring halo masses from stellar masses to compare simulated and observed halos, it is important to account for the scatter in the stellar-mass-halo-mass relation, especially at $\Mstellar \gtrsim 10^{10.5} \Msun$. 
Median \ion{Ne}{8} columns in the fiducial FIRE-2 model are about as high as observed upper limits allow, while the simulations analyzed do not reproduce the highest observed columns. 
This suggests that the median \ion{Ne}{8} profiles predicted by the simulations are consistent with observations, but that the simulations may underpredict the scatter. 
We find similar agreement with analytical models that assume a product of the halo gas fraction and metallicity (relative to solar) $\sim 0.1$, indicating that observations are consistent with plausible CGM temperatures, metallicities, and gas masses.
Variants of the FIRE simulations with a modified supernova feedback model and/or AGN feedback included (as well as some other cosmological simulations from the literature) more systematically underpredict \ion{Ne}{8} columns. 
The circumgalactic \ion{Ne}{8} observations therefore provide valuable constraints on simulations that otherwise predict realistic galaxy properties.
\end{abstract}




\section{Introduction}

The circumgalactic medium, a halo of gas surrounding galaxies, plays an import role in galaxy formation and evolution \citep[e.g., the review by][]{tumlinson_peeples_werk_2017_cgmreview}. If this gas accretes onto the central galaxy, it can form stars. Conversely, cutting off accretion onto a central galaxy can quench it \citep[e.g., the review by ][]{faucher-giguere_oh_2023}. This can occur due to the feedback from the central galaxy (stars and AGN), which can heat ISM gas and eject it into the CGM. It can also heat the CGM, and possibly generate outflows which escape the halo altogether \citep[e.g.,][]{mitchell_schaye_2022, hafen_faucher-giguere_etal_2020, pandya_somerville_etal_2020, muratov_keres_etal_2017, muratov_keres_etal_2015}.

This means that understanding the CGM can help us understand the processes which regulate galaxy formation. However, we currently do not know how much mass there is in the CGM around $\Lstar$ (halo mass $\sim 10^{12} \Msun$) galaxies \citep{werk_proschaska_etal_2014}. Halos around this mass, roughly $\sim 10^{12}$--$10^{13}\Msun$, are interesting, as these halos are where central galaxy quenching mostly occurs \citep[e.g.,][]{behroozi_wechsler_etal_2019}. This is the mass range we will investigate in this paper.

We have many observations of cool CGM gas ($\sim 10^{4}$--$10^{5}$~K) from UV absorption line observations \citep[e.g.,][]{tumlinson_peeples_werk_2017_cgmreview}, from which we can infer its temperature, density, and metallicity, and mass. There are also many detections of \ion{O}{6} absorption. This can come from warmer gas at $\sim 10^{5.5}$~K, but it can also trace cool, photo-ionized gas. 

\begin{figure*}
\centering
\includegraphics[width=\textwidth]{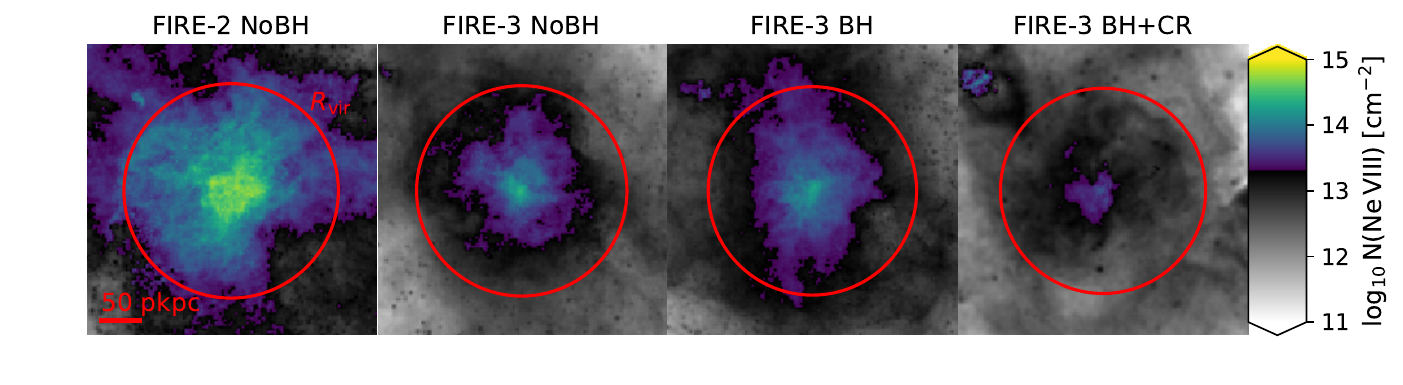}
\caption{\ion{Ne}{8} column density images for the m12f halo at $z=1$ ($\Mvir \approx 10^{11.8} \Msun$), simulated with four different physics models (different panels). The color scale switches from black and white to colors at a column density of $10^{13.3} \pcmsq$, the minimum column density \citet{burchett_tripp_etal_2018} (CASBaH survey) are sensitive to. 
\citet{qu_chen_etal_2024_preprint} (CUBS survey) have a higher detection limit, $\approx 10^{14} \pcmsq$. 
The different panels show the same physical area; the red circles show each halo's virial radius. 
The panels illustrate the observably different \ion{Ne}{8} column densities produced by the different FIRE physics models.}
\label{fig:m12maps}
\end{figure*}

The warm/hot, virialized CGM gas phase 
is where CGM mass estimates are most uncertain \citep{werk_proschaska_etal_2014}. Its absorption and emission lines are largely in the X-ray band \citep[e.g.,][]{perna_loeb_1998, hellsten_gnedin_miralda-escude_1998, chen_weinberg_etal_2003, cen_fang_2006, branchini_ursino_etal_2009}, where absorption line detections are rare and do not often reach high significance \citep[e.g.,][]{nicastro_etal_2016}. There is a larger sample of X-ray observations of the Milky Way halo \citep[{e.g.,}][]{bregman_lloyd-davies_edward_2007, gupta_mathur_etal_2014, miller_bregman_2015, gupta_mathur_etal_2017, das_mathur_etal_2019}, but as this is a single system, it can be difficult to draw conclusions about the CGM in general from these.

Planned and proposed future instruments such as 
the Athena X-IFU \citep{athena_white_paper, Athena_2018_07},
Arcus \citep{smith_abraham_etal_2016_arcus},
LEM \citep{kraft_markevitch_etal_2022},
HUBS \citep{cui_chen_etal_2020}, and
Lynx \citep{lynx_2018_08} 
can give us the X-ray absorption and emission line data we need \citep[e.g.,][]{wijers_schaye_oppenheimer_2020, wijers_schaye_2022, nelson_byrohl_etal_2023}. 
However, current X-ray instruments do not have the sensitivity and spectral resolution required to observe the CGM of a large set of $\sim \Lstar$ halos. Fast radio burst dispersion measures \citep[e.g.,][]{mcquinn_2014} and measurements of the Sunyaev-Zel'dovich (SZ) effect \citep[e.g., the review by][]{mroczkowski_nagai_etal_2018} could also provide future constraints on the CGM ionized gas content, but robust measurements of the CGM content of Milky-Way-like halos will require larger samples of localized FRBs \citep[e.g.,][]{ravi_2019} or instrumental improvements for SZ measurements such as a higher spatial resolution \citep{mroczkowski_nagai_etal_2018}.

Therefore, in this paper, we focus on an extreme UV (EUV) absorption line: the \ion{Ne}{8} doublet at $770, 780 \Ang$. \citet{burchett_tripp_etal_2018} and \citet{qu_chen_etal_2024_preprint} completed the CASBaH and CUBS surveys, respectively, for \ion{Ne}{8} absorption in quasar spectra, and obtained galaxy observations along the quasar sightlines. This allows them to associate the absorbers with the halos of specific galaxies. \ion{Ne}{8} ion fractions peak at $\approx10^{5.6}$--$10^{6.2}$~K in collisional ionisation equilibrium (CIE). We define this peak as the range of temperatures where the ion fraction is at least $0.1 \times$ the maximum CIE ion fraction. This peak temperature range makes \ion{Ne}{8} a good candidate to trace the warm/hot, volume-filling gas in $\sim 10^{12} \Msun$ halos \citep[e.g.,][]{verner_tytler_barthel_1994}. 

We use a multi-pronged approach to study the physical nature of \ion{Ne}{8} in the CGM.  
We start with a theoretical analysis of the properties of \ion{Ne}{8} in cosmological zoom-in simulations from the FIRE project (Feedback In Realistic Environments),\footnote{See the FIRE project website: \url{https://fire.northwestern.edu/}} compare different simulation variants with observations, and further explore more general insights that can be extracted using idealized analytic models. 
In \S\ref{sec:meth}, we describe the FIRE-2 and FIRE-3 simulations we analyse, and how we predict column densities from them. 
Next, in \S\ref{sec:simprop}, we explore the properties of \ion{Ne}{8} in the halos we study.
In \S\ref{sec:obscomp}, we then describe how the FIRE predictions compare to the observations.
Next, we explore a simple analytical model for \ion{Ne}{8} column densities in \S\ref{sec:anmodel}.
In \S\ref{sec:discussion}, we compare our results to others in the literature.
Finally, we summarize our results in \S\ref{sec:conclusions}.

We use a flat $\Lambda$CDM cosmology, with a redshift~0 Hubble parameter in units of $100 \us \kmps \mathrm{Mpc}^{-1}$ of $h \approx 0.7$, and cosmic mean densities of dark energy and total matter, normalized to the critical density, of $\Omega_{\Lambda} \approx 0.7$ and $\Omega_{m} \approx 0.3$, respectively. We analyse each simulation using the exact parameter set it was run with.
Distances measured in kpc use physical/proper kpc, unless specified otherwise.

\section{Methods}
\label{sec:meth}

\begin{figure*}
\centering
\includegraphics[width=0.8\textwidth]{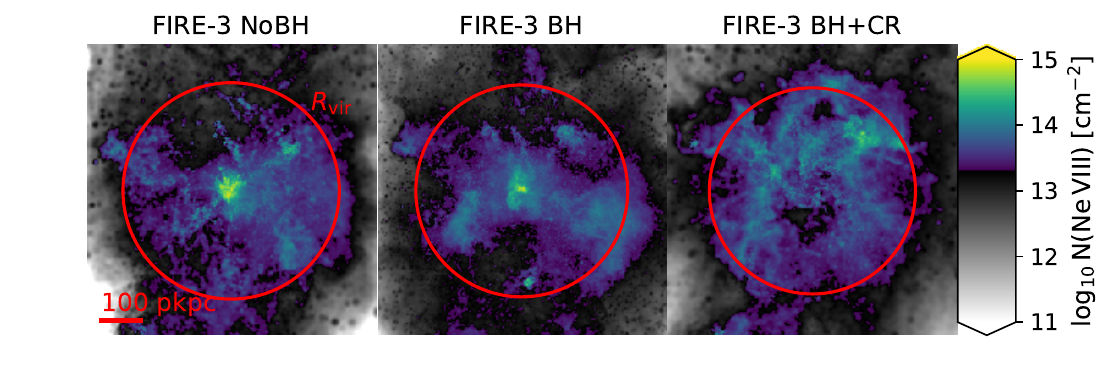}
\caption{As Fig.~\ref{fig:m12maps}, but for the m13h113 halo (also at $z=1$, $\Mvir \approx 10^{12.7} \Msun$) run with the three FIRE-3 physics models. The \ion{Ne}{8} in m13 halos has a clumpier distribution than in the m12 halos. This is because \ion{Ne}{8} ion fractions are low at the virial temperatures of these halos, meaning the ion preferentially traces somewhat cooler, denser, clumpier gas than the volume-filling phase (see also Figs.~\ref{fig:clumpiness} and~\ref{fig:3dprof}).}
\label{fig:m13maps}
\end{figure*}

We analyse simulations run with four different physics models: one fiducial FIRE-2 model, and three FIRE-3 models. The FIRE-3 models have either no AGN feedback, AGN feedback without cosmic rays, or AGN feedback with cosmic rays. 
Below, we describe these simulations in more detail, and briefly explain how we predict column densities from them.

\subsection{The FIRE-2 simulations}
\label{sec:simgenf2}

The FIRE-2 simulations are described in detail in \citet{hopkins_wetzel_etal_2018}. 
The simulations use the meshless finite-mass (MFM) magneto-hydrodynamics solver \citep{hopkins_2015_hydro}, in the GIZMO\footnote{The public GIZMO version is available at \url{http://www.tapir.caltech.edu/\~phopkins/Site/GIZMO.html}.} code \citep{hopkins_2015_gizmo}. 

In the FIRE-2 simulations, 11 elements are explicitly tracked, including neon. The radiative cooling and heating calculations \citep{hopkins_wetzel_etal_2018} include processes relevant to the cool and cold ISM, such as cooling from molecules and fine-structure lines. The cooling and heating processes are applicable to temperatures of  10--$10^{10} \us \mathrm{K}$.
For temperatures $ > 10^{4}$~K, metal-line cooling follows \citet{wiersema_etal_2009_tables}. The effects of ionizing photons from both a \citet{FG09} background and stars in the simulations are included. 
For radiation from stars, a radiative transfer method with a self-shielding approximation is used \citep[LEBRON;][]{hopkins_quataert_murray_2012, hopkins_wetzel_etal_2018}. 

Star formation can occur in gas if it is (1) self-gravitating (the potential energy is larger than the thermal and kinetic energy combined), (2) self-shielding (it contains molecular gas), (3) Jeans-unstable (the thermal Jeans mass is smaller than the resolution element mass), and (4) has a minimum density $\mathrm{n}_{\mathrm{H}} > 10^{3} \pcmsq$. 
The star formation rate is then simply the molecular gas density over the free-fall time of the resolution element. Star particles represent a single-age stellar population, and have the mass and element abundances of the gas particles they formed from. 

The feedback from each single-age stellar population is taken from \textsc{starburst99} \citep{starburst99}, assuming a \citet{kroupa_2001} stellar initial mass function and a stellar mass range of $0.1$--$100 \Msun$. 
This includes supernova rates, and abundances, mass, and energy in supernovae and stellar winds. 
The injected supernova momentum and energy depend on the Sedov-Taylor solution for the explosion, given the density and mass of the surrounding gas \citep{hopkins_wetzel_etal_2018_sne, martizzi_faucher-giguere_quataert_2015}.

The FIRE-2 simulations we analyse include a prescription for subgrid metal diffusion \citep{hopkins_wetzel_etal_2018, colbrook_ma_etal_2017}. 
The FIRE-2 model contains neither black holes, nor cosmic rays, so we will label it `FIRE-2 NoBH'.
These simulations have redshift~0 main halo masses of $\sim 10^{12} \Msun$, and are listed in Tab.~\ref{tab:sims}.

\subsection{The FIRE-3 simulations}
\label{sec:simgenf3}

The FIRE-3 simulations are a set of simulations run with similar stellar physics models but using the FIRE-3 code \citep{hopkins_wetzel_etal_2023}. Much of the modeling is the same as in the FIRE-2 simulations, with some updates. 
The FIRE-3 simulations include magnetic fields \citep{hopkins_raives_2016_mhd, hopkins_2016_mhd_divB_control}, unlike our FIRE-2 sample. 
In the FIRE-3 NoBH simulations, the magnetic pressure in the CGM is generally $<0.1$ times the thermal pressure (with fractions $<0.01$  typical; Sultan et al., in prep.), meaning its dynamical effect is likely small.
This is consistent with the small differences previously reported between the FIRE-2 runs with and without magnetic fields \citep{hopkins_chan_etal_2020_whatabout}, and the small effects of magnetic fields on the CGM \citet{su_hopkins_etal_2019} found for a FIRE-2 m12 halo. Note that this is different from what \citet{van-de-voort_bieri_etal_2021} found for zoom-in simulations run with the IllustrisTNG and Auriga feedback models. In those simulations, magnetic fields are dynamically more important in the CGM.

The UV/X-ray background was updated to that of \citet{FG20}.
When calculating ion fractions, we will use the \citet{ploeckinger_schaye_2020_tableref} tables\footnote{The tables are available at \url{https://dataverse.harvard.edu/dataset.xhtml?persistentId=doi:10.7910/DVN/GR3L5N}. We specifically use the UVB\_dust1\_CR1\_G1\_shield1.hdf5 table.}, 
produced by \citet{ploeckinger_schaye_2020}, who also assumed a \citet{FG20} UV/X-ray background.
Stellar evolution and feedback modeling were also updated, and the star formation criteria from FIRE-2 were adjusted for FIRE-3: the gas must be (1) self-gravitating, (2) Jeans-unstable, and (3) converging or not diverging too rapidly. (A strict minimum density is not required). 
As for the FIRE-2 simulations, subgrid metal diffusion is included in the simulations we analyze.

The FIRE-3 simulations also include updated stellar yields for supernovae (core-collapse and type Ia supernovae) and for stellar winds \citep{hopkins_wetzel_etal_2023}. 
The type Ia yields were updated from the \citet{iwamoto_brachwitz_etal_1999} values to the average of the \citet{leung_nomoto_2018} W7 and WDD2 models, and the core-collapse yields from \citet{nomoto_tominaga_etal_2006} to a modified \citet{sukhbold_ertl_etal_2016} model. 
The supernova yield updates result in lower neon yields in FIRE-3 than in FIRE-2: the ratio of total neon mass to total stellar mass is larger by a factor $\approx 2$ in the FIRE-2 simulations we analyse than in our FIRE-3 simulations.
The FIRE-2/3 model for stellar winds does not produce neon.

We consider a set of three FIRE-3 physics models. The first includes no AGN feedback (NoBH), the second includes AGN feedback without cosmic rays (BH), and the third includes AGN feedback with cosmic rays (BH+CR). Note that the BH+CR model contains cosmic rays from AGN and supernovae, while the other models do not contain cosmic rays from any source. For the FIRE-3 simulations, we consider halos with redshift~0 halo masses of $\sim 10^{12} \Msun$ (m12 halos). Unlike for the FIRE-2 model, we also include more massive m13 halos, with redshift~0 halo masses $\sim 10^{13} \Msun$.

In the BH and BH+CR models, black holes (BHs) are stochastically seeded from star-forming gas, especially at high surface densities and low metallicities \citep{wellons_faucher-giguere_etal_2023, hopkins_wetzel_etal_2018}.
They merge if they are close together and gravitationally bound.
The black hole accretion model is described by 
\citet{wellons_faucher-giguere_etal_2023} and \citet{hopkins_wetzel_etal_2023}.

AGN feedback follows \citet{wellons_faucher-giguere_etal_2023}, with the parameters given by \citet{byrne_faucher-giguere_etal_2023}.
Radiative feedback is injected at a rate of 0.1 times the accretion rate.
Winds are injected \citep{torrey_hopkins_etal_2020, su_hopkins_etal_2021} at a fixed velocity of $3000 \kmps$, parallel to the BH angular momentum \citep{hopkins_torrey_etal_2016}. The wind mass injection rate is equal to the black hole accretion rate.
In the BH+CR model, jets are modelled by adding CRs to the created wind resolution elements, with an energy efficiency of $10^{-3}$ relative to $\dot{M}_{\rm BH} c^{2}$, where $\dot{M}_{\rm BH}$ is the accretion rate.

In the BH+CR simulations, the cosmic ray (CR) model evolves the CR spectrum for protons and electrons. 
The CR scattering rate depends on local plasma properties \citep[][]{chan_hopkins_etal_2019, hopkins_butsky_etal_2022}. 
Specifically, the modified external driving model in \S5.3.2 of \citet{hopkins_squire_etal_2022} is used. 
This model
was calibrated to reproduce the Voyager and AMS-02 observations from MeV to TeV energies in Milky Way-like simulations.
CRs are injected by supernova explosions and stellar winds, making up 10\% of the initial kinetic energy in both, and by AGN as described above. 

We must stress an important point about the FIRE-3 simulations analyzed in this paper. 
The FIRE-3 runs we analyze implement a ``velocity-aware'' numerical method to deposit energy and momentum to gas resolution elements. 
This scheme improves conservation properties, as described in \cite{hopkins_wetzel_etal_2023}, but introduces significant differences in the net terminal momentum injected following SNe relative to FIRE-2, especially in the regime of locally converging gas flows. 
 In this limit, the momentum injected can be substantially larger than standard models for the terminal momentum in a static medium \citep[e.g.,][]{martizzi_faucher-giguere_quataert_2015, hopkins_wetzel_etal_2018_sne}. 
For the relatively massive halos we study in this paper, this results in galaxy stellar masses in FIRE-3 simulations that are lower than in FIRE-2 \citep[see][]{hopkins_wetzel_etal_2023}. 
This change in the momentum deposited by SNe in massive galaxies depends on a non-unique choice for how to implement the subgrid SN physics. 
 This subtlety was not described in the original FIRE-3 methods paper, but is explained in detail in \cite{hopkins_2024_preprint}. 

We note that the FIRE-3 SNe feedback model implemented in the present simulations is not a priori ``incorrect'' or ``unphysical.'' Rather, it reflects a different theoretical assumption for the unresolved, subgrid dynamics relative to FIRE-2. 
A priori, both the FIRE-2 model and this FIRE-3 feedback model are reasonable. However, as we will show, our comparison with \ion{Ne}{8} CGM observations empirically favor the FIRE-2 model. 
\cite{hopkins_2024_preprint} also discusses some theoretical issues with the particular implementation used in the present FIRE-3 runs that could make it less realistic. 
In future FIRE-3 simulations, we will likely use a modified implementation that produces injected SN momentum, stellar masses, and (presumably) CGM properties in closer agreement with FIRE-2 for the same basic physics (e.g., NoBH). 
For simplicity, we will simply refer to the present runs as FIRE-3, but caution that the results will not necessarily apply to future FIRE-3 runs with a modified SN model.

\subsection{The simulation sample}
\label{sec:simsel}

We analyse a set of cosmological zoom simulations of halos with masses $\sim 10^{12} \Msun$ (m12) and $\sim 10^{13} \Msun$ (m13).

\begin{table*}
\caption{The FIRE-2 and FIRE-3 simulations analysed in this work. We show initial conditions (ICs), physics model, resolution (for gas mass resolution elements), and the halo mass, stellar mass, and virial radius for redshifts 1.0 and~0.5.}
\label{tab:sims}
\begin{tabular}{l l l l l l l l l}                                                                                                                     
\hline                                                                                                                                          
         &       &       & \multicolumn{3}{c}{$z=1.0$}   & \multicolumn{3}{c}{$z=0.5$}   \\                                 
ICs      & model         & resolution    & $\mathrm{M}_{\mathrm{vir}}$   & $\mathrm{M}_{\star}$          & $\mathrm{R}_{                \mathrm{vir}}$   & $\mathrm{M}_{\mathrm{vir}}$   & $\mathrm{M}_{\star}$          & $\mathrm{R}_{\mathrm{vir}}$   \\                                    
         &       & $[\mathrm{M}_{\odot}]$        & $[\mathrm{M}_{\odot}]$        & $[\mathrm{M}_{\odot}]$        & [pkpc    ]        & $[\mathrm{M}_{\odot}]$        & $[\mathrm{M}_{\odot}]$        & [pkpc]        \\                                     
\hline                                                                                                                                          
\hline                                                                                                                                          
m12r     & FIRE-2 NoBH   & 7e3   & 2.9e11        & 2.6e9         & 103   & 3.5e11        & 4.6e9         & 139   \\                         
m12z     & FIRE-2 NoBH   & 4e3   & 3.2e11        & 9.1e8         & 107   & 5.2e11        & 2.7e9         & 160   \\                         
m12w     & FIRE-2 NoBH   & 7e3   & 4.1e11        & 3.6e9         & 115   & 8.7e11        & 1.1e10        & 188   \\                         
m12c     & FIRE-2 NoBH   & 7e3   & 6.0e11        & 5.1e9         & 132   & 6.9e11        & 1.9e10        & 176   \\                         
m12b     & FIRE-2 NoBH   & 7e3   & 6.8e11        & 2.5e10        & 138   & 1.2e12        & 4.7e10        & 212   \\                         
m12f     & FIRE-2 NoBH   & 7e3   & 7.2e11        & 9.6e9         & 141   & 1.1e12        & 2.4e10        & 203   \\                         
m12m     & FIRE-2 NoBH   & 7e3   & 8.1e11        & 5.3e9         & 147   & 1.1e12        & 2.7e10        & 206   \\                         
m12i     & FIRE-2 NoBH   & 7e3   & 8.4e11        & 1.0e10        & 149   & 8.3e11        & 2.7e10        & 187   \\
\hline
\hline
m12r     & FIRE-3 NoBH   & 7e3   & 2.8e11        & 1.6e9         & 101   & 3.5e11        & 2.7e9         & 139   \\
m12z     & FIRE-3 NoBH   & 4e3   & 2.8e11        & 6.8e8         & 103   & 4.6e11        & 8.0e8         & 153   \\
m12w     & FIRE-3 NoBH   & 7e3   & 3.8e11        & 2.4e9         & 112   & 7.6e11        & 3.4e9         & 180   \\
m12f     & FIRE-3 NoBH   & 7e3   & 6.8e11        & 4.0e9         & 138   & 9.9e11        & 6.7e9         & 199   \\
m12q     & FIRE-3 NoBH   & 7e3   & 8.3e11        & 5.1e9         & 148   & 1.2e12        & 9.2e9         & 212   \\
m13h206          & FIRE-3 NoBH   & 3e5   & 4.3e12        & 1.2e11        & 253   & 6.1e12        & 2.2e11        & 363  \\
m13h113          & FIRE-3 NoBH   & 3e5   & 5.6e12        & 1.0e11        & 276   & 8.2e12        & 2.6e11        & 400  \\
m13h236          & FIRE-3 NoBH   & 3e5   & 6.0e12        & 1.1e11        & 282   & 8.3e12        & 2.2e11        & 401  \\
m13h007          & FIRE-3 NoBH   & 3e5   & 8.4e12        & 9.7e10        & 316   & 1.4e13        & 1.6e11        & 477  \\
m13h029          & FIRE-3 NoBH   & 3e5   & 9.1e12        & 1.3e11        & 324   & 1.2e13        & 3.0e11        & 453  \\
m13h002          & FIRE-3 NoBH   & 3e5   & 1.5e13        & 1.4e11        & 379   & 2.2e13        & 2.3e11        & 552  \\
m13h223          & FIRE-3 NoBH   & 3e5   & 1.6e13        & 1.1e11        & 395   & 2.1e13        & 2.5e11        & 546  \\
\hline
\hline
m12r     & FIRE-3 BH     & 7e3   & 2.8e11        & 2.0e9         & 102   & 3.5e11        & 3.1e9         & 139   \\
m12w     & FIRE-3 BH     & 7e3   & 3.8e11        & 3.5e9         & 112   & 8.2e11        & 6.3e9         & 185   \\
m12b     & FIRE-3 BH     & 7e3   & 6.0e11        & 4.5e9         & 132   & 1.1e12        & 8.0e9         & 205   \\
m12m     & FIRE-3 BH     & 7e3   & 6.3e11        & 3.5e9         & 135   & 9.9e11        & 6.9e9         & 199   \\
m12f     & FIRE-3 BH     & 7e3   & 6.6e11        & 5.6e9         & 137   & 1.0e12        & 8.5e9         & 201   \\
m12i     & FIRE-3 BH     & 7e3   & 7.8e11        & 4.9e9         & 144   & 7.3e11        & 8.1e9         & 179   \\
m12q     & FIRE-3 BH     & 7e3   & 8.3e11        & 4.9e9         & 148   & 1.1e12        & 7.3e9         & 209   \\
m13h206          & FIRE-3 BH     & 3e4   & 4.3e12        & 4.8e10        & 252   & 6.1e12        & 1.2e11        & 362  \\
m13h113          & FIRE-3 BH     & 3e4   & 5.3e12        & 3.4e10        & 270   & 8.1e12        & 8.9e10        & 398  \\
\hline
\hline
m12f     & FIRE-3 BHCR   & 6e4   & 6.3e11        & 3.5e9         & 135   & 1.0e12        & 3.9e9         & 200   \\
m12m     & FIRE-3 BHCR   & 6e4   & 6.5e11        & 5.1e9         & 136   & 8.4e11        & 8.6e9         & 188   \\
m12i     & FIRE-3 BHCR   & 6e4   & 7.5e11        & 4.1e9         & 143   & 6.9e11        & 5.1e9         & 176   \\
m12q     & FIRE-3 BHCR   & 6e4   & 8.1e11        & 3.1e9         & 147   & 1.2e12        & 3.8e9         & 209   \\
m13h206          & FIRE-3 BHCR   & 3e5   & 3.9e12        & 3.8e10        & 244   & 5.7e12        & 4.3e10        & 354  \\
m13h113          & FIRE-3 BHCR   & 3e5   & 4.9e12        & 2.9e10        & 263   & 7.4e12        & 3.6e10        & 386  \\
m13h236          & FIRE-3 BHCR   & 3e5   & 5.1e12        & 2.4e10        & 268   & 7.2e12        & 2.9e10        & 383  \\
m13h007          & FIRE-3 BHCR   & 3e5   & 7.5e12        & 4.1e10        & 304   & 1.2e13        & 4.7e10        & 457  \\
m13h029          & FIRE-3 BHCR   & 3e5   & 8.5e12        & 3.7e10        & 318   & 1.0e13        & 5.2e10        & 433  \\
m13h002          & FIRE-3 BHCR   & 3e5   & 1.2e13        & 2.7e10        & 353   & 1.7e13        & 3.7e10        & 513  \\
m13h037          & FIRE-3 BHCR   & 3e5   & 1.4e13        & 4.8e10        & 374   & 1.9e13        & 5.9e10        & 531  \\
m13h009          & FIRE-3 BHCR   & 3e5   & 1.9e13        & 3.2e10        & 418   & 2.5e13        & 3.6e10        & 584  \\
\hline
\end{tabular}
\end{table*}

The set of FIRE-2 simulations we analyse comes from the public release of FIRE-2 data by \citet{wetzel_hayward_etal_2023}. We specifically consider the halos with redshift~0 masses of $\sim 10^{12} \Msun$ in the `core' suite\footnote{One minor difference is that the m12f and m12i halos were simulated using a code update \citep[\S2.4.1]{wetzel_hayward_etal_2023}. The effect of this difference should be minimal.}. These halos were selected from a cosmological volume based on their mass, and being relatively isolated from halos of similar or higher masses.

Most of the FIRE-3 halos we analyse were presented in \citet{byrne_faucher-giguere_etal_2023}. 
The FIRE-3 NoBH m12i simulation was also presented by \citet{hopkins_wetzel_etal_2023}. 
For a subset of the halos we analyse here, FIRE-3 simulations have been analyzed for central galaxy properties by \cite{{byrne_faucher-giguere_etal_2023}} and for other CGM properties by Sultan et al. (in prep.).

We list the halo mass, virial radius, and stellar mass of the central galaxy of each simulation in Tab.~\ref{tab:sims}. (See \S\ref{sec:halogaldef} for definitions.) We show these values at redshifts $0.5$ and~$1.0$, since we analyse simulations between these redshifts. This range was selected to roughly match the redshift range searched by \citet{burchett_tripp_etal_2018} as part of the CASBaH survey; the \citet{qu_chen_etal_2024_preprint} data from the CUBS survey cover a smaller redshift range. We similarly analysed m12 and m13 halos, since the \citet{burchett_tripp_etal_2018} and \citet{qu_chen_etal_2024_preprint} galaxies likely reside in halos of roughly these masses (see \S\ref{sec:compmeth}).

For each physics model and halo mass category,
we analysed the halos from simulations which had reached redshift~0.5. This ensured each simulation covered our target redshift range 0.5--1.0. Within this range, we specifically analyse simulation outputs at redshifts 0.5, 0.6, 0.7, 0.8, 0.9, and~1.0. 

Of our halos, only m12f was run
for all m12 physics models, and only m13h113 and m13h206 reached redshift~0.5 for all three m13 physics models. In comparisons between physics models, we will therefore often focus on these halos.

\subsection{Finding and centering our halos}
\label{sec:halogaldef}

The halos we analyse are the main halos in each zoom-in volume. We find the halo centers using a shrinking-spheres method \citep{power_navarro_etal_2003}, starting at the center of mass of the zoom-in region. We use all zoom region resolution elements (high-resolution dark matter, gas, stars, and black holes if included) to determine the halo center. We then find the halo virial radius and mass using the \citet{bryan_norman_1998} mean enclosed halo density.

For the central galaxy stellar masses (Tab.~\ref{tab:sims}), we first find the galaxy center by taking all stellar resolution elements within $0.3 \Rvir$ of the halo center, then recalculating the shrinking spheres center using only this stellar mass. We then measure the stellar mass within $0.1 \Rvir$ of the galaxy center.

\subsection{Ion abundance calculation}

As mentioned above, we use the \citet{ploeckinger_schaye_2020} tables to calculate the ion fractions. We assume no dust depletion as dust is expected to be sputtered at the temperatures we are interested in \citep[e.g.,][]{tsai_mathews_1995}.
These use the same photo-ionizing UV/X-ray background as assumed for the radiative cooling in the FIRE-3 simulations we study. 
For consistency, we use these same tables for our FIRE-2 analysis. 
We interpolate these tables (linearly for ion fraction and redshift, and linearly in log space for temperature, density, and metallicity) to calculate ion fractions for a given resolution element. 

\subsection{Column density calculation}
\label{sec:cdmeth}

Given the number of ions in each resolution element, we now want to calculate the column density along various lines of sight. We do this by creating `column density maps'. We first define a volume and a line-of-sight direction. For the volume, we use a cube with side lengths (diameter) of $4 \Rvir$, centered on the halo center. For the line-of-sight directions, we simply use the $x$-, $y$-, and $z$-axes of the simulation. We note that these directions are random with respect to the galaxy and halo orientation.

We divide the plane perpendicular to the line of sight into pixels of size $(3 \pkpc)^2$. We then distribute the ions in each resolution element in the volume across these pixels based on each resolution element's position and size. This size is the MFM smoothing length.
We assume the ion distribution in each resolution element is described by a spherical Wendland C2 kernel \citep{wendland_1995}. 
Figs.~\ref{fig:m12maps} and~\ref{fig:m13maps} show examples of these column density maps.

\section{Ne~\textsc{viii} in FIRE}
\label{sec:simprop}

\subsection{Column densities and clumpiness}

We start our analysis by considering how \ion{Ne}{8} is distributed in the FIRE halos. To do this, we show images of some representative example m12 and m13 halos in Figs.~\ref{fig:m12maps} and~\ref{fig:m13maps}, respectively. For each halo mass sample, we show one halo, simulated using the different labelled physics models. These plots show the simulations at $z=1.0$, with the line of sight along the $z$-axis. All column densities are measured to $2 \Rvir$ in either direction along the line of sight from the halo center.

Comparing the m12 and m13 halos, the \ion{Ne}{8} in the m12 halos is fairly smoothly distributed, while this ion has a clumpier distribution in the m13 halos. Differences between the panels for the different physics models reflect some broader differences, but some of these differences are just a coincidence for these halos and this particular redshift. 

For the m12 halos, the FIRE-2 NoBH model generally predicts higher column densities than the FIRE-3 models analyzed here, and the FIRE-3 NoBH and BH models predict similar column density profiles across halos and redshifts. The FIRE-3 BH+CR model generally predicts lower column densities than the other two FIRE-3 models, but with only two halos for this physics model, this is somewhat uncertain.
In general, the FIRE-3 NoBH and BH models also predict similar column densities in m13 halos, while FIRE-3 BH+CR predicts lower column densities in the halo center. Note that the `hole' in the center of the m13h113 FIRE-3 BH+CR image is not a common feature in the m13 BH+CR model.

\begin{figure}
\centering
\includegraphics[width=\columnwidth]{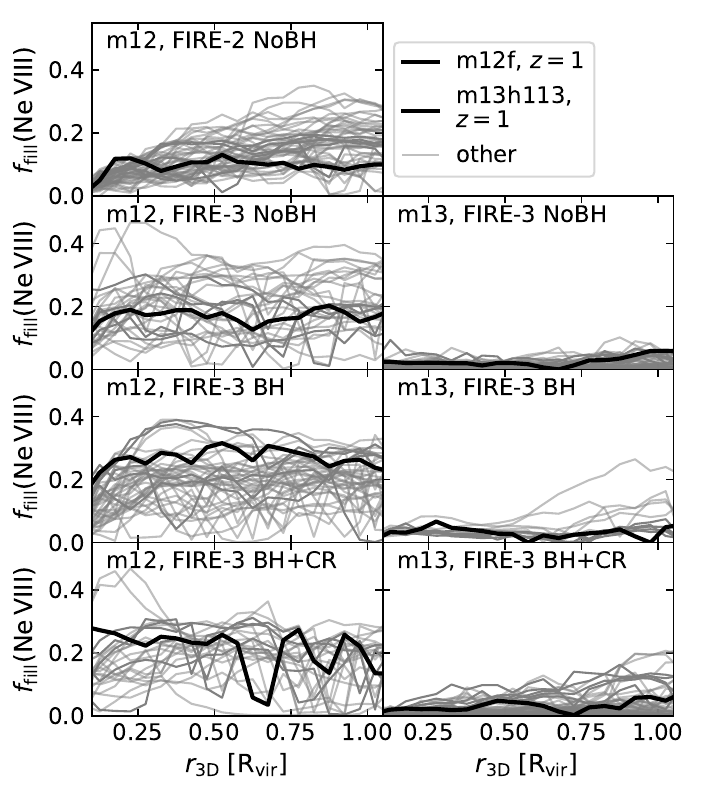}
\caption{Filling fraction of \ion{Ne}{8} in the m12 and m13 halos at redshifts~0.5, 0.6, 0.7, 0.8, 0.9, and~1.0 with different physics models (different panels). We calculate the filling fraction as the volume-weighted \ion{Ne}{8} number density, divided by the \ion{Ne}{8}-weighted \ion{Ne}{8} number density (this is the inverse of the clumping factor in eq.~\ref{eq:clumpiness}). Values lie between 0 and 1; higher values mean the distribution is smoother. The black curves show the halos from Figs.~\ref{fig:m12maps} and~\ref{fig:m13maps}. The \ion{Ne}{8} distribution in the m13 halos is typically clumpier than that in m12 halos. This is because the \ion{Ne}{8} ion fraction peaks roughly at the temperature of the volume-filling gas in the m12 halos, but this ion traces cooler, denser gas than the volume-filling phase in the hotter m13 halos.}
\label{fig:clumpiness}
\end{figure}

We explore the clumpiness of the m12 and m13 halos in more detail in Fig.~\ref{fig:clumpiness}. We show all the halos from our sample, and highlight the halos and redshift from Figs.~\ref{fig:m12maps} and~\ref{fig:m13maps} in black. We measure the volume-weighted \ion{Ne}{8} `smoothness' using the clumping factor
\begin{align}
\begin{split}
f_{\mathrm{c, V}} &= 
\frac{\langle \mathrm{n}(\mathrm{Ne}^{7+})^{2}\rangle_{\mathrm{V}}}{\langle \mathrm{n}(\mathrm{Ne}^{7+})\rangle_{\mathrm{V}}^{2}}
= \frac{\langle \mathrm{n}(\mathrm{Ne}^{7+})\rangle_{\mathcal{N}(\mathrm{Ne}^{7+})}}{\langle \mathrm{n}(\mathrm{Ne}^{7+})\rangle_{\mathrm{V}}},
\label{eq:clumpiness}
\end{split}
\end{align}
where the angle brackets indicate averages over resolution elements, weighted by volume $\mathrm{V}$ or number of \ion{Ne}{8} ions $\mathcal{N}(\mathrm{Ne}^{7+})$. We indicate the \ion{Ne}{8} number density as $\mathrm{n}(\mathrm{Ne}^{7+})$. The second equality shows that $f_{\mathrm{c, V}}$ is equal to the \ion{Ne}{8}-weighted \ion{Ne}{8} density divided by the volume-weighted \ion{Ne}{8} density.

In order to keep our range of values limited (between~0 and~1), we show $f_{\mathrm{fill}} = 1 \,/\, f_{\mathrm{c, V}}$ in Fig.~\ref{fig:clumpiness}. We note that this can be interpreted as the fraction of the CGM volume occupied by \ion{Ne}{8} in a simple case: if some fraction of the resolution elements all have the same ion density $\mathrm{n}_{\mathrm{Ne\,VIII}}$, and the rest contain no \ion{Ne}{8}, $1 \,/\, f_{\mathrm{c, V}}$ is the fraction of the volume which contains \ion{Ne}{8} ions.

We calculate $f_{\mathrm{fill}}$ in radial bins of size $0.05 \Rvir$ and show the smoothness as a function of (3D) distance to the halo center. We restrict this analysis to radial bins because the halos are significantly denser in their centers than in their outskirts. A clumping factor measured over the whole halo would therefore not only be sensitive to how smoothly the \ion{Ne}{8} is distributed, but also to how concentrated the ions are in the halo center. 

As suggested by the maps in Figs.~\ref{fig:m12maps} and~\ref{fig:m13maps}, the \ion{Ne}{8} in the m12 halos is more smoothly distributed than in the m13 halos. A similar plot to Fig.~\ref{fig:clumpiness} showing the volume-weighted gas density clumping factor shows that the m13 CGM is not intrinsically clumpier than the m12 CGM, meaning that the differences here are a result of how well (or poorly) \ion{Ne}{8} probes the smooth, virialized gas in the halo.

\subsection{Ne~\textsc{viii} and CGM properties}
\label{sec:ne8_cgm_props}

\begin{figure}
    \centering
    \includegraphics[width=\columnwidth]{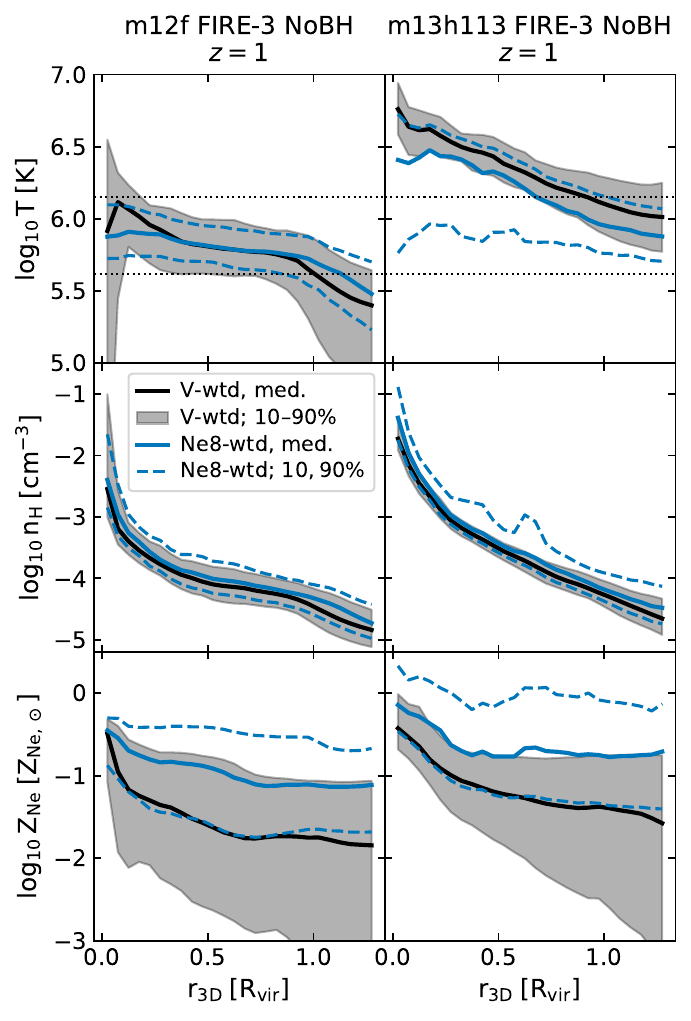}
    \caption{Temperature, density, and neon abundance profiles for the m12f FIRE-3 NoBH halo and the m13h113 FIRE-3 NoBH halo, both at $z=1$. These halos are shown in the second panel from the left in Fig.~\ref{fig:m12maps} and the left-hand panel in Fig.~\ref{fig:m13maps}, respectively. Blue curves show \ion{Ne}{8}-weighted medians (solid lines) and 10$^{\mathrm{th}}$ and 90$^{\mathrm{th}}$ percentiles of the \ion{Ne}{8}-weighted distributions. Black curves and shading show the same percentiles of the volume-weighted distributions. The dotted, horizontal lines in the top panels show the range of temperatures where, for gas in collisional ionization equilibrium (CIE), the \ion{Ne}{8} ion fraction is at least 0.1 times the maximum ion fraction in CIE.}
    \label{fig:3dprof}
\end{figure}

In Fig.~\ref{fig:3dprof}, we explore the temperature, density and metallicity\footnote{We normalize the neon mass fraction by the \citet{asplund_grevesse_etal_2009} neon solar mass fraction $10^{-2.90}$ here, following \citet{ploeckinger_schaye_2020}.} of the \ion{Ne}{8} in the CGM. We consider how \ion{Ne}{8} traces the warm-hot CGM in these halos, and whether the ionization of the neon is driven primarily by electron collisions or by photo-ionization.

In the top panels of Fig.~\ref{fig:3dprof}, we show the volume- and \ion{Ne}{8}-weighted temperature of an example m12 (left) and m13 (right) halo. We show the FIRE-3 NoBH model for the m12f and m13h113 halos at $z=1$; these halos are also shown in Figs.~\ref{fig:m12maps} and~\ref{fig:m13maps}. The volume-weighted median and 10$^\mathrm{th}$--90$^\mathrm{th}$ percentile temperatures (black) are shown in these top panels. The gas in the higher-mass m13 halos, is, as expected, hotter than the gas in the m12 halos. In the blue lines, we show the same percentiles of the \ion{Ne}{8}-weighted temperature distribution. In the m12 halos, the \ion{Ne}{8} traces roughly the same temperatures as the volume-filling gas phase. In the m13 halos, \ion{Ne}{8} preferentially traces gas that is cooler than the volume-filling phase, although some \ion{Ne}{8} is found in the volume-filling gas. 

The horizontal dotted lines indicate why. For gas in collisional ionization equilibrium (CIE), these indicate the temperature range where the \ion{Ne}{8} fraction is at least $0.1$ times the maximum ionization fraction in CIE. In the m13 halos, \ion{Ne}{8} makes up a relatively small fraction of the neon in the volume-filling phase, especially in the halo center. Though there is less gas at temperatures where the \ion{Ne}{8} fraction is high, what gas there is contains relatively many \ion{Ne}{8} ions. 
In the m12 halos, the volume-filling, virialized gas is at temperatures close to that where the \ion{Ne}{8} fraction reaches its maximum, so this ion traces that volume-filling gas well. 

In the middle panels of Fig.~\ref{fig:3dprof}, we show the volume- and \ion{Ne}{8}-weighted density percentiles in these same halos. In the m12 halos, the \ion{Ne}{8} traces the density of the volume-filling gas quite well, with only a slight bias to higher densities. This bias towards high densities is stronger in the m13 halos. This is in line with the clumpier \ion{Ne}{8} distribution in m13 halos relative to the m12 halos. 

Finally, we show the same distributions for metallicity (neon mass fraction) in the bottom panels of Fig.~\ref{fig:3dprof}. Unsurprisingly, the \ion{Ne}{8} preferentially traces neon-enriched gas. Though different halos and physics models have different metallicity profiles, the median \ion{Ne}{8}-weighted metallicity consistently lies roughly along the 90$^{\mathrm{th}}$ percentile of the volume-weighted metallicity profile. 

The top panels show that much of the \ion{Ne}{8} in both the m12 and m13 halos is at temperatures where \ion{Ne}{8} can be found in collisional ionisation equilibrium (CIE). 
(For m13 halos, a lot of the \ion{Ne}{8} is at temperatures above the CIE peak, but collisional ionization also dominates over photoionization in this regime.)
However, some gas around $\Rvir$ in some m12 halos is photo-ionized, and gas outside $\Rvir$ in m12 halos is typically photo-ionized.
Additionally, some of the CIE-temperature gas has low densities, where photo-ionization affects the fraction of neon in the Ne$^{7+}$ ionisation state. This occurs in m12 halos and in the outskirts of m13 halos.
In other words, much of the \ion{Ne}{8} in m12 halos, and in the outskirts of m13 halos, is not clearly in photo-ionization equilibrium (PIE) or CIE, but instead ionized by both photons and electrons. In both m12 and m13 halos, the gas closer to the halo center is densest, and most likely to be in CIE, while the lower-density gas near $\Rvir$ is most likely to be photo-ionized.

\subsection{What determines the CGM Ne~\textsc{viii} content?}

We have seen that \ion{Ne}{8} broadly traces warm-hot gas in the CGM, and is a
good tracer of the volume-filling gas in $\Mvir \sim 10^{12} \Msun$ halos. Now we explore which CGM properties determine the \ion{Ne}{8} content of the halos most strongly. We focus on the m12 halos here, since we this is where most of the observational constraints are (see \S\ref{sec:obscomp}).

\begin{figure}
\centering
\includegraphics[width=\columnwidth]{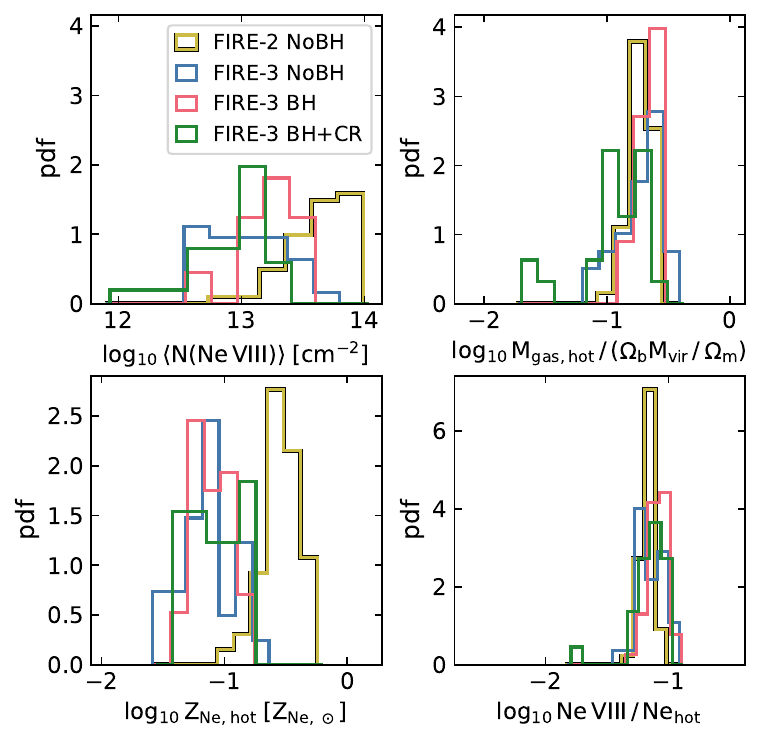}
\caption{Histograms of various CGM properties determining the \ion{Ne}{8} content of m12 halos. Different colors show different physics models. For each physics model, the histograms show probability density functions calculated over each halo run with that physics model, including data at redshifts~0.5, 0.6, 0.7, 0.8, 0.9, and~1.0 for each halo. The top left panel shows the mean CGM \ion{Ne}{8} column density.
As shown in Fig.~\ref{fig:m12maps}, the FIRE-2 NoBH mean column densities are higher than those of the FIRE-3 models. The top right panel shows the hot ($>10^{5}$~K) gas mass between 0.1 and 1~$\Rvir$, normalized to the halo baryon budget. The bottom left panel shows the mass-weighted metallicity of the ($>10^{5}$~K) gas between 0.1 and 1~$\Rvir$. The FIRE-2 NoBH model has a markedly higher metallicity than the FIRE-3 models, and the difference is large enough to explain the column density differences. The bottom right panel shows the \ion{Ne}{8} ion mass between 0.1 and 1~$\Rvir$, divided by the $\mathrm{T} > 10^{5}$~K neon mass in that same radial range. Note that for the FIRE-3 BH+CR model, we only have two halos available.}
\label{fig:cgmprophist}
\end{figure}

In the top left panel of Fig.~\ref{fig:cgmprophist}, we show the average \ion{Ne}{8} column densities in the halos simulated with the different physics models. We calculated these by dividing the number of \ion{Ne}{8} ions with a (3D) distance to the central galaxy between 0.1 and 1~$\Rvir$ by $\pi \Rvir^2$. This illustrates the magnitude of the column density differences we want to explain. The clearest difference is between the FIRE-2 NoBH model, and the three FIRE-3 models: the FIRE-2 model predicts higher \ion{Ne}{8} column densities than the FIRE-3 models (see also Fig.~\ref{fig:m12maps}). 

In the top right panel, we show histograms of the warm-hot ($> 10^5$~K) CGM gas mass, normalized to the halo baryon budget, for our m12 sample. We find warm-hot CGM masses of roughly 0.1--0.3 times the halo baryon budget. The warm-hot CGM mass distributions for the different physics models are similar.

In the bottom left panel of Fig.~\ref{fig:cgmprophist}, we compare the (mass-weighted mean) metallicity of the warm-hot CGM ($> 10^{5}$~K, 0.1--1~$\Rvir$) in our different physics models. Here, we see clear differences: the FIRE-2 NoBH simulations, with the highest \ion{Ne}{8} column densities, also have clearly higher metallicities than the FIRE-3 models. The differences are of the right size to explain the column density differences.

Finally, in the bottom right panel of Fig.~\ref{fig:cgmprophist}, we check the \ion{Ne}{8} ion fractions: the fraction of neon ions in the \ion{Ne}{8} ionization state. Note this is, at most, 0.24 in CIE. We specifically compare the \ion{Ne}{8} mass to the warm/hot neon mass: we divide all \ion{Ne}{8} at 0.1--1~$\Rvir$ by the neon at 0.1--1~$\Rvir$ and $T > 10^{5}$~K. We have checked that most \ion{Ne}{8} is at temperatures $>10^{5}$~K, so we are essentially measuring the \ion{Ne}{8} ion fraction in the warm/hot gas phase. We see a slightly lower \ion{Ne}{8} fraction in the FIRE-2 NoBH model, compared to the FIRE-3 models. The differences are, however, small compared to the metallicity and column density differences between the models.

We conclude that the \ion{Ne}{8} column densities are higher in the FIRE-2 simulations than in the FIRE-3 simulations due to the higher metallicity of the FIRE-2 NoBH warm-hot CGM in the halos we considered. We have checked that the neon abundance differences reflect the total metallicity differences between the models. The different models also include halos with a similar $\Mvir$ distribution, except that the two FIRE-3 BH+CR halos are at the higher end of the $\Mvir$ distribution. 

The (neon) metallicity differences are in part due to the $\approx 0.3 \dex$ lower neon yields in the FIRE-3 simulations, compared to FIRE-2 (see \S \ref{sec:simgenf3}). However, this is not enough of a change to fully explain the differences. The FIRE-3 models additionally produce m12 central galaxies with lower stellar masses than the FIRE-2 simulations at fixed halo mass.  
Lower stellar masses also mean lower metal production, and therefore a lower halo metal budget. 
In Appendix~\ref{app:m12plus}, we show that FIRE-3 NoBH halos with higher halo masses, and stellar masses comparable to those of the FIRE-2 NoBH halos, still produce lower \ion{Ne}{8} column densities than those  FIRE-2 NoBH halos.

\section{Comparison to Ne~\textsc{viii} observations}
\label{sec:obscomp}
\subsection{How to compare observed and simulated halos}
\label{sec:compmeth}

\begin{figure}
\centering
\includegraphics[width=\columnwidth]{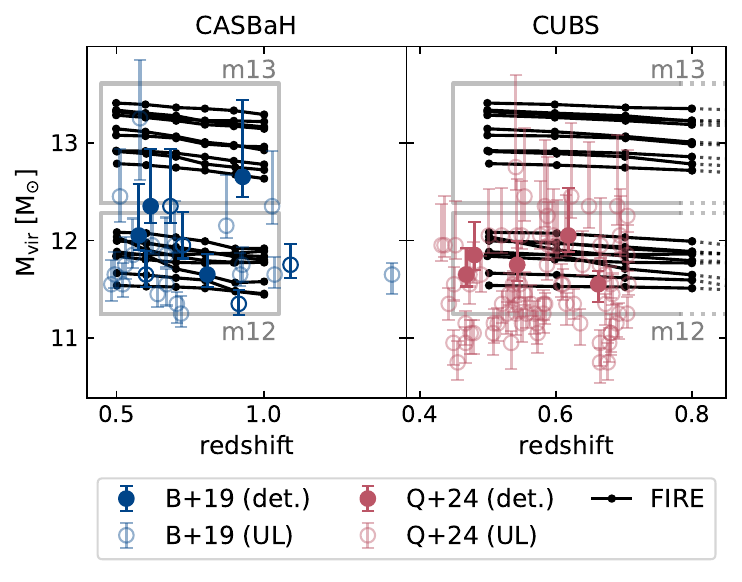}
\caption{Halo masses and redshifts for the simulated halos (black lines), compared to the distribution of halo masses and redshifts of the galaxies \citet{burchett_tripp_etal_2018} (left, B+19, CASBaH survey) and \citet{qu_chen_etal_2024_preprint} (right, Q+24, CUBS survey) found within 450~physical~kpc of their quasar sightlines. 
We calculate the observed galaxy halo masses as described in \S\ref{sec:compmeth}. Filled points indicate detections (det.), while lighter-colored, open points are upper limits (UL). Darker-colored, open points indicate CASBaH \ion{Ne}{8} detections that do not meet the stricter \citet{qu_chen_etal_2024_preprint} detection criteria.  The points show the halo mass with the highest probability, and error bars show the $1\sigma$ uncertainties.
The boxes indicate the ranges of mass and redshift where we consider data points good matches to the m12 and m13 halo samples.
}
\label{fig:obscomp_mzsel}
\end{figure}

\subsubsection{The CASBaH data}
\label{sec:CASBaH_data} 
We compare our FIRE predictions for \ion{Ne}{8} column densities to the observations of \citet{burchett_tripp_etal_2018}. Specifically, we use some of the quantities in their tab.~1. Briefly, their data combines a set of absorption spectra, which they searched for \ion{Ne}{8} absorption, with the CASBaH (COS Absorption Survey of Baryon Harbors) galaxy catalogue \citep{prochaska_burchett_etal_2019}. 
This catalogue includes galaxy spectroscopic redshifts, and galaxy stellar masses determined with spectral energy distribution (SED) fits and those redshifts.

These galaxies are matched to \ion{Ne}{8} absorption systems\footnote{\citet{burchett_tripp_etal_2018} grouped individual absorbers into absorption systems if they were within $\approx 600 \kmps$ of each other. The redshift attributed to the absorption system is its central velocity.} found along the sightline based on impact parameter ($< 450 \pkpc$) and velocity difference ($< 500 \kmps$). They select the closest galaxy to the line of sight if there are different options. 

For galaxies without corresponding absorbers, a $3\sigma$ upper limit on the \ion{Ne}{8} column density was determined. That means we can treat this as a survey where any sightline passing close enough to a galaxy was searched for absorption by that galaxy's halo. Therefore, medians and percentiles of the column densities around halos at a given impact parameter are a good comparison for the measurements and upper limits in the data.

\subsubsection{The CUBS data}
\label{sec:CUBS_data} 

We also compare our \ion{Ne}{8} column densities to the observations of \citet{qu_chen_etal_2024_preprint}. Their analysis is part of the CUBS survey \citep{chen_zahedi_etal_2020_cubs1}.
\citet{qu_chen_etal_2024_preprint} searched for \ion{Ne}{8} absorption at redshifts $0.43 \lesssim z \lesssim 0.72$ in quasar spectra, and compiled a catalogue of galaxies close to their quasar sightlines. 
The CUBS and CASBaH quasar sightline samples do not overlap \citep{burchett_tripp_etal_2018, chen_zahedi_etal_2020_cubs1}, meaning the two datasets are independent of each other.

\citet{qu_chen_etal_2024_preprint} determined their galaxy redshifts spectroscopically. Stellar masses were calculated based on three photometric bands. The relation \citet{qu_chen_etal_2024_preprint} used for this was fitted to a calibration sample of galaxies, with stellar masses derived from spectral energy distribution fits. Scatter in the fitted relation causes an estimated systematic error in the stellar masses of $\approx 0.2 \dex$. \citet{qu_chen_etal_2024_preprint} calculate halo masses $\mathrm{M}_{\mathrm{200c}}$\footnote{The halo mass $\mathrm{M}_{\mathrm{200c}}$ and its radius $\mathrm{R}_{\mathrm{200c}}$ are defined such that the mean density in the halo is 200 times the critical density at the halo redshift.} from their stellar masses.

Galaxies were matched to absorbers in two steps. First, all galaxies within  within $3 \us \mathrm{R}_{\mathrm{200c}}$ of the sightline were selected. Galaxies were then grouped if they were within 1~Mpc and $500 \kmps$ of each other.
Second, absorbers were matched to (groups of) galaxies if they had a velocity separation $\leq 1000 \kmps$. 
For galaxies and groups without measured column densities, $2\sigma$ upper limits on the \ion{Ne}{8} column densities were determined.
Within groups, a single galaxy is selected as the absorber counterpart. Based on correlations with \ion{O}{6} absorption, \citet{qu_chen_etal_2024_preprint} chose the galaxy with the smallest impact parameter in units of $\mathrm{R}_{\mathrm{200c}}$ as the counterpart.

\citet{qu_chen_etal_2024_preprint} compared their data to the CASBaH \ion{Ne}{8} data. They examined the CASBaH absorbers, and flag five of them as being significantly contaminated by interlopers. These interlopers are other absorption lines at other redshifts, which overlap with the reported \ion{Ne}{8} absorption lines. 
Specifically, the difference is in the treatment of absorption systems with significant contamination in one of the \ion{Ne}{8} 770, 780~{\Ang} doublet components. \citet{burchett_tripp_etal_2018} consider the candidate absorption systems to be detections, based in part on the other absorption lines detected at the same redshift as the candidate \ion{Ne}{8}. \citet{qu_chen_etal_2024_preprint} instead treat these as upper limits, based on a more conservative detection criterion where both doublet lines must be more clearly identified.

In their further analysis, \citet{qu_chen_etal_2024_preprint} treat the `flagged' measurements from \citet{burchett_tripp_etal_2018} as upper limits in order to get a sample with consistent detection criteria. We will indicate these data points with open symbols in our plots. 
Conversely, some of the 2$\sigma$ upper limits in the CUBS data might have been considered detections at a similar column density by \citet{burchett_tripp_etal_2018}.
In this work, we do not attempt to resolve the observational question of exactly how robust the flagged detections are. Instead, below we consider how our main conclusions depend on whether they are treated as detections vs.\ upper limits. 
As we will see, our overall conclusions regarding the median profiles are largely unaffected, though the quantitative degree by which the simulations underpredict the scatter is increased if all the \cite{burchett_tripp_etal_2018} detections are assumed to be robust.

\begin{figure*}
\centering
 \includegraphics[width=\textwidth]{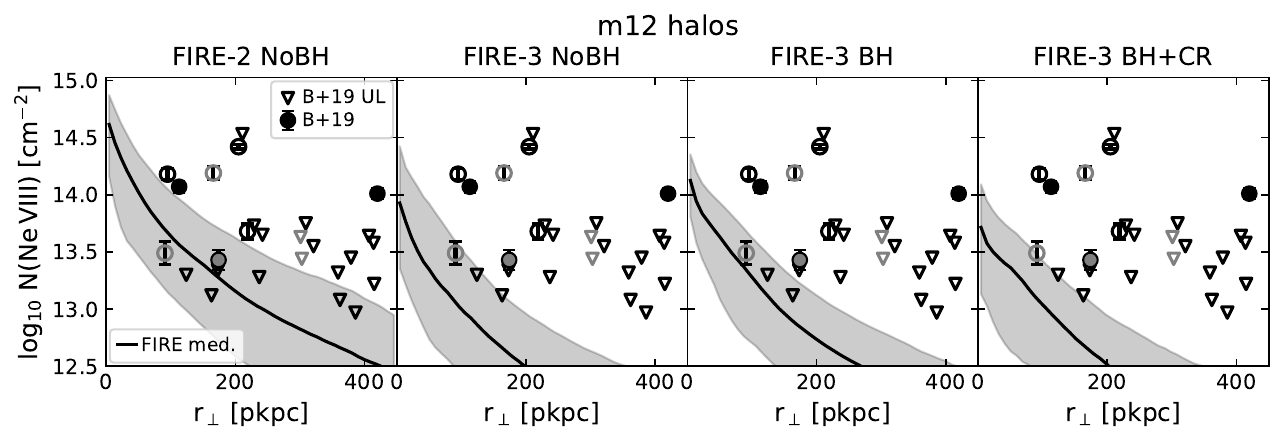}
\includegraphics[width=0.75\textwidth]{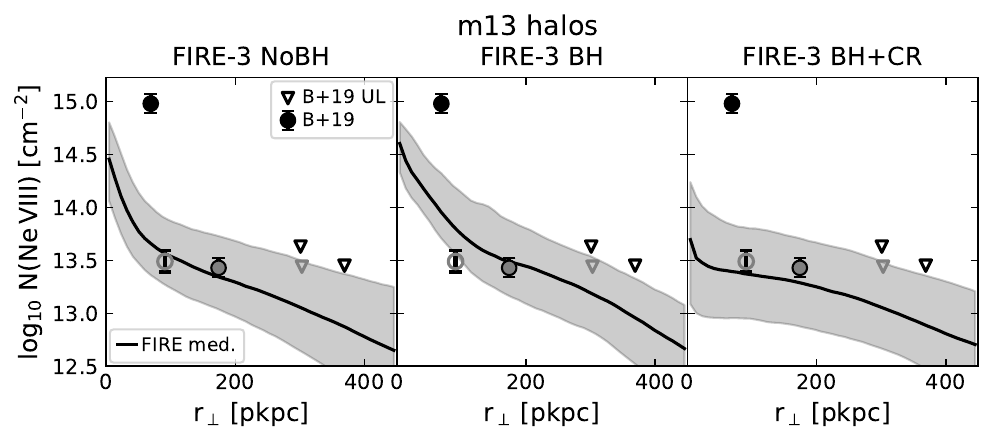}
\caption{\ion{Ne}{8} column densities predicted from our simulations, compared to the values (points with $1\sigma$ error bars) and upper limits ($3 \sigma$, open triangles) found by \citet[B+19]{burchett_tripp_etal_2018} in the CASBaH survey. The black points show observed galaxies which match the m12 (top panels) and m13 (bottom panels) halo mass and redshift range well, and gray points show plausible matches (see text for the precise selections).
Open circles indicate CASBaH \ion{Ne}{8} detections that do not meet the stricter \citet{qu_chen_etal_2024_preprint} detection criteria; \citet{qu_chen_etal_2024_preprint} treat these measurements as upper limits. For the simulation data, solid lines show median column densities as a function of impact parameter, while the shaded region shows the percentile 10--90 range of the column density at a given impact parameter. Different panels show different physics models, indicated above each panel. The m12 FIRE-2 NoBH model medians are consistent with the data, i.e., the median lies above $\lesssim 50$~per~cent of the detections and upper limits. The data do not allow for much higher median column densities than this model predicts within $\approx 200$~kpc.  We note that some of these highest values exceed groups of upper limits at similar impact parameters and halo masses. However, the FIRE-2 NoBH model may underpredict the highest measured values, depending on how we interpret the open circles. All three m12 FIRE-3 models underpredict the measured values. The m13 halos match two of the observed column densities, but underpredict the largest measured value.}
\label{fig:ne8_obscomp_b19}
\end{figure*}

\begin{figure*}
\centering
\includegraphics[width=\textwidth]{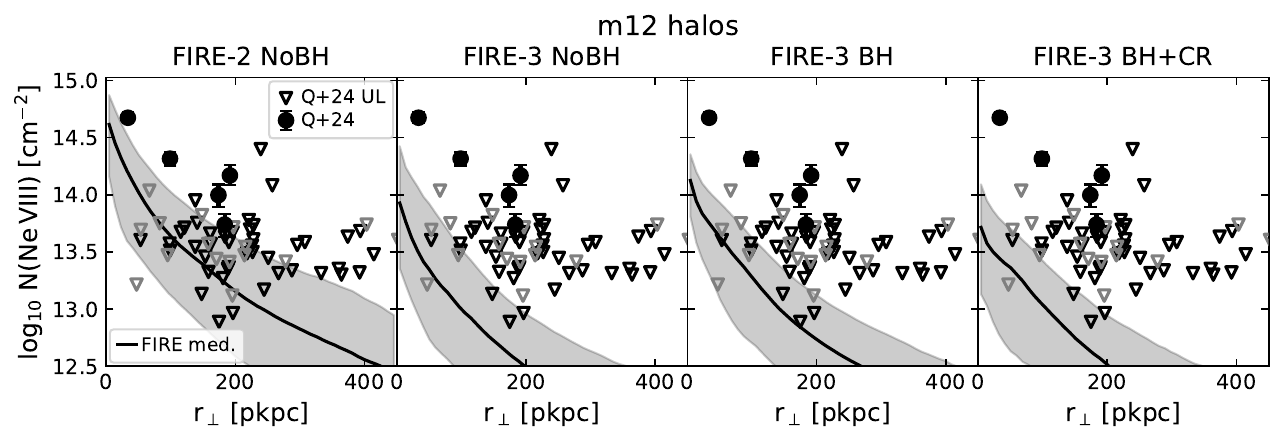}
\includegraphics[width=0.75\textwidth]{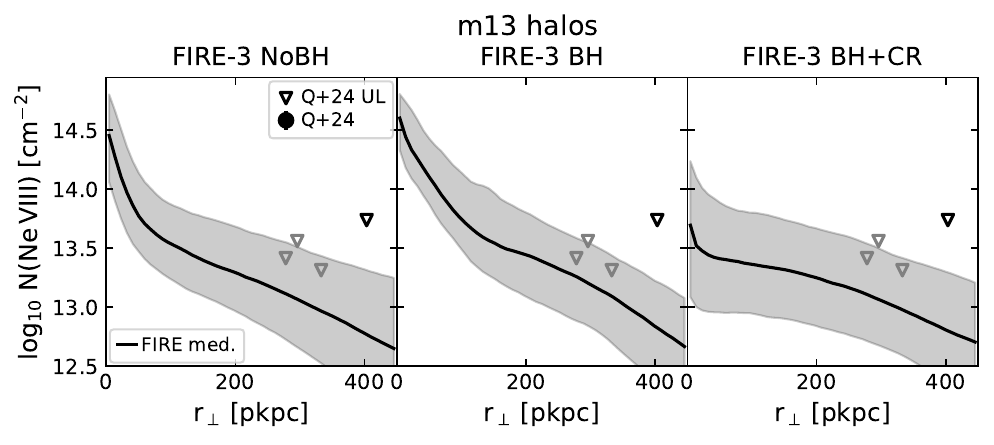}
\caption{As. Fig.~\ref{fig:ne8_obscomp_b19}, but comparing the m12 (top row) and m13 (bottom row) halos to the \citet[Q+24]{qu_chen_etal_2024_preprint} CUBS survey observations. The error bars on their measurements and their upper limits are both $2\sigma$. Again, the FIRE-2 NoBH m12 medians are consistent with the data, and about at large as the data allow within $\approx 100$~kpc. 
The FIRE-3 models underpredict the highest measured column densities. 
The m13 halos are consistent with the measured upper limits.}
\label{fig:ne8_obscomp_q23}
\end{figure*}

\subsubsection{Halo masses}
\label{sec:hmcalc}

Our goal is to compare the FIRE CGM to the CGM observations of \citet{burchett_tripp_etal_2018} and \citet{qu_chen_etal_2024_preprint}.
The \citet{burchett_tripp_etal_2018} observations are of \ion{Ne}{8} CGM absorbers around isolated galaxies with stellar masses $\sim 10^9$--$10^{11} \Msun$. For CGM comparisons to observations, matching the halo masses is important, because the halo mass determines the CGM baryon budget, extent, and virial temperature, i.e., three important factors determining the ion content of the warm/hot phase. \citet{burchett_tripp_etal_2018} measured the stellar masses of their galaxies, and used a modified version \citep{burchett_tripp_etal_2016} of a function fit to abundance matching simulations \citep{moster_naab_white_2013} to obtain halo masses based on these stellar masses. 

\citet{burchett_tripp_etal_2018} do not, however, report errors on the halo masses. Since there is substantial scatter in the stellar-mass-halo-mass relation, these could be quite large. Therefore, we used a UniverseMachine stellar and halo mass catalog\footnote{\url{https://halos.as.arizona.edu/UniverseMachine/DR1/SMDPL_SFR/}, accessed 2023-06-15} \citep{behroozi_wechsler_etal_2019} to reconstruct the stellar-mass-halo-mass relation and look into some sources of errors. 
Specifically, we used the catalogues based on the Small MultiDark–Planck (SMDPL) dark-matter-only simulation \citep{klypin_yepes_etal_2016, rodriguez-puebla_behroozi_etal_2016}, which has a volume of $(0.4\us\mathrm{Gpc} \,/\,h)^3$. \citet{behroozi_wechsler_etal_2019} analysed these using the \textsc{rockstar} halo finder \citep{behroozi_wechsler_wu_2013} and \textsc{Consistent Trees} \citet{behroozi_wechsler_etal_2013} merger trees. 
We only used UniverseMachine galaxies flagged as centrals for this analysis.
From this catalogue, we estimate the true halo mass probability distribution given an observed galaxy stellar mass and redshift. We do this by finding the distribution of halo masses in the catalogue for a given stellar mass, and then combine that with the probabilities of different true stellar masses from the observations. 

This different method makes a difference for the inferred most likely halo masses (i.e., those with the highest probability density), which is quite large for central galaxies with stellar masses $\gtrsim 10^{10.5} \Msun$. Accounting for scatter in the stellar-mass-halo-mass relation also leads to larger uncertainty estimates, again mostly at $ \Mstellar \gtrsim 10^{10.5} \Msun$.
We give a more detailed explanation of our method, and our recommendations for how to estimate these halo masses, in Appendix~\ref{app:smhm}.

\citet{qu_chen_etal_2024_preprint} did determine their halo masses using a stellar-mass-halo-mass relation from \citet{behroozi_wechsler_etal_2019}. However, they use an overdensity of 200 relative to the critical density to define halos, where we use the \citet{bryan_norman_1998} density definition, and they do not report errors on their halo masses. We therefore apply the same method as described above to the CUBS stellar masses to calculate the most likely halo mass, and its errors, as we applied to the CASBaH data.

\subsection{Column density profiles}
\label{sec:column_profiles}

In Fig.~\ref{fig:obscomp_mzsel}, we show the masses 
(\citeauthor{bryan_norman_1998} \citeyear{bryan_norman_1998} definition)
and redshifts of the halos in our simulated sample, and the observed samples of  \citet[CASBaH]{burchett_tripp_etal_2018} and \citet[CUBS]{qu_chen_etal_2024_preprint}. 
The black lines indicate the masses and redshifts of different halos. If a halo was simulated with different physics models, we only show the masses and redshifts for one of them. This is for legibility, and is reasonable because the halo masses differ little between these simulations with different physics models. 
For the observed data points, we show the most likely halo masses (mode of the probability density function) with points. 
The $1\sigma$ uncertainty ranges show the log halo mass range enclosing a probability of 68\%.

We only show data points with impact parameters $<450 \pkpc$ in Fig.~\ref{fig:obscomp_mzsel}. For the CASBaH data, this matched their own galaxy selection, but this excludes some of the CUBS upper limits, especially at high halo masses. We focus on these impact parameters because, for the m12 halos, this $450 \pkpc$ is $\approx 2$--$3\Rvir$, which is towards the edge of the zoom-in region of the simulations. For the m13 halos, larger impact parameters might be acceptable given the volume of the zoom region, but the CUBS upper limits there are not very constraining, so we focus on the impact parameters where \ion{Ne}{8} was detected.

The gray boxes are drawn at the minimum and maximum halo mass and redshift in the m12 and m13 samples, with a redshift margin of 0.05, and a log halo mass margin of 0.2.
An observed galaxy is considered a good match to the m12 or m13 FIRE sample if its best-estimate halo mass and redshift fall within the solid box. In Figs.\ \ref{fig:ne8_obscomp_b19} and~\ref{fig:ne8_obscomp_q23}, these halos are shown with black markers.
We will also compare FIRE predictions to data points in larger sections of this plot, using larger margins of 0.1 in redshift and 0.4 in log halo mass. 
We consider data points in these larger regions to be plausible matches. In Figs.\ \ref{fig:ne8_obscomp_b19} and~\ref{fig:ne8_obscomp_q23}, these halos are shown with grey markers.
These margins for the most likely halo masses are fairly consistent with whether the halo mass uncertainties include the simulated ranges.

In Figs.~\ref{fig:ne8_obscomp_b19} and~\ref{fig:ne8_obscomp_q23}, we compare the m12 and m13 halo column densities in the FIRE-2/3 simulations to the measurements of \citet{burchett_tripp_etal_2018} and \citet{qu_chen_etal_2024_preprint}. 
For all physics models and for the m12 and m13 halos, the median column densities are consistent with the observed upper limits. Though some upper limits lie below the median curves, at least about half the upper limits and measurements lie above the median curves in all physics models we consider, in both the m12 and m13 halos. However, the median and 90$^{\rm th}$ percentile profiles for all models underpredict the highest reported column density values.

\begin{figure*}
\centering
\includegraphics[width=0.85\textwidth]{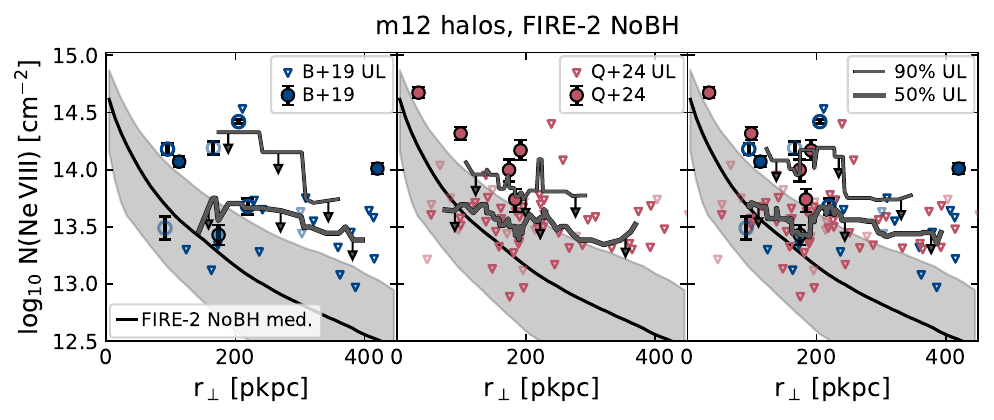}
\caption{As Figs.~\ref{fig:ne8_obscomp_b19} and~\ref{fig:ne8_obscomp_q23}, but focusing on the FIRE-2 NoBH m12 halos and estimating upper limits to the median and 90$^{\mathrm{th}}$ percentile profiles. The left-hand panel compares the CASBaH \citep{burchett_tripp_etal_2018} \ion{Ne}{8} column density measurements (circles) and $3\sigma$ upper limits (triangles) as a function of impact parameter to the median (solid, black line) and 10$^{\mathrm{th}}$--90$^{\mathrm{th}}$ percentile range (gray shaded area) predicted from our FIRE-2 NoBH m12 simulations. 
Open/filled and lighter/darker symbols have the same meaning as in Fig.~\ref{fig:ne8_obscomp_b19}. 
The middle panel shows the same as the left-hand panel, but instead of the CASBaH data, we show the CUBS data \citep{qu_chen_etal_2024_preprint} with $2\sigma$ upper limits.
In the right-hand panel, we show both datasets together.
We estimate conservative upper limits on the median and 90$^{\mathrm{th}}$ percentile of the observed column densities as a function of impact parameter in each panel (see the text in \S \ref{sec:column_profiles} for details). 
We show these running percentiles of the measured column densities and upper limits with the dark gray lines. 
From left to right, we use 8 (15), 10 (20), and 15 (30) points to calculate the running median (90$^{\mathrm{th}}$ percentile).}
\label{fig:ne8_obscomp_fire2}
\end{figure*}

For the FIRE-3 m12 halos, the observed (non-upper-limit) column densities are mostly above the 90\textsuperscript{th} percentile of the simulated values at the same impact parameter. For the FIRE-3 models, overall more than 10~per~cent of the observed column densities in the CUBS and CASBaH data (including sufficiently tight upper limits) within $\sim 0.5$--$1\Rvir$ lie above the 90\textsuperscript{th} percentile of the simulation predictions. Because most observations are upper limits, we cannot rule out the lower median column densities predicted by these models, but it is clear that these FIRE-3 models do not produce enough high \ion{Ne}{8} column densities in m12 halos to match the detections.

The FIRE-3 m13 halos match the two measured \citet{burchett_tripp_etal_2018} column densities that plausibly match the m13 halo mass range, but all models underpredict the column density of the one measurement which clearly matches the m13 halo mass range (the one with \ion{Ne}{8} column $\approx 10^{15}$ cm$^{-2}$, which is the highest measured value in the \citet{burchett_tripp_etal_2018} sample). 

Overall, the FIRE-2 NoBH simulations match the observations better than the FIRE-3 simulations. 
We show a more detailed comparison of the FIRE-2 halos with observations in Fig.~\ref{fig:ne8_obscomp_fire2}. 
The left panel compares the simulations to the data from \cite{burchett_tripp_etal_2018} only, the middle panel to the data from \cite{qu_chen_etal_2024_preprint} only, and the right panel to the combined dataset.
In that figure, we quantify the comparison between the FIRE-2 simulations and observations further by estimating conservative upper limits on the median and 90$^{\mathrm{th}}$ percentile of \ion{Ne}{8} column densities as a function of impact parameter (dark gray curves). We do this by calculating percentiles from the measurements and upper limits together, treating all upper limits as actual detections at the highest column densities allowed. 
Although the treatment is simple, it is instructive because it quantifies how high the observed profile percentiles can be in the most optimistic case. 
This exercise also allows us to assess the impact of treating the detections from \cite{burchett_tripp_etal_2018} that did not meet the stricter \cite{qu_chen_etal_2024_preprint} detection citeria (see \S\ref{sec:CUBS_data}) as detections vs. upper limits.

More specifically, we calculate running percentiles. We order the data by impact parameter, then take the first $n$ points, and calculate the desired percentile of their column densities and the median of their impact parameters. This is the first point on the upper limit curve. We add points by taking the same percentiles of data points 1 through $n + 1$, 2 through $n + 2$, etc.
For these calculations, we used all data points shown in each panel, i.e., data that matches the simulated halo mass range well and reasonably. We chose the number of points $n$ to use for each dataset and percentile by looking for values that produced reasonably smooth curves, without sacrificing too much of the impact parameter range represented in the data.

From this, we draw the following conclusions:
\begin{itemize}
\item In the inner impact parameter bins, where the observations are most constraining as this is where we expect the highest columns, the conservative upper limits to the observed medians are similar to the predicted medians from FIRE-2 for the m12 halos. 
Therefore, the true median profile cannot be significantly higher than predicted by FIRE-2. 
This is so regardless of whether we consider only the data from \cite{burchett_tripp_etal_2018}, only the data from \cite{qu_chen_etal_2024_preprint}, or the combined dataset. 
Furthermore, this conclusion holds independent of whether all the detections reported by \cite{burchett_tripp_etal_2018}, including the ones that did not meet the stricter detection criteria of \cite{qu_chen_etal_2024_preprint}, are robust or best interpreted as upper limits.

\item If we focus on the \cite{burchett_tripp_etal_2018} data, the innermost running median point from the CASBaH data in Fig.~\ref{fig:ne8_obscomp_fire2} is calculated from 5 column densities reported as detections by the authors, plus three upper limits that lie below the 5 detections. 
Therefore, the estimated ``upper limit'' on the running median in that region is actually a direct estimate of the median itself. 
This indicates that if all the detections reported by \cite{burchett_tripp_etal_2018} are robust, the true median column densities cannot be substantially lower than the FIRE-2 predictions at impact parameter $\approx 150$~kpc. 
Thus, the FIRE-2 simulations analyzed predict a median \ion{Ne}{8} column density profile consistent with the observations at that impact parameter.
This also means that if the FIRE-2 model underpredicts the 90$^\mathrm{th}$ percentile column density profile, it is the column density scatter that is underestimated, not the median.

\item Focusing on the estimated upper limits to the 90$^\mathrm{th}$ percentiles, the conclusions depend more sensitively on whether we use the \cite{burchett_tripp_etal_2018} vs. the \cite{qu_chen_etal_2024_preprint} data. 
For the \cite{qu_chen_etal_2024_preprint} data in the middle panel, we see that in the inner halo, the upper limit to the 90$^\mathrm{th}$ percentile is similar to the  90$^\mathrm{th}$ percentile from FIRE-2. 
This suggests that the FIRE-2 simulations may be entirely consistent with the CUBS data for the \ion{Ne}{8} column density profile, including its scatter. 
For the \cite{burchett_tripp_etal_2018} data, the 90$^\mathrm{th}$ percentile upper limits are higher due to the high column detections. If all the reported detections are assumed to be robust, the data imply that while FIRE-2 predicts a consistent median profile, the simulations significantly underestimate the scatter of the distribution. 
\end{itemize}

\section{Analytical modeling}
\label{sec:anmodel}

Some of the \citet{burchett_tripp_etal_2018} and \citet{qu_chen_etal_2024_preprint} column densities are higher than predicted in any of our FIRE simulations. Some of these, such as the $\approx 10^{14} \pcmsq$ absorption system at an impact parameter of $\approx 400$~pkpc, might be outliers or mismatched with their galaxies: this data point lies above many upper limits at smaller impact parameters. However, the simulations may also be imperfect models.

To better understand the halo properties needed to explain the observed \ion{Ne}{8} column densities,
we predict column densities using a simple analytical model. The goals here are (1) to see how we expect the column densities to depend on the halo mass, size, and temperature, and (2) to check whether the high \citet{burchett_tripp_etal_2018} and \citet{qu_chen_etal_2024_preprint} column densities can be explained
with reasonable CGM masses and metallicities. 

In our phenomenological model, derived in more detail in Appendix~\ref{app:plmodel}, we assume the various thermodynamical properties of the halo gas are 
power law functions of distance to the halo center.
We describe the gravitational potential as 
\begin{equation}
v_{\mathrm{c}} \propto r^{m},
\label{eq:mdef}
\end{equation}
where  $v_{\mathrm{c}}$ is the circular velocity, $r$ is the distance to the halo center, and $m$ is an exponent to be determined.
We additionally try a few different values of the entropy profile logarithmic slope 
\begin{equation}
l \equiv \frac{\mathrm{d}\, \ln K}{\mathrm{d}\, \ln r},
\label{eq:ldef}
\end{equation}
where $K$ is the entropy $K = k \mathrm{T}\mathrm{n}^{-\frac{2}{3}}$, $k$ is the Boltzmann constant, and $\mathrm{n}$ is the total free particle number density.  

We also assume hydrostatic equilibrium, yielding a power-law temperature profile $\mathrm{T} \propto r^{2m}$, with a normalization depending on $l$, $m$, and $v_{\mathrm{c}}$ at the virial radius: $v_{\mathrm{c}}(\Rvir) = \sqrt{\frac{G \Mvir}{\Rvir}}$. The implied density 
profile is $\mathrm{n}_{\mathrm{H}} \propto r^{-\frac{3}{2} l + 3m}$, and we set the normalization in the fiducial model by requiring that the mass of this CGM gas is the cosmic baryon fraction of the halo mass $\Mvir$ (at  $0.1$--$1\Rvir$).

\begin{figure*}
\centering
\includegraphics[width=\textwidth]{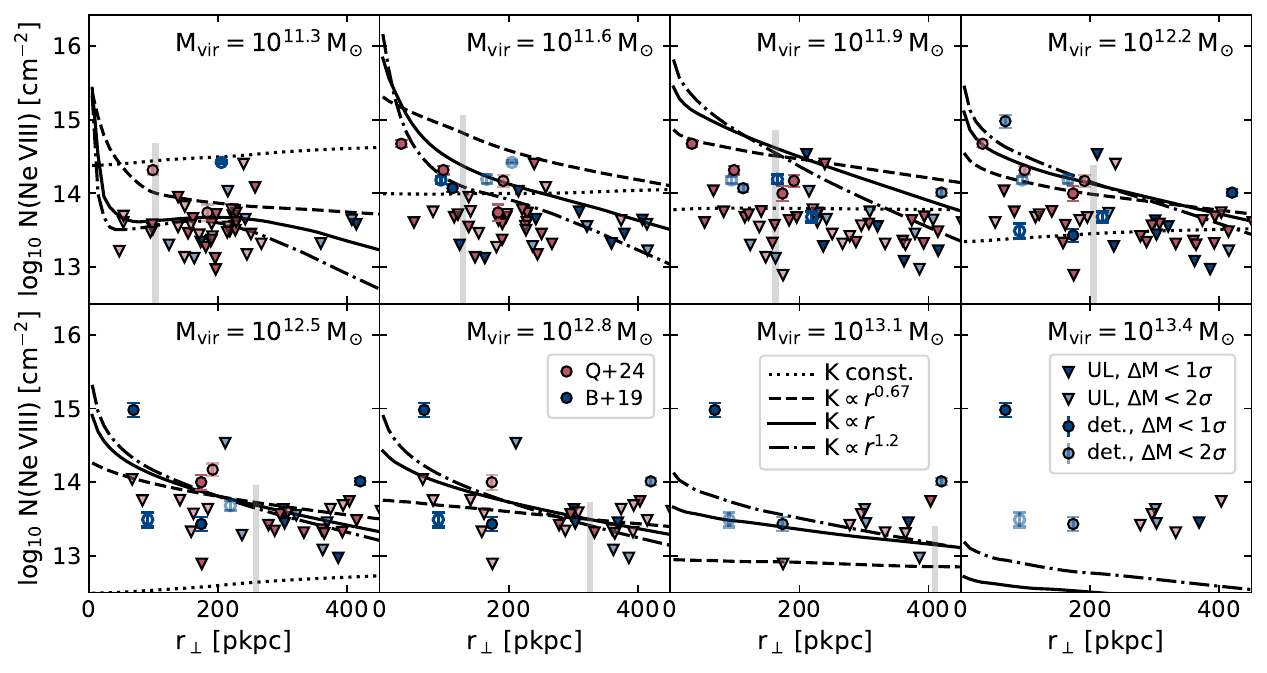}
\caption{\ion{Ne}{8} column density as a function of impact parameter for some power law CGM models with a warm/hot CGM mass budget $\mathrm{f}_{\mathrm{CGM}} = 1$ and a uniform metallicity of $0.3 \us \mathrm{Z}_{\odot}$. Different logarithmic entropy slopes are shown in different line styles. We use a circular velocity slope $m=-0.1$ (eq.~\ref{eq:mdef}). Each panel shows one halo mass, indicated in the upper right of the panel. We show how the measurements (points with error bars) and upper limits (downward-facing triangles) of \citet[CASBaH, blue]{burchett_tripp_etal_2018} and \citet[CUBS, red]{qu_chen_etal_2024_preprint} compare to these models. The brighter measurement and upper limit points have halo masses within $1 \sigma$ of the halo mass shown in the panel, while points with lighter colors have halo masses between $1$ and $2 \sigma$ from the model halo masses. The open, blue circles indicate CASBaH \ion{Ne}{8} detections that do not meet the more conservative detection criteria of \citet{qu_chen_etal_2024_preprint}. Vertical gray lines indicate the virial radius; for $\Mvir = 10^{13.4} \Msun$, this radius is outside the plotted range.
All measurements but one are achievable in some variation of a power law model with a baryonically closed CGM and a uniform metallicity of $Z=0.3 \us \mathrm{Z}_{\odot}$, and many upper limits require lower CGM gas fractions and/or metallicities.}
\label{fig:plmodels}
\end{figure*}

Assuming the warm/hot CGM contains the cosmic baryon fraction of the halo in mass is an optimistic (but not extreme) assumption.
However,
we want to explore an optimistic model here to assess the consistency of the high column density measurements with the cosmic baryon budget. 
We will also explore lower values of 
$\mathrm{f}_\mathrm{CGM} = \mathrm{M}_{\mathrm{CGM}} \,/\, (\Omega_{\mathrm{b}} \Mvir \,/\, \Omega_{\mathrm{m}})$, where $\mathrm{M}_{\mathrm{CGM}}$ is the (warm/hot) CGM mass between 0.1~and $1 \Rvir$.

In calculating the column densities for these profiles, we assume a metallicity of 0.3~$\mathrm{Z}_{\odot}$ in our fiducial model. 
We calculate ion fractions using the \citet{ploeckinger_schaye_2020} ionization tables, in the same way as for the FIRE predictions.

Next, we need to decide on a circular velocity profile slope $m$. We base these choices on the $v_{\mathrm{c}}$ slopes in NFW halo mass profiles \citep{NFW_1997}. In these profiles, the circular velocity profile slope $\frac{\mathrm{d} \log v_{\mathrm{c}}}{\mathrm{d} \log r}$ depends on $\frac{r}{r_{s}}$, where $r_s$ is a scale radius. We roughly estimate the relation between $r_s$ and $\Rvir$ using tab.~3 of \citet{dutton_maccio_2014}; this yields concentrations  $\Rvir\,/\,r_{s} \approx 6$--$9$ for halos with $\Mvir = 10^{11.5}$--$10^{13.5} \Msun$ around redshifts $0.5$ and~$1$, decreasing with halo mass. This yields slopes  $\frac{\mathrm{d} \log v_{\mathrm{c}}}{\mathrm{d} \log r} \approx -0.2$ to $0.25$ in the radial range $0.1$--$1 \Rvir$. The slopes decrease (become more negative) with halo center distance and increase (become more positive) with halo mass.
We choose a fiducial circular velocity slope $m=-0.1$, which is reasonable for the large radii (in virial radius units) where many of the measurements most likely lie.

For the entropy slopes, we show a range of plausible values. 
\citet{zhu_xu_2021} find outer halo entropy slopes $\approx 1.2$ from an observational group/cluster sample.
\citet{babyk_mcnamara_etal_2018} study entropy profiles in halos over a range of masses, using observations around early-type galaxies and clusters. At small radii, their findings are consistent with a slope of $\frac{2}{3}$, and at larger radii with $1.1$.
An entropy slope of zero can occur in e.g., cluster centers \citep[e.g.,][]{donahue_horner_etal_2006}.
Although the halo mass range relevant for \ion{Ne}{8} observations is mostly below groups and clusters, we note that an entropy slope $\approx 1$ is representative of cooling flow models for Milky Way-mass halos \citep[][Sultan et al., in prep.]{stern_fielding_etal_2019}. We show show this slope as well.

We note that not all entropy and circular velocity profile combinations are possible. If both slopes are zero, there is no pressure gradient, the assumption of hydrostatic equilibrium becomes meaningless, and the temperature profile normalization is not constrained. If $l$ is too large and/or $m$ is too negative, the density slope becomes too steep and the mass integral within the virial radius diverges.

We show these model \ion{Ne}{8} profiles in Fig.~\ref{fig:plmodels}, alongside the observations. We note that the upper limits are $2\sigma$ for \citet[CUBS, red]{qu_chen_etal_2024_preprint} and $3\sigma$ for \citet[CASBaH, blue]{burchett_tripp_etal_2018}. All measured column densities but one can be achieved in a power-law CGM model, containing the cosmic baryon fraction of the halo mass in the warm-hot CGM phase, and with a uniform metallicity of $0.3 \us \mathrm{Z}_{\odot}$. Though this is a somewhat optimistic model, it does not require unphysical CGM gas masses or extreme CGM metallicities. We note that the validity of these power-law models beyond $\Rvir$ is uncertain.

We also note that the halos producing the highest column densities are within (though at the high end of) the m12 mass range (see Fig.~\ref{fig:obscomp_mzsel}), meaning these halos are represented in our simulation sample.

\begin{figure*}
\centering
\includegraphics[width=0.7\textwidth]{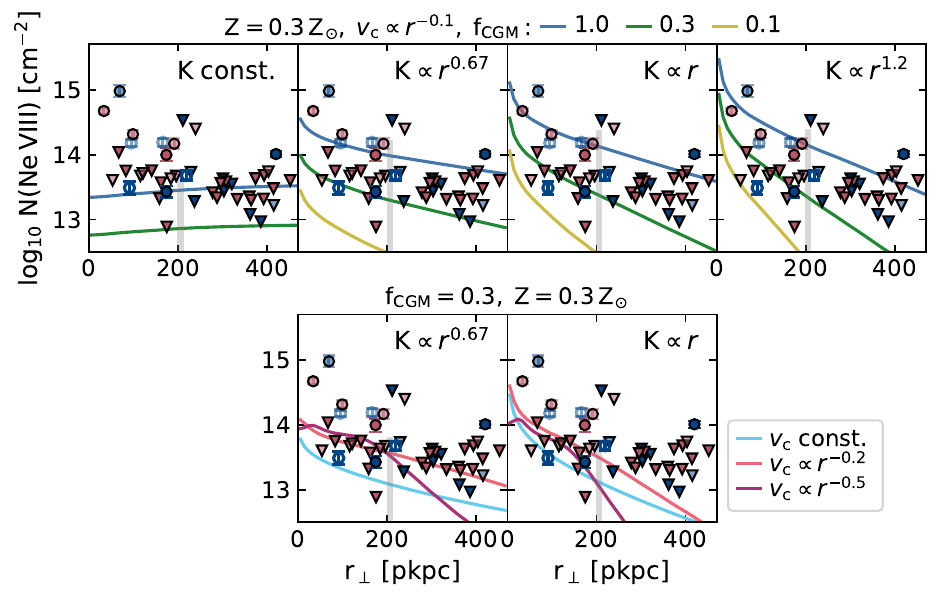}
\caption{\emph{Top row}: A comparison of analytical models with different CGM baryon fractions $\mathrm{f}_\mathrm{CGM} = \mathrm{M}_{\mathrm{CGM}} \,/\, (\Omega_{\mathrm{b}} \Mvir \,/\, \Omega_{\mathrm{m}})$, where $\mathrm{M}_{\mathrm{CGM}}$ is the (warm/hot) CGM mass. We show power law models with different entropy slopes for a halo with $\Mvir = 10^{12.2} \Msun$ and a metallicity of $0.3 \us \mathrm{Z}_{\odot}$ at redshift $z=0.75$. Different entropy slopes are shown in different panels and slope values are given in the upper right of each plot. Model column densities are shown for $\mathrm{f}_\mathrm{CGM}=1, 0.3,$ and $0.1$. 
The measurements are shown in the same way as Fig.~\ref{fig:plmodels}, and vertical, gray lines indicate the virial radius. While the data generally favor $\mathrm{f}_\mathrm{CGM} \times (\mathrm{Z} \,/\, \mathrm{Z}_{\odot} ) \approx 0.1$ ($\mathrm{f}_\mathrm{CGM} \approx 0.3$ for the $\mathrm{Z} \approx 0.3 \us \mathrm{Z}_{\odot}$ assumed in the figure), 
some measurements are only consistent with our power law model with $\mathrm{f}_\mathrm{CGM} \times ( \mathrm{Z} \,/\, \mathrm{Z}_{\odot} ) \approx 0.3$ ($\mathrm{f}_\mathrm{CGM} \approx 1$ in the figure), and a high entropy slope. 
\emph{Bottom row}: For two entropy slopes, we show the effect of varying the circular velocity profile slope. The differences between the profiles are generally smaller than those over the plausible range of entropy slopes.}
\label{fig:plmodels_fcgm}
\end{figure*}

In Fig.~\ref{fig:plmodels_fcgm} we further explore these power law models in relation to the measured column densities. We choose the halo mass from Fig.~\ref{fig:plmodels} consistent with the most data points, $\Mvir = 10^{12.2} \Msun$, and explore whether the data are consistent with lower CGM mass fractions than the cosmic baryon fraction of Fig.~\ref{fig:plmodels}.

We see that some data points are indeed consistent with lower CGM masses $\sim 0.3 \Omega_{\mathrm{b}} \,/\, \Omega_{\mathrm{m}} \Mvir$, although we note that we are still assuming a constant metallicity of $0.3$ times solar throughout the halo. Metallicity and CGM mass are not precisely degenerate in this column density model, as the CGM mass (along with the chosen entropy and circular velocity slopes) determines the density profile. Lower densities can affect the ionization balance if they are low enough for photo-ionization to become important. In practice, though, a model with CGM mass $0.3 \Omega_{\mathrm{b}} \,/\, \Omega_{\mathrm{m}} \Mvir$ and solar metallicity looks similar to the model with CGM mass $\Omega_{\mathrm{b}} \,/\, \Omega_{\mathrm{m}} \Mvir$ and 0.3 times solar metallicity shown in blue in Fig.~\ref{fig:plmodels_fcgm}. The largest differences are furthest from the halo center and with the steepest entropy slope.

However, some data points can only be explained with this power-law model of the volume-filling phase if $\mathrm{f}_{\mathrm{CGM}} \times ( \mathrm{Z} \, /\, \mathrm{Z}_{\odot}) \approx 0.3$, and if the entropy slope is steep.
This implies that these high measured column densities might be probing baryon-complete halos, or perhaps particularly dense or metal-enriched warm/hot gas within a halo. 
The highest measured column density can only be explained with a relatively high CGM metallicity or a non-uniform metallicity or density distribution.
We expect \ion{Ne}{8} to broadly trace the volume-filling gas in a halo of the mass in Fig.~\ref{fig:plmodels_fcgm}, but Figs.~\ref{fig:m12maps} and~\ref{fig:clumpiness} show that some denser-than-average \ion{Ne}{8} can exist in these halos.

The metallicities and CGM warm-hot gas content in the FIRE halos (Fig.~\ref{fig:cgmprophist}) paint a similar picture to our analytical model: the FIRE-2 NoBH model has halo metallicities $\sim 0.3$ times solar, and $\mathrm{f}_{\mathrm{CGM}}$ parameters in the $\approx 0.1$--0.3 range. As the analytical models would suggest, these halos are consistent with the upper limits, but the 90$^{\mathrm{th}}$ percentile of the simulated column densities lies below the largest measured values. Note that these large measured values might represent high column density values within typical halos, rather than halos with uniformly high densities and metallicities.

The m12 halos simulated with the FIRE-3 models have similar $\mathrm{f}_{\mathrm{CGM}}$ to the FIRE-2 NoBH model, but they have metallicities around and below $\approx 0.1$ times solar. They indeed produce lower column densities than the FIRE-2 NoBH halos, and their median column densities are consistent with the upper limits, but lie below all the measured column densities. 
As discussed in \S \ref{sec:simgenf3} and \S \ref{sec:ne8_cgm_props}, the lower metallicities in the FIRE-3 halos analyzed here are likely due to the lower metal yields in FIRE-3, and the lower stellar masses at fixed halo mass resulting from the enhanced supernova momentum injected in converging flows in the FIRE-3 runs included in this work.

The CASBaH data allow higher warm/hot CGM metal masses than the CUBS data. Although the CASBaH and CUBS upper limits on the median column density are similar, the CASBaH data imply higher 90$^\mathrm{th}$ percentile column densities within $\Rvir$ than the CUBS data allow. As the amount of \ion{Ne}{8} in the halo sets the mean column density, rather than the median, the higher 90$^\mathrm{th}$ percentiles are relevant here. 

The \ion{Ne}{8} data broadly favor CGM mass and metal contents $\mathrm{f}_\mathrm{CGM} \times (\mathrm{Z} \,/\, \mathrm{Z}_{\odot}) \sim 0.1$ for $\sim 10^{12} \Msun$ halos. 
We have checked that the implied total metal masses are consistent with being produced by nucleosynthesis from observed stars. 
Below we compare the \ion{Ne}{8} result with constraints from other observations on the CGM mass and metallicity.

The Milky Way hot CGM gas fraction is difficult to constrain, because X-ray absorption and emission line measurements are mostly sensitive to gas in the inner $\approx 50$~kpc, as are the dispersion measures towards LMC pulsars. Indeed, redshift~0 X-ray absorption in extragalactic sources may come from hot interstellar medium rather than circumgalactic gas \citep{yao_wang_2005}. However, \citet{bregman_anderson_etal_2018} favor $\mathrm{f}_{\mathrm{CGM}} \sim 0.1$--$0.2$ based on constraints on the density profile slope, and they find metallicities $\mathrm{Z} \gtrsim 0.3 \us \mathrm{Z}_{\odot}$ within $\approx 50$~kpc.
\citet{miller_bregman_2015} found similar $\mathrm{f}_{\mathrm{CGM}}$ and metallicities $\mathrm{Z} = 0.3$--$1 \us \mathrm{Z}_{\odot}$ from \ion{O}{7} and \ion{O}{8} absorption and emission.
\citet{gupta_mathur_etal_2012} have also used \ion{O}{7} and \ion{O}{8} absorption and emission lines to constrain the Milky Way hot CGM mass. Assuming a uniform hot CGM mass density (i.e., a flat density profile), they found higher values of $\mathrm{f}_{\mathrm{CGM}} > 0.5$ at $\mathrm{Z} = 0.3 \mathrm{Z}_{\odot}$. 
In their review, \citet{donahue_voit_etal_2022} give a Milky-Way value of $\mathrm{f}_{\mathrm{CGM}} \lesssim 0.5$ from X-ray absorption and emission, while observations of ram-pressure stripping of Milky Way satellite galaxies favour similar or larger electron densities (their fig.~10).

For external galaxies, \citet{bregman_hodges-kluck_etal_2022} constrained warm/hot CGM gas masses by stacking observations of the thermal Sunyaev-Zel'dovich effect around 12 nearby $\Lstar$ galaxies.
They found $\mathrm{f}_{\mathrm{CGM}} \approx 0.3$ for their default assumed gas temperature. Their $\mathrm{f}_{\mathrm{CGM}}$ is consistent with our findings, although we do not claim to meaningfully constrain $\mathrm{f}_{\mathrm{CGM}}$ or metallicity separately.
\citet{zhang_comparat_etal_2024_preprint} measure X-ray emission from the CGM by stacking eROSITA data around galaxies in bins of stellar mass. Their measurements are sensitive to $\mathrm{f}_{\mathrm{CGM}}$ and metallicity, and they find values of 
$\mathrm{f}_{\mathrm{CGM}} \times (\mathrm{Z} \,/\, \mathrm{Z}_{\odot}) \approx 0.03$--0.15 for
roughly Milky-Way-mass galaxies with different metallicity assumptions, and a larger statistical error range (their fig.~9). Larger values than 0.15 are allowed if the hot gas temperature is lower than they assumed. 

Overall, measuring the hot metal or gas mass in the Milky Way and other $\Lstar$ galaxies is difficult, and comes with systematic uncertainties from assumptions necessary in the analysis. Nonetheless, the values implied by most other observations than those of \ion{Ne}{8} absorption are similar to those we estimate here.

\section{Discussion}
\label{sec:discussion}

\subsection{FIRE column densities vs.\ observations}

We have compared various FIRE simulations of $\sim 10^{12} \Msun$ (m12) and $\sim 10^{13} \Msun$ (m13) halos at $z=0.5$--$1.0$ to observations of CGM \ion{Ne}{8} absorbers by \citet[CASBaH]{burchett_tripp_etal_2018} and \citet[CUBS]{qu_chen_etal_2024_preprint}.  For the m12 halos, we find that the FIRE-2 NoBH model produces higher \ion{Ne}{8} column densities in the CGM than the three FIRE-3 models we examine, including one that similarly has no AGN feedback. This is largely due to the higher CGM metallicity in the FIRE-2 halos. The FIRE-2 predictions for the median m12 \ion{Ne}{8} column densities are consistent with the data, but the 90\textsuperscript{th} percentile of the column densities as a function of impact parameter is lower than the CASBaH data imply, and tentatively lower than the CUBS data imply. The FIRE-3 m12 halo \ion{Ne}{8} column densities underpredict the data more systematically. 
For the m13 halos, the data is limited, but the FIRE predictions are mostly consistent with the data available.

Overall, within the FIRE samples we analysed, the m12 FIRE-2 NoBH model reproduces the \ion{Ne}{8} observations better than the FIRE-3 models. As we discuss in Appendix~\ref{app:m12plus}, differences in central galaxy stellar masses 
contribute to these differences. 
Generally, the FIRE-2 m12 halos have higher stellar masses than FIRE-3 halos with the same halo mass, as discussed in the previous section. 
The higher stellar masses and the higher stellar metal yields (per unit stellar mass) in FIRE-2 result in higher CGM metallicities.

In \S \ref{sec:anmodel} we analyzed idealized, power-law analytic models of the CGM to gain further insight into the physical conditions required to explain the observed \ion{Ne}{8} columns. 
Using this analytic framework, we found that 
the observations favor models with a product of the  halo gas fraction and metallicity of roughly  
$f_{\rm CGM} \times (\mathrm{Z} \, /\, \mathrm{Z}_{\odot}) \sim 0.1$. 
Reassuringly, this is consistent with the FIRE-2 m12 halos which we find to be the best match to the observations in simulation sample, as well as with independent empirical constraints including X-ray observations.

There are some caveats to our comparison with observations. 
The small sample sizes (in both FIRE halos and measured absorbers) imply significant statistical uncertainties. 
Furthermore, the large uncertainties in some of the observed galaxies' halo masses and the lack of `contamination' from nearby halos and/or IGM in the FIRE zoom regions introduce some systematic uncertainties into the comparisons of zoom-in simulations with observations. 

There are also some purely observational uncertainties. 
Notably, the \ion{Ne}{8} $770, 780 \Ang$ doublet falls in a crowded part of the spectrum \citep[e.g.][fig.~6]{burchett_tripp_etal_2018}. This makes it more difficult to measure the equivalent width and line width of an absorber, and even to identify the doublet. 
This might explain some of the differences between the \citet{qu_chen_etal_2024_preprint} and \citet{burchett_tripp_etal_2018} data: within 450~pkpc of galaxies, \citet{burchett_tripp_etal_2018} measure column densities $\geq 10^{14} \pcmsq$ in 6 out of 28 sightlines, while \citet{qu_chen_etal_2024_preprint}
 measure absorption of this strength in 3 out of 65 sightlines. (We excluded sightlines with column density upper limits $\geq 10^{14} \pcmsq$ from our counts.)
 These are small numbers, impacted by e.g., a  \citet{qu_chen_etal_2024_preprint} absorber with a \ion{Ne}{8} column density barely below $10^{14} \pcmsq$. However, the incidence of high column density CGM absorbers is clearly higher in the \citet{burchett_tripp_etal_2018} data. 

 Finally, we note that the highest observed column density we compare to is $\approx 10^{15} \pcmsq$, measured by \citet{burchett_tripp_etal_2018} and most consistent with an m13 host halo. \citet{tripp_meiring_etal_2011} had previously measured this absorption system; it consists of many absorption components, some of which have large velocity offsets from the host galaxy or group: $\approx 200$--$400 \kmps$ \citep[e.g.,][bottom two panels of the rightmost column in fig.~2]{burchett_tripp_etal_2018}. \citet{tripp_meiring_etal_2011} concluded that \ion{Ne}{8} was tracing an outflow in this system.
 In principle, FIRE simulations include galaxy/halo-scale outflows. However, if such systems are rare, 
 it is reasonable that our set of halos did not capture an outflow event like this.

\subsection{Can cool gas cause the observed \ion{Ne}{8} absorption?}
In this paper, we have presented evidence that the \ion{Ne}{8} absorption profiles observed around $z \lesssim 1$ galaxies are generally well explained by a volume-filling, hot phase. 
Here we briefly comment on the alternate possibility that the \ion{Ne}{8} absorption instead arises in a cool ($T \sim 10^{4}$ K) phase. 
This scenario was addressed by \cite{burchett_tripp_etal_2018}, who concluded this is unlikely for the bulk of the observed \ion{Ne}{8} on CGM scales. 
The reason is that \ion{Ne}{8} cannot be produced by collisional ionization in cool gas, so it would have to be photoionized. 
However, for \ion{Ne}{8} to be produced by photoionization by the cosmic UV/X-ray background, the hydrogren densities must be very low since the \ion{Ne}{8} fraction peaks at $n_{\rm H} \lesssim 10^{-5}$ cm$^{-3}$ for photoionized gas. 
There are two issues with this. 
The first is that, in order to explain the observed columns, the path lengths $L = N_{\rm H}/n_{\rm H}$ must be comparable to (or larger than) the virial radius of the halos. 
The second is that these low densities are comparable to (or lower than) the densities in the hot phase, which is both predicted to exist and observed in X-rays in $\sim 10^{12}$ M$_{\odot}$ halos (e.g., Fig. \ref{fig:3dprof}). 
Since $P \propto n T$, this would imply not only that the cool phase be volume-filling, but it would also be under-pressurized by $\sim 100 \times$ relative to the $T\sim 10^{6}$ K phase.  
Outside the virialized CGM, i.e. at large impact parameters not yet heated by accretion shocks or feedback, some \ion{Ne}{8} could arise from low-density, photoionized gas \cite[e.g.,][]{stern_faucher-giguere_etal_2018}, but this cannot explain the main CGM observations considered in this paper. 
In summary, a cool phase interpretation has severe difficulties explaining the CGM observations.
Local sources of ionization are unlikely to change this overall conclusion as their effects are typically limited to relatively small impact parameters \citep[e.g.,][]{upton-sanderbeck_mcquinn_etal_2018, zhu_springel_2024_preprint}.

\subsection{How do other simulations fare?}

Here, we consider whether other simulations can produce \ion{Ne}{8} column densities that match these observations. 

\citet{ji_chan_etal_2020} had previously compared FIRE-2 simulations to the \citet{burchett_tripp_etal_2018} data, focusing on differences between a model with cosmic rays and one without them. They find the model without cosmic rays agrees well with the \ion{Ne}{8} observations, although their CR model reasonably matches \ion{Ne}{8} as well, and compares better to observations of some other ion column densities.
We find that the median column density of the FIRE-2 NoBH model is consistent with the data, but this model may underpredict the higher percentiles of the column density at a given impact parameter. We note that the comparisons are somewhat different. For example, \citet{ji_chan_etal_2020} use a single halo simulated with a different FIRE-2 physics model, assume a different UV/X-ray background, compare data in virial radius units, and show the mean column density while we plot the median.

\citet{wijers_schaye_oppenheimer_2020} discuss \ion{Ne}{8} column densities around halos of different masses in the EAGLE simulations \citep{eagle_paper, eagle_calibration, mcalpine_helly_etal_2016, eagle-team_2017}. 
We compare the CUBS and CASBaH data in e.g., Fig.~\ref{fig:plmodels} to their fig.~C1, $z=0.5$ column density profiles. This is the highest-redshift data in that paper.
The median column densities in EAGLE are consistent with the \citet[CUBS]{qu_chen_etal_2024_preprint} data (mostly upper limits), but the 90\textsuperscript{th} percentile column density profile lies below many of CUBS and CASBaH measurements. 
The discrepancy is larger at $\mathrm{M}_{200\mathrm{c}} = 10^{11.5}$--$10^{12.0} \Msun$, and smaller at $\mathrm{M}_{200\mathrm{c}} \approx 10^{12.5}$--$10^{13.0} \Msun$.

\citet{liang_kravtsov_agertz_2016} predict \ion{Ne}{8} column densities from zoom-in simulations of a $\sim 10^{12} \Msun$ halo using the \textsc{RAMSES} code, comparing different feedback prescriptions. The \ion{Ne}{8} column densities are $\lesssim 10^{14} \pcmsq$ for the two models shown in their fig.~15.
This is consistent with the CUBS upper limits, but the higher end of the column density distribution at a given impact parameter is underpredicted for both the \citet{burchett_tripp_etal_2018} and \citet{qu_chen_etal_2024_preprint} measurements for that halo mass. In comparisons of other CGM absorption lines to data, they conclude that models which produce realistic galaxies do not necessarily predict a realistic CGM.

\citet{ford_oppenheimer_etal_2013} predict \ion{Ne}{8} column densities from a smooth particle hydrodynamic simulation in a cosmological volume. This model produces reasonable $\lesssim \Lstar$ galaxy properties, including the redshift~0 stellar mass function. They show median column densities at impact parameters of 10, 100, and 1000~pkpc, for halos of $10^{11}$, $10^{12}$, and $10^{13} \Msun$. The column densities do not exceed $\approx 10^{13} \pcmsq$ across these impact parameter and halo mass ranges, well below the measured values.

As part of the AGORA collaboration, \citet{strawn_roca-fabrega_etal_2024} recently compared the CGM properties of a halo that reaches a mass of $\sim 10^{12} \Msun$ at redshift~0, simulated with many different codes. In their fig.~15, they show predicted \ion{Ne}{8} column densities as a function of impact parameter. These are for redshift~1, which the ART-I, ENZO, GADGET-3, GEAR, and AREPO-T simulations reach. These simulations show a wide range of median and 16$^\mathrm{th}$ and 84$^\mathrm{th}$ percentile profiles. The ENZO simulation column density profile is similar to that of the FIRE-2 NoBH model, and ART-I predicts somewhat higher values, with a median closer to the FIRE-2 NoBH 90$^\mathrm{th}$ percentile. The other simulations predict lower column densities.
Therefore, this ENZO simulation might be consistent with the data, while ART-I is the only simulation to overpredict the median observed column density profile. However, we note that with a single simulated halo not carefully matched to the observational sample, these comparisons are highly uncertain.

In summary, some of the \ion{Ne}{8} predictions from simulations we have found in the literature may match median column density profiles; the data mostly provide an upper limit on this median. However, many simulations have trouble producing the higher  \ion{Ne}{8} column density values from the CUBS and CASBaH surveys, similar to what we found for the FIRE-2 simulations.
Comparing cosmological hydrodynamical simulations with CGM observations is particularly valuable as different models that produce realistic galaxy populations can differ significantly in their CGM predictions. 
For example, \citet{davies_crain_etal_2019_tngcomp} show that different simulations which produce realistic $z\approx 0$ galaxy populations can have meaningfully different CGM gas fractions at the $\Mvir \sim 10^{12} \Msun$ mass scale probed by \ion{Ne}{8}. 

\subsection{How do idealized models compare to the observations?}

Next, we consider whether previously-published idealized models for the CGM can produce \ion{Ne}{8} column densities that match these observations.

We first estimate \ion{Ne}{8} in the power-law cooling-flow model described in \citet{stern_fielding_etal_2019}, in which the CGM mass is such that the inflow rate induced by radiative losses equals the star formation rate (SFR). In such inflows radiative losses are balanced by compressive heating, so the temperature remains $\approx T_{\mathrm{vir}}$ down to the galaxy radius, and entropy increases roughly linearly with radius. We further assume $Z=0.3{\rm Z}_\odot$, and SFRs equal  the $16^{\rm th}-84^{\rm th}$ percentiles in the  UniverseMachine catalogues for the appropriate halo mass.We find that
at all halo masses, cooling flow solutions (not shown here for brevity) that assume median SFRs predict \ion{Ne}{8} column density profiles consistent with the median column density profiles allowed by the upper limits. At
$\Mvir \approx 10^{11.5}$--$10^{12} \Msun$ 
detected \ion{Ne}{8} columns are consistent with predictions of cooling flow solutions with $84^{\rm th}$-percentile SFRs, while at higher halo masses \ion{Ne}{8} detections are under-predicted, typically by a factor of $3-10$. These results hold for different assumed 
circular velocity profiles ($v_{\rm c}\propto r^0$, $r^{-0.1}$, or $r^{-0.2}$) and different assumed metallicities ($0.1-1\,{\rm Z}_\odot$). We conclude that at $\lesssim10^{12}\,\Msun$ \ion{Ne}{8} observations are consistent with the hot CGM phase forming a cooling flow, while a different origin is required to explain \ion{Ne}{8} detections at higher halo masses.

A number of other groups have predicted \ion{Ne}{8} column densities from analytical models. For example, \citet{faerman_pandya_etal_2022} combine CGM gas masses for $10^{12} \Msun$ halos from the Santa Cruz semi-analytical model (SAM) with an analytical model for CGM density, metallicity, etc.\ profiles. They find a large range of CGM parameters in the SAM, and use Milky-Way-like halos with $0.05< \mathrm{f}_{\mathrm{CGM}} \lesssim 1$ and CGM metallicities $\approx 0.1$--$2 \us \mathrm{Z}_{\odot}$. (They limit the total halo baryon mass, including the galaxy, to the halo baryon budget.) They predict column densities as seen from inside a galaxy (as would be measured for the Milky Way halo). They predict column densities of $0.4$--$2 \times 10^{14} \pcmsq$, depending on the halo CGM mass fraction. This would translate to column densities of $10^{13.9}$--$10^{14.6} \pcmsq$ in sightlines through the halo center, increasing with gas fraction. Those values are plausibly consistent with the \citet{burchett_tripp_etal_2018} and \citet{qu_chen_etal_2024_preprint} measurements, although most measurements consistent with this halo mass are at impact parameters $\gtrsim 0.5 \Rvir$.

\citet{voit_2019} predicts \ion{Ne}{8} column densities at an impact parameter of 50~pkpc, as a function of halo mass, for different variations of a precipitation-limited CGM model. These models give $\mathrm{f}_{\mathrm{CGM}} \approx 0.1$--0.3. Depending on the exact model variation, these column densities reach $\approx 10^{14}$--$10^{14.5} \pcmsq$ in halos with $\mathrm{M}_{200\mathrm{c}} \approx 1$--$2 \times 10^{12} \Msun$, with lower column densities at lower halo masses. These predictions are somewhat high compared to the observations at $\approx 10^{12} \Msun$, though we note they assume the CGM at these masses has solar metallicity.

\citet{qu_bregman_2018} compare their analytical CGM model predictions to a different set of \ion{Ne}{8} observations and find agreement. They find $\mathrm{f}_{\mathrm{CGM}} \sim 0.05$ for a metallicity of $0.3 \us \mathrm{Z}_{\odot}$ in $\sim 10^{12} \Msun$ halos. They predict column densities $\sim 10^{14} \pcmsq$ for halos with a mass $\sim 10^{12} \Msun$, and column densities up to $\approx 2 \times 10^{14} \pcmsq$ outside $0.3$~virial radii in the different variations of their model. This is roughly the range of the \citet{burchett_tripp_etal_2018} observations, and about as high as the \citet{qu_chen_etal_2024_preprint} data allow for the median column density.

\section{Conclusions}
\label{sec:conclusions}

We have predicted \ion{Ne}{8} column densities in $\Mvir \sim 10^{12}$ and $\sim 10^{13} \Msun$ halos based on a set of FIRE-2 and FIRE-3 simulations, run with different physics models, and compared these predictions to observations from the CASBaH \citep{burchett_tripp_etal_2018} and CUBS \citep{qu_chen_etal_2024_preprint} surveys. Our main conclusions are:
\begin{itemize}
\item{In $\Mvir \sim 10^{12} \Msun$ FIRE halos at $z = 0.5$--$1.0$, \ion{Ne}{8} traces the relatively smooth, volume-filling phase of the CGM (Figs.~\ref{fig:m12maps} and~\ref{fig:clumpiness}). At $\Mvir \sim 10^{13} \Msun$, where the volume-filling phase is typically hotter than optimal for \ion{Ne}{8}, this ion has a clumpier distribution (Figs.~\ref{fig:m13maps} and~\ref{fig:clumpiness}).}
\item{Both in $\sim10^{12} \Msun$ and $\sim 10^{13} \Msun$ halos in FIRE, \ion{Ne}{8} is mostly collisionally ionized. 
Around and beyond $\Rvir$, some \ion{Ne}{8} is in photo-ionization equilibrium.
(Fig.~\ref{fig:3dprof}).} 
\item{In FIRE-2, $\Mvir \sim 10^{12} \Msun$ halos without black hole feedback produce higher \ion{Ne}{8} column densities than in FIRE-3 (Figs.~\ref{fig:ne8_obscomp_b19} and~\ref{fig:ne8_obscomp_q23}). This is largely driven by the higher hot CGM metallicity in FIRE-2 (Fig.~\ref{fig:cgmprophist}).}
\item{When comparing CGM observations to predictions from simulations and idealized models, it is important to account for the large uncertainties in halo mass estimates based on galaxy stellar masses. The main source of this uncertainty is scatter in the stellar-mass-halo-mass relation at stellar masses $\gtrsim 10^{10.5} \Msun$ (appendix~\ref{app:smhm} and Figs.~\ref{fig:msmh}--\ref{fig:mherr}).}
\item{
The CUBS and CASBaH surveys report measured \ion{
Ne}{8} column densities, but most of the data are upper limits. 
Given the large number of upper limits, the FIRE-2 NoBH model for $\sim 10^{12} \Msun$ halos appears broadly consistent with the distribution of measured column densities within an impact parameter of 450~kpc (Fig.~\ref{fig:ne8_obscomp_fire2}). 
We note that the median \ion{Ne}{8} column densities cannot be substantially higher than this model predicts, because this would be in tension with many upper limits. However, since there are some reported detections above the FIRE-2 NoBH 90$^{\mathrm{th}}$ percentile column densities, it is possible that the simulations under-predict the scatter in \ion{Ne}{8} column densities.
\item{The FIRE-3 models analyzed in this paper (which use a modified supernovae feedback model and/or include AGN feedback) more clearly underpredict these high column densities.
Some other cosmological simulations from the literature (\citet{ford_oppenheimer_etal_2013}, \citet[RAMSES]{liang_kravtsov_agertz_2016}, and \citet[EAGLE]{wijers_schaye_oppenheimer_2020}), also underpredict the 
highest measured \ion{Ne}{8} column densities to varying degrees, though one simulation from the AGORA comparison project (ART-I) might overpredict \ion{Ne}{8} column densities \citep{strawn_roca-fabrega_etal_2024}.
}
\item{Overall, we find a consistent picture from our analysis of FIRE simulations and idealized, analytic power-law CGM models. 
Namely, the observed \ion{Ne}{8} column densities can be mostly reproduced by a CGM with a warm/hot gas fraction and a metallicity whose product is roughly $\mathrm{M}_{\mathrm{CGM}} / (\mathrm{M}_{\mathrm{vir}} \Omega_{\mathrm{b}} \,/\, \Omega_{\mathrm{m}}) \times (\mathrm{Z} \, /\, \mathrm{Z}_{\odot}) \sim 0.1$.} 
These are physically plausible values, realized in particular in FIRE-2 halos.}
\end{itemize}

As they become available, a larger set of simulated halos and/or measured absorbers could improve the robustness of these comparisons. We have also not used all the information in the observations: \citet{burchett_tripp_etal_2018} and 
\citet{qu_chen_etal_2024_preprint} also measure the velocities of their absorption systems as a whole, and their component velocities. This kinematic and spatial information will contain information on ongoing gas flows, as well as the current warm/hot halo gas content. We intend to study these absorber kinematics in future work.

\section*{Data availability}

A number of the FIRE-2 simulations analysed here are publicly available at \url{http://flathub.flatironinstitute.org/fire} \citep{wetzel_hayward_etal_2023}. 
Additional data, including initial conditions and derived data products, are available at \url{https://fire.northwestern.edu/data/}.
A public version of the GIZMO code is available at \url{http://www.tapir.caltech.edu/~phopkins/Site/GIZMO.html}.
The scripts used to analyse the simulations are available at \url{https://github.com/nastasha-w/ne8abs_paper}. Data shown in the plots is available on reasonable request to the authors. 

\section*{Acknowledgements}
We thank Zhijie Qu and Hsiao-Wen Chen for sharing their CUBS data, and Phil Hopkins for helpful comments.
N.A.W.\ was supported by a CIERA Postdoctoral Fellowship.
JS was supported by the Israel Science Foundation (grant No. 2584/21). 
C.-A.F.-G.\ was supported by NSF through grants AST-2108230, AST-2307327, and CAREER award AST-1652522; by NASA through grants 17-ATP17-0067 and 21-ATP21-0036; by STScI through grants HST-GO-16730.016-A and JWST-AR-03252.001-A; and by CXO through grant TM2-23005X.
L.B.\ was supported by the DOE Computer Science
Graduate Fellowship through grant DE-SC0020347.
I.S.\ was supported by the NSF Graduate Research Fellowship under Grant No.\ DGE-2234667.

Numerical calculations were run on the Northwestern computer cluster Quest, the Caltech computer cluster Wheeler, Frontera allocation FTA-Hopkins/AST20016 supported by the NSF and TACC, XSEDE/ACCESS allocations ACI-1548562, TGAST140023, and TG-AST140064 also supported by the NSF and TACC, and NASA HEC allocations SMD-16-7561, SMD-17-1204, and SMD-16-7592.

The SMDPL simulation and halo finding has been performed at LRZ Munich within the project pr87yi. The CosmoSim database (\url{www.cosmosim.org}) providing the file access is a service by the Leibniz-Institute for Astrophysics Potsdam (AIP).

We made use of \textsc{Python} and \textsc{C} code for this analysis. In particular, we used the following \textsc{Python} modules: \textsc{astropy} \citep{astropy_2013, astropy_2018, astropy_2022}, \textsc{h5py} \citep{h5py},  \textsc{matplotlib} \citep{matplotlib}, \textsc{numpy} \citep{numpy_2020}, \textsc{pandas} \citep{reback_2020_pandas, mckinney_2010_pandas}, and \textsc{scipy} \citep{scipy_2020}, and the \textsc{ipython} command-line interface \citep{ipython}.



\bibliographystyle{aasjournal}
\bibliography{bibliography}

\begin{thebibliography}{}
\expandafter\ifx\csname natexlab\endcsname\relax\def\natexlab#1{#1}\fi
\providecommand{\url}[1]{\href{#1}{#1}}
\providecommand{\dodoi}[1]{doi:~\href{http://doi.org/#1}{\nolinkurl{#1}}}
\providecommand{\doeprint}[1]{\href{http://ascl.net/#1}{\nolinkurl{http://ascl.net/#1}}}
\providecommand{\doarXiv}[1]{\href{https://arxiv.org/abs/#1}{\nolinkurl{https://arxiv.org/abs/#1}}}

\bibitem[{Asplund {et~al.}(2009)Asplund, Grevesse, Sauval, \&
  Scott}]{asplund_grevesse_etal_2009}
Asplund, M., Grevesse, N., Sauval, A.~J., \& Scott, P. 2009, Annual Review of
  Astronomy and Astrophysics, 47, 481,
  \dodoi{10.1146/annurev.astro.46.060407.145222}

\bibitem[{{Astropy Collaboration} {et~al.}(2013){Astropy Collaboration},
  {Robitaille}, {Tollerud}, {Greenfield}, {Droettboom}, {Bray}, {Aldcroft},
  {Davis}, {Ginsburg}, {Price-Whelan}, {Kerzendorf}, {Conley}, {Crighton},
  {Barbary}, {Muna}, {Ferguson}, {Grollier}, {Parikh}, {Nair}, {Unther},
  {Deil}, {Woillez}, {Conseil}, {Kramer}, {Turner}, {Singer}, {Fox}, {Weaver},
  {Zabalza}, {Edwards}, {Azalee Bostroem}, {Burke}, {Casey}, {Crawford},
  {Dencheva}, {Ely}, {Jenness}, {Labrie}, {Lim}, {Pierfederici}, {Pontzen},
  {Ptak}, {Refsdal}, {Servillat}, \& {Streicher}}]{astropy_2013}
{Astropy Collaboration}, {Robitaille}, T.~P., {Tollerud}, E.~J., {et~al.} 2013,
  \aap, 558, A33, \dodoi{10.1051/0004-6361/201322068}

\bibitem[{{Astropy Collaboration} {et~al.}(2018){Astropy Collaboration},
  {Price-Whelan}, {Sip{H{o}}cz}, {G{"u}nther}, {Lim}, {Crawford}, {Conseil},
  {Shupe}, {Craig}, {Dencheva}, {Ginsburg}, {Vand erPlas}, {Bradley},
  {P{'e}rez-Su{'a}rez}, {de Val-Borro}, {Aldcroft}, {Cruz}, {Robitaille},
  {Tollerud}, {Ardelean}, {Babej}, {Bach}, {Bachetti}, {Bakanov}, {Bamford},
  {Barentsen}, {Barmby}, {Baumbach}, {Berry}, {Biscani}, {Boquien}, {Bostroem},
  {Bouma}, {Brammer}, {Bray}, {Breytenbach}, {Buddelmeijer}, {Burke},
  {Calderone}, {Cano Rodr{'i}guez}, {Cara}, {Cardoso}, {Cheedella}, {Copin},
  {Corrales}, {Crichton}, {D'Avella}, {Deil}, {Depagne}, {Dietrich}, {Donath},
  {Droettboom}, {Earl}, {Erben}, {Fabbro}, {Ferreira}, {Finethy}, {Fox},
  {Garrison}, {Gibbons}, {Goldstein}, {Gommers}, {Greco}, {Greenfield},
  {Groener}, {Grollier}, {Hagen}, {Hirst}, {Homeier}, {Horton}, {Hosseinzadeh},
  {Hu}, {Hunkeler}, {Ivezi{'c}}, {Jain}, {Jenness}, {Kanarek}, {Kendrew},
  {Kern}, {Kerzendorf}, {Khvalko}, {King}, {Kirkby}, {Kulkarni}, {Kumar},
  {Lee}, {Lenz}, {Littlefair}, {Ma}, {Macleod}, {Mastropietro}, {McCully},
  {Montagnac}, {Morris}, {Mueller}, {Mumford}, {Muna}, {Murphy}, {Nelson},
  {Nguyen}, {Ninan}, {N{"o}the}, {Ogaz}, {Oh}, {Parejko}, {Parley}, {Pascual},
  {Patil}, {Patil}, {Plunkett}, {Prochaska}, {Rastogi}, {Reddy Janga},
  {Sabater}, {Sakurikar}, {Seifert}, {Sherbert}, {Sherwood-Taylor}, {Shih},
  {Sick}, {Silbiger}, {Singanamalla}, {Singer}, {Sladen}, {Sooley},
  {Sornarajah}, {Streicher}, {Teuben}, {Thomas}, {Tremblay}, {Turner},
  {Terr{'o}n}, {van Kerkwijk}, {de la Vega}, {Watkins}, {Weaver}, {Whitmore},
  {Woillez}, {Zabalza}, \& {Astropy Contributors}}]{astropy_2018}
{Astropy Collaboration}, {Price-Whelan}, A.~M., {Sip{H{o}}cz}, B.~M., {et~al.}
  2018, aj, 156, 123, \dodoi{10.3847/1538-3881/aabc4f}

\bibitem[{{Astropy Collaboration} {et~al.}(2022){Astropy Collaboration},
  {Price-Whelan}, {Lim}, {Earl}, {Starkman}, {Bradley}, {Shupe}, {Patil},
  {Corrales}, {Brasseur}, {N{"o}the}, {Donath}, {Tollerud}, {Morris},
  {Ginsburg}, {Vaher}, {Weaver}, {Tocknell}, {Jamieson}, {van Kerkwijk},
  {Robitaille}, {Merry}, {Bachetti}, {G{"u}nther}, {Aldcroft},
  {Alvarado-Montes}, {Archibald}, {B{'o}di}, {Bapat}, {Barentsen}, {Baz{'a}n},
  {Biswas}, {Boquien}, {Burke}, {Cara}, {Cara}, {Conroy}, {Conseil}, {Craig},
  {Cross}, {Cruz}, {D'Eugenio}, {Dencheva}, {Devillepoix}, {Dietrich},
  {Eigenbrot}, {Erben}, {Ferreira}, {Foreman-Mackey}, {Fox}, {Freij}, {Garg},
  {Geda}, {Glattly}, {Gondhalekar}, {Gordon}, {Grant}, {Greenfield}, {Groener},
  {Guest}, {Gurovich}, {Handberg}, {Hart}, {Hatfield-Dodds}, {Homeier},
  {Hosseinzadeh}, {Jenness}, {Jones}, {Joseph}, {Kalmbach}, {Karamehmetoglu},
  {Ka{l}uszy{'n}ski}, {Kelley}, {Kern}, {Kerzendorf}, {Koch}, {Kulumani},
  {Lee}, {Ly}, {Ma}, {MacBride}, {Maljaars}, {Muna}, {Murphy}, {Norman},
  {O'Steen}, {Oman}, {Pacifici}, {Pascual}, {Pascual-Granado}, {Patil},
  {Perren}, {Pickering}, {Rastogi}, {Roulston}, {Ryan}, {Rykoff}, {Sabater},
  {Sakurikar}, {Salgado}, {Sanghi}, {Saunders}, {Savchenko}, {Schwardt},
  {Seifert-Eckert}, {Shih}, {Jain}, {Shukla}, {Sick}, {Simpson},
  {Singanamalla}, {Singer}, {Singhal}, {Sinha}, {Sip{H{o}}cz}, {Spitler},
  {Stansby}, {Streicher}, {{{S}}umak}, {Swinbank}, {Taranu}, {Tewary},
  {Tremblay}, {Val-Borro}, {Van Kooten}, {Vasovi{'c}}, {Verma}, {de Miranda
  Cardoso}, {Williams}, {Wilson}, {Winkel}, {Wood-Vasey}, {Xue}, {Yoachim},
  {Zhang}, {Zonca}, \& {Astropy Project Contributors}}]{astropy_2022}
{Astropy Collaboration}, {Price-Whelan}, A.~M., {Lim}, P.~L., {et~al.} 2022,
  apj, 935, 167, \dodoi{10.3847/1538-4357/ac7c74}

\bibitem[{{Babyk} {et~al.}(2018){Babyk}, {McNamara}, {Nulsen}, {Russell},
  {Vantyghem}, {Hogan}, \& {Pulido}}]{babyk_mcnamara_etal_2018}
{Babyk}, I.~V., {McNamara}, B.~R., {Nulsen}, P.~E.~J., {et~al.} 2018, \apj,
  862, 39, \dodoi{10.3847/1538-4357/aacce5}

\bibitem[{{Barret} {et~al.}(2018){Barret}, {Lam Trong}, {den Herder}, {Piro},
  {Cappi}, {Houvelin}, {Kelley}, {Mas-Hesse}, {Mitsuda}, {Paltani}, {Rauw},
  {Rozanska}, {Wilms}, {Bandler}, {Barbera}, {Barcons}, {Bozzo}, {Ceballos},
  {Charles}, {Costantini}, {Decourchelle}, {den Hartog}, {Duband}, {Duval},
  {Fiore}, {Gatti}, {Goldwurm}, {Jackson}, {Jonker}, {Kilbourne}, {Macculi},
  {Mendez}, {Molendi}, {Orleanski}, {Pajot}, {Pointecouteau}, {Porter},
  {Pratt}, {Pr{\^e}le}, {Ravera}, {Sato}, {Schaye}, {Shinozaki}, {Thibert},
  {Valenziano}, {Valette}, {Vink}, {Webb}, {Wise}, {Yamasaki}, {Douchin},
  {Mesnager}, {Pontet}, {Pradines}, {Branduardi-Raymont}, {Bulbul}, {Dadina},
  {Ettori}, {Finoguenov}, {Fukazawa}, {Janiuk}, {Kaastra}, {Mazzotta},
  {Miller}, {Miniutti}, {Naze}, {Nicastro}, {Scioritino}, {Simonescu},
  {Torrejon}, {Frezouls}, {Geoffray}, {Peille}, {Aicardi}, {Andr{\'e}},
  {Daniel}, {Cl{\'e}net}, {Etcheverry}, {Gloaguen}, {Hervet}, {Jolly}, {Ledot},
  {Paillet}, {Schmisser}, {Vella}, {Damery}, {Boyce}, {Dipirro}, {Lotti},
  {Schwander}, {Smith}, {Van Leeuwen}, {van Weers}, {Clerc}, {Cobo}, {Dauser},
  {Kirsch}, {Cucchetti}, {Eckart}, {Ferrando}, \& {Natalucci}}]{Athena_2018_07}
{Barret}, D., {Lam Trong}, T., {den Herder}, J.-W., {et~al.} 2018, in Society
  of Photo-Optical Instrumentation Engineers (SPIE) Conference Series, Vol.
  10699, Space Telescopes and Instrumentation 2018: Ultraviolet to Gamma Ray,
  106991G, \dodoi{10.1117/12.2312409}

\bibitem[{{Behroozi} {et~al.}(2019){Behroozi}, {Wechsler}, {Hearin}, \&
  {Conroy}}]{behroozi_wechsler_etal_2019}
{Behroozi}, P., {Wechsler}, R.~H., {Hearin}, A.~P., \& {Conroy}, C. 2019,
  \mnras, 488, 3143, \dodoi{10.1093/mnras/stz1182}

\bibitem[{{Behroozi} {et~al.}(2013{\natexlab{a}}){Behroozi}, {Wechsler}, \&
  {Wu}}]{behroozi_wechsler_wu_2013}
{Behroozi}, P.~S., {Wechsler}, R.~H., \& {Wu}, H.-Y. 2013{\natexlab{a}}, \apj,
  762, 109, \dodoi{10.1088/0004-637X/762/2/109}

\bibitem[{{Behroozi} {et~al.}(2013{\natexlab{b}}){Behroozi}, {Wechsler}, {Wu},
  {Busha}, {Klypin}, \& {Primack}}]{behroozi_wechsler_etal_2013}
{Behroozi}, P.~S., {Wechsler}, R.~H., {Wu}, H.-Y., {et~al.} 2013{\natexlab{b}},
  \apj, 763, 18, \dodoi{10.1088/0004-637X/763/1/18}

\bibitem[{{Branchini} {et~al.}(2009){Branchini}, {Ursino}, {Corsi}, {Martizzi},
  {Amati}, {den Herder}, {Galeazzi}, {Gendre}, {Kaastra}, {Moscardini},
  {Nicastro}, {Ohashi}, {Paerels}, {Piro}, {Roncarelli}, {Takei}, \&
  {Viel}}]{branchini_ursino_etal_2009}
{Branchini}, E., {Ursino}, E., {Corsi}, A., {et~al.} 2009, \apj, 697, 328,
  \dodoi{10.1088/0004-637X/697/1/328}

\bibitem[{{Bregman} {et~al.}(2018){Bregman}, {Anderson}, {Miller},
  {Hodges-Kluck}, {Dai}, {Li}, {Li}, \& {Qu}}]{bregman_anderson_etal_2018}
{Bregman}, J.~N., {Anderson}, M.~E., {Miller}, M.~J., {et~al.} 2018, \apj, 862,
  3, \dodoi{10.3847/1538-4357/aacafe}

\bibitem[{{Bregman} {et~al.}(2022){Bregman}, {Hodges-Kluck}, {Qu}, {Pratt},
  {Li}, \& {Yun}}]{bregman_hodges-kluck_etal_2022}
{Bregman}, J.~N., {Hodges-Kluck}, E., {Qu}, Z., {et~al.} 2022, \apj, 928, 14,
  \dodoi{10.3847/1538-4357/ac51de}

\bibitem[{{Bregman} \& {Lloyd-Davies}(2007)}]{bregman_lloyd-davies_edward_2007}
{Bregman}, J.~N., \& {Lloyd-Davies}, E.~J. 2007, \apj, 669, 990,
  \dodoi{10.1086/521321}

\bibitem[{{Bryan} \& {Norman}(1998)}]{bryan_norman_1998}
{Bryan}, G.~L., \& {Norman}, M.~L. 1998, \apj, 495, 80, \dodoi{10.1086/305262}

\bibitem[{{Burchett} {et~al.}(2016){Burchett}, {Tripp}, {Bordoloi}, {Werk},
  {Prochaska}, {Tumlinson}, {Willmer}, {O'Meara}, \&
  {Katz}}]{burchett_tripp_etal_2016}
{Burchett}, J.~N., {Tripp}, T.~M., {Bordoloi}, R., {et~al.} 2016, \apj, 832,
  124, \dodoi{10.3847/0004-637X/832/2/124}

\bibitem[{{Burchett} {et~al.}(2019){Burchett}, {Tripp}, {Prochaska}, {Werk},
  {Tumlinson}, {Howk}, {Willmer}, {Lehner}, {Meiring}, \&
  {Bowen}}]{burchett_tripp_etal_2018}
{Burchett}, J.~N., {Tripp}, T.~M., {Prochaska}, J.~X., {et~al.} 2019, \apj,
  877, L20, \dodoi{10.3847/2041-8213/ab1f7f}

\bibitem[{{Byrne} {et~al.}(2023){Byrne}, {Faucher-Gigu{\`e}re}, {Wellons},
  {Hopkins}, {Angl{\'e}s-Alc{\'a}zar}, {Sultan}, {Wijers}, {Moreno}, \&
  {Ponnada}}]{byrne_faucher-giguere_etal_2023}
{Byrne}, L., {Faucher-Gigu{\`e}re}, C.-A., {Wellons}, S., {et~al.} 2023, arXiv
  e-prints, arXiv:2310.16086, \dodoi{10.48550/arXiv.2310.16086}

\bibitem[{{Cen} \& {Fang}(2006)}]{cen_fang_2006}
{Cen}, R., \& {Fang}, T. 2006, \apj, 650, 573, \dodoi{10.1086/506506}

\bibitem[{{Chan} {et~al.}(2019){Chan}, {Kere{\v{s}}}, {Hopkins}, {Quataert},
  {Su}, {Hayward}, \& {Faucher-Gigu{\`e}re}}]{chan_hopkins_etal_2019}
{Chan}, T.~K., {Kere{\v{s}}}, D., {Hopkins}, P.~F., {et~al.} 2019, \mnras, 488,
  3716, \dodoi{10.1093/mnras/stz1895}

\bibitem[{{Chen} {et~al.}(2020){Chen}, {Zahedy}, {Boettcher}, {Cooper},
  {Johnson}, {Rudie}, {Chen}, {Walth}, {Cantalupo}, {Cooksey},
  {Faucher-Gigu{\`e}re}, {Greene}, {Lopez}, {Mulchaey}, {Penton}, {Petitjean},
  {Putman}, {Rafelski}, {Rauch}, {Schaye}, {Simcoe}, \&
  {Weiner}}]{chen_zahedi_etal_2020_cubs1}
{Chen}, H.-W., {Zahedy}, F.~S., {Boettcher}, E., {et~al.} 2020, \mnras, 497,
  498, \dodoi{10.1093/mnras/staa1773}

\bibitem[{{Chen} {et~al.}(2003){Chen}, {Weinberg}, {Katz}, \&
  {Dav{\'e}}}]{chen_weinberg_etal_2003}
{Chen}, X., {Weinberg}, D.~H., {Katz}, N., \& {Dav{\'e}}, R. 2003, \apj, 594,
  42, \dodoi{10.1086/376751}

\bibitem[{{Colbrook} {et~al.}(2017){Colbrook}, {Ma}, {Hopkins}, \&
  {Squire}}]{colbrook_ma_etal_2017}
{Colbrook}, M.~J., {Ma}, X., {Hopkins}, P.~F., \& {Squire}, J. 2017, \mnras,
  467, 2421, \dodoi{10.1093/mnras/stx261}

\bibitem[{Collette(2013)}]{h5py}
Collette, A. 2013, Python and HDF5 (Sebastopol CA, USA: O'Reilly).
\newblock \url{http://www.h5py.org/}

\bibitem[{{Crain} {et~al.}(2015){Crain}, {Schaye}, {Bower}, {Furlong},
  {Schaller}, {Theuns}, {Dalla Vecchia}, {Frenk}, {McCarthy}, {Helly},
  {Jenkins}, {Rosas-Guevara}, {White}, \& {Trayford}}]{eagle_calibration}
{Crain}, R.~A., {Schaye}, J., {Bower}, R.~G., {et~al.} 2015, \mnras, 450, 1937,
  \dodoi{10.1093/mnras/stv725}

\bibitem[{{Cui} {et~al.}(2020){Cui}, {Chen}, {Gao}, {Guo}, {Jin}, {Wang},
  {Wang}, {Wang}, {Wang}, {Wang}, {Wang}, {Yuan}, \&
  {Zhang}}]{cui_chen_etal_2020}
{Cui}, W., {Chen}, L.~B., {Gao}, B., {et~al.} 2020, Journal of Low Temperature
  Physics, 199, 502, \dodoi{10.1007/s10909-019-02279-3}

\bibitem[{{Das} {et~al.}(2019){Das}, {Mathur}, {Gupta}, {Nicastro}, \&
  {Krongold}}]{das_mathur_etal_2019}
{Das}, S., {Mathur}, S., {Gupta}, A., {Nicastro}, F., \& {Krongold}, Y. 2019,
  \apj, 887, 257, \dodoi{10.3847/1538-4357/ab5846}

\bibitem[{Davies {et~al.}(2020)Davies, Crain, Oppenheimer, \&
  Schaye}]{davies_crain_etal_2019_tngcomp}
Davies, J.~J., Crain, R.~A., Oppenheimer, B.~D., \& Schaye, J. 2020, Monthly
  Notices of the Royal Astronomical Society, 491, 4462

\bibitem[{{Donahue} {et~al.}(2006){Donahue}, {Horner}, {Cavagnolo}, \&
  {Voit}}]{donahue_horner_etal_2006}
{Donahue}, M., {Horner}, D.~J., {Cavagnolo}, K.~W., \& {Voit}, G.~M. 2006,
  \apj, 643, 730, \dodoi{10.1086/503270}

\bibitem[{{Donahue} \& {Voit}(2022)}]{donahue_voit_etal_2022}
{Donahue}, M., \& {Voit}, G.~M. 2022, \physrep, 973, 1,
  \dodoi{10.1016/j.physrep.2022.04.005}

\bibitem[{{Dutton} \& {Macci{\`o}}(2014)}]{dutton_maccio_2014}
{Dutton}, A.~A., \& {Macci{\`o}}, A.~V. 2014, \mnras, 441, 3359,
  \dodoi{10.1093/mnras/stu742}

\bibitem[{{Faerman} {et~al.}(2022){Faerman}, {Pandya}, {Somerville}, \&
  {Sternberg}}]{faerman_pandya_etal_2022}
{Faerman}, Y., {Pandya}, V., {Somerville}, R.~S., \& {Sternberg}, A. 2022,
  \apj, 928, 37, \dodoi{10.3847/1538-4357/ac4ca6}

\bibitem[{{Faucher-Gigu{\`e}re}(2020)}]{FG20}
{Faucher-Gigu{\`e}re}, C.-A. 2020, \mnras, 493, 1614,
  \dodoi{10.1093/mnras/staa302}

\bibitem[{{Faucher-Gigu{\`e}re} {et~al.}(2009){Faucher-Gigu{\`e}re}, {Lidz},
  {Zaldarriaga}, \& {Hernquist}}]{FG09}
{Faucher-Gigu{\`e}re}, C.-A., {Lidz}, A., {Zaldarriaga}, M., \& {Hernquist}, L.
  2009, \apj, 703, 1416, \dodoi{10.1088/0004-637X/703/2/1416}

\bibitem[{{Faucher-Gigu{\`e}re} \& {Oh}(2023)}]{faucher-giguere_oh_2023}
{Faucher-Gigu{\`e}re}, C.-A., \& {Oh}, S.~P. 2023, \araa, 61, 131,
  \dodoi{10.1146/annurev-astro-052920-125203}

\bibitem[{{Ford} {et~al.}(2013){Ford}, {Oppenheimer}, {Dav{\'e}}, {Katz},
  {Kollmeier}, \& {Weinberg}}]{ford_oppenheimer_etal_2013}
{Ford}, A.~B., {Oppenheimer}, B.~D., {Dav{\'e}}, R., {et~al.} 2013, \mnras,
  432, 89, \dodoi{10.1093/mnras/stt393}

\bibitem[{{Gupta} {et~al.}(2014){Gupta}, {Mathur}, {Galeazzi}, \&
  {Krongold}}]{gupta_mathur_etal_2014}
{Gupta}, A., {Mathur}, S., {Galeazzi}, M., \& {Krongold}, Y. 2014, \apss, 352,
  775, \dodoi{10.1007/s10509-014-1958-z}

\bibitem[{{Gupta} {et~al.}(2017){Gupta}, {Mathur}, \&
  {Krongold}}]{gupta_mathur_etal_2017}
{Gupta}, A., {Mathur}, S., \& {Krongold}, Y. 2017, \apj, 836, 243,
  \dodoi{10.3847/1538-4357/836/2/243}

\bibitem[{{Gupta} {et~al.}(2012){Gupta}, {Mathur}, {Krongold}, {Nicastro}, \&
  {Galeazzi}}]{gupta_mathur_etal_2012}
{Gupta}, A., {Mathur}, S., {Krongold}, Y., {Nicastro}, F., \& {Galeazzi}, M.
  2012, \apjl, 756, L8, \dodoi{10.1088/2041-8205/756/1/L8}

\bibitem[{{Hafen} {et~al.}(2020){Hafen}, {Faucher-Gigu{\`e}re},
  {Angl{\'e}s-Alc{\'a}zar}, {Stern}, {Kere{\v{s}}}, {Esmerian}, {Wetzel},
  {El-Badry}, {Chan}, \& {Murray}}]{hafen_faucher-giguere_etal_2020}
{Hafen}, Z., {Faucher-Gigu{\`e}re}, C.-A., {Angl{\'e}s-Alc{\'a}zar}, D.,
  {et~al.} 2020, \mnras, 494, 3581, \dodoi{10.1093/mnras/staa902}

\bibitem[{Harris {et~al.}(2020)Harris, Millman, van~der Walt, Gommers,
  Virtanen, Cournapeau, Wieser, Taylor, Berg, Smith, Kern, Picus, Hoyer, van
  Kerkwijk, Brett, Haldane, del R{\'{i}}o, Wiebe, Peterson,
  G{\'{e}}rard-Marchant, Sheppard, Reddy, Weckesser, Abbasi, Gohlke, \&
  Oliphant}]{numpy_2020}
Harris, C.~R., Millman, K.~J., van~der Walt, S.~J., {et~al.} 2020, Nature, 585,
  357, \dodoi{10.1038/s41586-020-2649-2}

\bibitem[{{Hellsten} {et~al.}(1998){Hellsten}, {Gnedin}, \&
  {Miralda-Escud{\'e}}}]{hellsten_gnedin_miralda-escude_1998}
{Hellsten}, U., {Gnedin}, N.~Y., \& {Miralda-Escud{\'e}}, J. 1998, \apj, 509,
  56, \dodoi{10.1086/306499}

\bibitem[{{Hopkins}(2015{\natexlab{a}})}]{hopkins_2015_hydro}
{Hopkins}, P.~F. 2015{\natexlab{a}}, \mnras, 450, 53,
  \dodoi{10.1093/mnras/stv195}

\bibitem[{{Hopkins}(2015{\natexlab{b}})}]{hopkins_2015_gizmo}
---. 2015{\natexlab{b}}, \mnras, 450, 53, \dodoi{10.1093/mnras/stv195}

\bibitem[{{Hopkins}(2016)}]{hopkins_2016_mhd_divB_control}
---. 2016, \mnras, 462, 576, \dodoi{10.1093/mnras/stw1578}

\bibitem[{{Hopkins}(2024)}]{hopkins_2024_preprint}
---. 2024, arXiv e-prints, arXiv:2404.16987.
\newblock \doarXiv{2404.16987}

\bibitem[{{Hopkins} {et~al.}(2022{\natexlab{a}}){Hopkins}, {Butsky},
  {Panopoulou}, {Ji}, {Quataert}, {Faucher-Gigu{\`e}re}, \&
  {Kere{\v{s}}}}]{hopkins_butsky_etal_2022}
{Hopkins}, P.~F., {Butsky}, I.~S., {Panopoulou}, G.~V., {et~al.}
  2022{\natexlab{a}}, \mnras, 516, 3470, \dodoi{10.1093/mnras/stac1791}

\bibitem[{{Hopkins} {et~al.}(2012){Hopkins}, {Quataert}, \&
  {Murray}}]{hopkins_quataert_murray_2012}
{Hopkins}, P.~F., {Quataert}, E., \& {Murray}, N. 2012, \mnras, 421, 3488,
  \dodoi{10.1111/j.1365-2966.2012.20578.x}

\bibitem[{{Hopkins} \& {Raives}(2016)}]{hopkins_raives_2016_mhd}
{Hopkins}, P.~F., \& {Raives}, M.~J. 2016, \mnras, 455, 51,
  \dodoi{10.1093/mnras/stv2180}

\bibitem[{{Hopkins} {et~al.}(2022{\natexlab{b}}){Hopkins}, {Squire}, {Butsky},
  \& {Ji}}]{hopkins_squire_etal_2022}
{Hopkins}, P.~F., {Squire}, J., {Butsky}, I.~S., \& {Ji}, S.
  2022{\natexlab{b}}, \mnras, 517, 5413, \dodoi{10.1093/mnras/stac2909}

\bibitem[{{Hopkins} {et~al.}(2016){Hopkins}, {Torrey}, {Faucher-Gigu{\`e}re},
  {Quataert}, \& {Murray}}]{hopkins_torrey_etal_2016}
{Hopkins}, P.~F., {Torrey}, P., {Faucher-Gigu{\`e}re}, C.-A., {Quataert}, E.,
  \& {Murray}, N. 2016, \mnras, 458, 816, \dodoi{10.1093/mnras/stw289}

\bibitem[{{Hopkins} {et~al.}(2018{\natexlab{a}}){Hopkins}, {Wetzel},
  {Kere{\v{s}}}, {Faucher-Gigu{\`e}re}, {Quataert}, {Boylan-Kolchin}, {Murray},
  {Hayward}, {Garrison-Kimmel}, {Hummels}, {Feldmann}, {Torrey}, {Ma},
  {Angl{\'e}s-Alc{\'a}zar}, {Su}, {Orr}, {Schmitz}, {Escala}, {Sanderson},
  {Grudi{\'c}}, {Hafen}, {Kim}, {Fitts}, {Bullock}, {Wheeler}, {Chan},
  {Elbert}, \& {Narayanan}}]{hopkins_wetzel_etal_2018}
{Hopkins}, P.~F., {Wetzel}, A., {Kere{\v{s}}}, D., {et~al.} 2018{\natexlab{a}},
  \mnras, 480, 800, \dodoi{10.1093/mnras/sty1690}

\bibitem[{{Hopkins} {et~al.}(2018{\natexlab{b}}){Hopkins}, {Wetzel},
  {Kere{\v{s}}}, {Faucher-Gigu{\`e}re}, {Quataert}, {Boylan-Kolchin}, {Murray},
  {Hayward}, \& {El-Badry}}]{hopkins_wetzel_etal_2018_sne}
---. 2018{\natexlab{b}}, \mnras, 477, 1578, \dodoi{10.1093/mnras/sty674}

\bibitem[{{Hopkins} {et~al.}(2020){Hopkins}, {Chan}, {Garrison-Kimmel}, {Ji},
  {Su}, {Hummels}, {Kere{\v{s}}}, {Quataert}, \&
  {Faucher-Gigu{\`e}re}}]{hopkins_chan_etal_2020_whatabout}
{Hopkins}, P.~F., {Chan}, T.~K., {Garrison-Kimmel}, S., {et~al.} 2020, \mnras,
  492, 3465, \dodoi{10.1093/mnras/stz3321}

\bibitem[{{Hopkins} {et~al.}(2023){Hopkins}, {Wetzel}, {Wheeler}, {Sanderson},
  {Grudi{\'c}}, {Sameie}, {Boylan-Kolchin}, {Orr}, {Ma}, {Faucher-Gigu{\`e}re},
  {Kere{\v{s}}}, {Quataert}, {Su}, {Moreno}, {Feldmann}, {Bullock}, {Loebman},
  {Angl{\'e}s-Alc{\'a}zar}, {Stern}, {Necib}, {Choban}, \&
  {Hayward}}]{hopkins_wetzel_etal_2023}
{Hopkins}, P.~F., {Wetzel}, A., {Wheeler}, C., {et~al.} 2023, \mnras, 519,
  3154, \dodoi{10.1093/mnras/stac3489}

\bibitem[{Hunter(2007)}]{matplotlib}
Hunter, J.~D. 2007, Computing in Science \& Engineering, 9, 90,
  \dodoi{10.1109/MCSE.2007.55}

\bibitem[{Iwamoto {et~al.}(1999)Iwamoto, Brachwitz, Nomoto, Kishimoto, Umeda,
  Hix, \& Thielemann}]{iwamoto_brachwitz_etal_1999}
Iwamoto, K., Brachwitz, F., Nomoto, K., {et~al.} 1999, The Astrophysical
  Journal Supplement Series, 125, 439, \dodoi{10.1086/313278}

\bibitem[{{Ji} {et~al.}(2020){Ji}, {Chan}, {Hummels}, {Hopkins}, {Stern},
  {Kere{\v{s}}}, {Quataert}, {Faucher-Gigu{\`e}re}, \&
  {Murray}}]{ji_chan_etal_2020}
{Ji}, S., {Chan}, T.~K., {Hummels}, C.~B., {et~al.} 2020, \mnras, 496, 4221,
  \dodoi{10.1093/mnras/staa1849}

\bibitem[{{Klypin} {et~al.}(2016){Klypin}, {Yepes}, {Gottl{\"o}ber}, {Prada},
  \& {He{\ss}}}]{klypin_yepes_etal_2016}
{Klypin}, A., {Yepes}, G., {Gottl{\"o}ber}, S., {Prada}, F., \& {He{\ss}}, S.
  2016, \mnras, 457, 4340, \dodoi{10.1093/mnras/stw248}

\bibitem[{{Kraft} {et~al.}(2022){Kraft}, {Markevitch}, {Kilbourne}, {Adams},
  {Akamatsu}, {Ayromlou}, {Bandler}, {Barbera}, {Bennett}, {Bhardwaj}, {Biffi},
  {Bodewits}, {Bogdan}, {Bonamente}, {Borgani}, {Branduardi-Raymont},
  {Bregman}, {Burchett}, {Cann}, {Carter}, {Chakraborty}, {Churazov}, {Crain},
  {Cumbee}, {Dave}, {DiPirro}, {Dolag}, {Bertrand Doriese}, {Drake}, {Dunn},
  {Eckart}, {Eckert}, {Ettori}, {Forman}, {Galeazzi}, {Gall}, {Gatuzz}, {Hell},
  {Hodges-Kluck}, {Jackman}, {Jahromi}, {Jennings}, {Jones}, {Kaaret},
  {Kavanagh}, {Kelley}, {Khabibullin}, {Kim}, {Koutroumpa}, {Kovacs}, {Kuntz},
  {Lau}, {Lee}, {Leutenegger}, {Lin}, {Lisse}, {Lo Cicero}, {Lovisari},
  {McCammon}, {McEntee}, {Mernier}, {Miller}, {Nagai}, {Negro}, {Nelson},
  {Ness}, {Nulsen}, {Ogorzalek}, {Oppenheimer}, {Oskinova}, {Patnaude},
  {Pfeifle}, {Pillepich}, {Plucinsky}, {Pooley}, {Porter}, {Randall}, {Rasia},
  {Raymond}, {Ruszkowski}, {Sakai}, {Sarkar}, {Sasaki}, {Sato},
  {Schellenberger}, {Schaye}, {Simionescu}, {Smith}, {Steiner}, {Stern}, {Su},
  {Sun}, {Tremblay}, {Truong}, {Tutt}, {Ursino}, {Veilleux}, {Vikhlinin},
  {Vladutescu-Zopp}, {Vogelsberger}, {Walker}, {Weaver}, {Weigt}, {Werk},
  {Werner}, {Wolk}, {Zhang}, {Zhang}, {Zhuravleva}, \&
  {ZuHone}}]{kraft_markevitch_etal_2022}
{Kraft}, R., {Markevitch}, M., {Kilbourne}, C., {et~al.} 2022, arXiv e-prints,
  arXiv:2211.09827, \dodoi{10.48550/arXiv.2211.09827}

\bibitem[{Kroupa(2001)}]{kroupa_2001}
Kroupa, P. 2001, Monthly Notices of the Royal Astronomical Society, 322, 231,
  \dodoi{10.1046/j.1365-8711.2001.04022.x}

\bibitem[{{Leitherer} {et~al.}(1999){Leitherer}, {Schaerer}, {Goldader},
  {Delgado}, {Robert}, {Kune}, {de Mello}, {Devost}, \&
  {Heckman}}]{starburst99}
{Leitherer}, C., {Schaerer}, D., {Goldader}, J.~D., {et~al.} 1999, \apjs, 123,
  3, \dodoi{10.1086/313233}

\bibitem[{{Leung} \& {Nomoto}(2018)}]{leung_nomoto_2018}
{Leung}, S.-C., \& {Nomoto}, K. 2018, \apj, 861, 143,
  \dodoi{10.3847/1538-4357/aac2df}

\bibitem[{{Liang} {et~al.}(2016){Liang}, {Kravtsov}, \&
  {Agertz}}]{liang_kravtsov_agertz_2016}
{Liang}, C.~J., {Kravtsov}, A.~V., \& {Agertz}, O. 2016, \mnras, 458, 1164,
  \dodoi{10.1093/mnras/stw375}

\bibitem[{{Martizzi} {et~al.}(2015){Martizzi}, {Faucher-Gigu{\`e}re}, \&
  {Quataert}}]{martizzi_faucher-giguere_quataert_2015}
{Martizzi}, D., {Faucher-Gigu{\`e}re}, C.-A., \& {Quataert}, E. 2015, \mnras,
  450, 504, \dodoi{10.1093/mnras/stv562}

\bibitem[{{McAlpine} {et~al.}(2016){McAlpine}, {Helly}, {Schaller}, {Trayford},
  {Qu}, {Furlong}, {Bower}, {Crain}, {Schaye}, {Theuns}, {Dalla Vecchia},
  {Frenk}, {McCarthy}, {Jenkins}, {Rosas-Guevara}, {White}, {Baes}, {Camps}, \&
  {Lemson}}]{mcalpine_helly_etal_2016}
{McAlpine}, S., {Helly}, J.~C., {Schaller}, M., {et~al.} 2016, Astronomy and
  Computing, 15, 72, \dodoi{10.1016/j.ascom.2016.02.004}

\bibitem[{{McQuinn}(2014)}]{mcquinn_2014}
{McQuinn}, M. 2014, \apjl, 780, L33, \dodoi{10.1088/2041-8205/780/2/L33}

\bibitem[{{Miller} \& {Bregman}(2015)}]{miller_bregman_2015}
{Miller}, M.~J., \& {Bregman}, J.~N. 2015, \apj, 800, 14,
  \dodoi{10.1088/0004-637X/800/1/14}

\bibitem[{{Mitchell} \& {Schaye}(2022)}]{mitchell_schaye_2022}
{Mitchell}, P.~D., \& {Schaye}, J. 2022, \mnras, 511, 2948,
  \dodoi{10.1093/mnras/stab3339}

\bibitem[{{Moster} {et~al.}(2013){Moster}, {Naab}, \&
  {White}}]{moster_naab_white_2013}
{Moster}, B.~P., {Naab}, T., \& {White}, S. D.~M. 2013, \mnras, 428, 3121,
  \dodoi{10.1093/mnras/sts261}

\bibitem[{{Mroczkowski} {et~al.}(2019){Mroczkowski}, {Nagai}, {Basu}, {Chluba},
  {Sayers}, {Adam}, {Churazov}, {Crites}, {Di Mascolo}, {Eckert},
  {Macias-Perez}, {Mayet}, {Perotto}, {Pointecouteau}, {Romero}, {Ruppin},
  {Scannapieco}, \& {ZuHone}}]{mroczkowski_nagai_etal_2018}
{Mroczkowski}, T., {Nagai}, D., {Basu}, K., {et~al.} 2019, \ssr, 215, 17,
  \dodoi{10.1007/s11214-019-0581-2}

\bibitem[{{Muratov} {et~al.}(2015){Muratov}, {Kere{\v{s}}},
  {Faucher-Gigu{\`e}re}, {Hopkins}, {Quataert}, \&
  {Murray}}]{muratov_keres_etal_2015}
{Muratov}, A.~L., {Kere{\v{s}}}, D., {Faucher-Gigu{\`e}re}, C.-A., {et~al.}
  2015, \mnras, 454, 2691, \dodoi{10.1093/mnras/stv2126}

\bibitem[{{Muratov} {et~al.}(2017){Muratov}, {Kere{\v{s}}},
  {Faucher-Gigu{\`e}re}, {Hopkins}, {Ma}, {Angl{\'e}s-Alc{\'a}zar}, {Chan},
  {Torrey}, {Hafen}, {Quataert}, \& {Murray}}]{muratov_keres_etal_2017}
---. 2017, \mnras, 468, 4170, \dodoi{10.1093/mnras/stx667}

\bibitem[{{Nandra} {et~al.}(2013){Nandra}, {Barret}, {Barcons}, {Fabian}, {den
  Herder}, {Piro}, {Watson}, {Adami}, {Aird}, {Afonso}, {Alexander},
  {Argiroffi}, {Amati}, {Arnaud}, {Atteia}, {Audard}, {Badenes}, {Ballet},
  {Ballo}, {Bamba}, {Bhardwaj}, {Stefano Battistelli}, {Becker}, {De Becker},
  {Behar}, {Bianchi}, {Biffi}, {B{\^\i}rzan}, {Bocchino}, {Bogdanov}, {Boirin},
  {Boller}, {Borgani}, {Borm}, {Bouch{\'e}}, {Bourdin}, {Bower}, {Braito},
  {Branchini}, {Branduardi-Raymont}, {Bregman}, {Brenneman}, {Brightman},
  {Br{\"u}ggen}, {Buchner}, {Bulbul}, {Brusa}, {Bursa}, {Caccianiga},
  {Cackett}, {Campana}, {Cappelluti}, {Cappi}, {Carrera}, {Ceballos},
  {Christensen}, {Chu}, {Churazov}, {Clerc}, {Corbel}, {Corral}, {Comastri},
  {Costantini}, {Croston}, {Dadina}, {D'Ai}, {Decourchelle}, {Della Ceca},
  {Dennerl}, {Dolag}, {Done}, {Dovciak}, {Drake}, {Eckert}, {Edge}, {Ettori},
  {Ezoe}, {Feigelson}, {Fender}, {Feruglio}, {Finoguenov}, {Fiore}, {Galeazzi},
  {Gallagher}, {Gandhi}, {Gaspari}, {Gastaldello}, {Georgakakis},
  {Georgantopoulos}, {Gilfanov}, {Gitti}, {Gladstone}, {Goosmann}, {Gosset},
  {Grosso}, {Guedel}, {Guerrero}, {Haberl}, {Hardcastle}, {Heinz}, {Alonso
  Herrero}, {Herv{\'e}}, {Holmstrom}, {Iwasawa}, {Jonker}, {Kaastra}, {Kara},
  {Karas}, {Kastner}, {King}, {Kosenko}, {Koutroumpa}, {Kraft}, {Kreykenbohm},
  {Lallement}, {Lanzuisi}, {Lee}, {Lemoine-Goumard}, {Lobban}, {Lodato},
  {Lovisari}, {Lotti}, {McCharthy}, {McNamara}, {Maggio}, {Maiolino}, {De
  Marco}, {de Martino}, {Mateos}, {Matt}, {Maughan}, {Mazzotta}, {Mendez},
  {Merloni}, {Micela}, {Miceli}, {Mignani}, {Miller}, {Miniutti}, {Molendi},
  {Montez}, {Moretti}, {Motch}, {Naz{\'e}}, {Nevalainen}, {Nicastro}, {Nulsen},
  {Ohashi}, {O'Brien}, {Osborne}, {Oskinova}, {Pacaud}, {Paerels}, {Page},
  {Papadakis}, {Pareschi}, {Petre}, {Petrucci}, {Piconcelli}, {Pillitteri},
  {Pinto}, {de Plaa}, {Pointecouteau}, {Ponman}, {Ponti}, {Porquet}, {Pounds},
  {Pratt}, {Predehl}, {Proga}, {Psaltis}, {Rafferty}, {Ramos-Ceja}, {Ranalli},
  {Rasia}, {Rau}, {Rauw}, {Rea}, {Read}, {Reeves}, {Reiprich}, {Renaud},
  {Reynolds}, {Risaliti}, {Rodriguez}, {Rodriguez Hidalgo}, {Roncarelli},
  {Rosario}, {Rossetti}, {Rozanska}, {Rovilos}, {Salvaterra}, {Salvato}, {Di
  Salvo}, {Sanders}, {Sanz-Forcada}, {Schawinski}, {Schaye}, {Schwope},
  {Sciortino}, {Severgnini}, {Shankar}, {Sijacki}, {Sim}, {Schmid}, {Smith},
  {Steiner}, {Stelzer}, {Stewart}, {Strohmayer}, {Str{\"u}der}, {Sun}, {Takei},
  {Tatischeff}, {Tiengo}, {Tombesi}, {Trinchieri}, {Tsuru}, {Ud-Doula},
  {Ursino}, {Valencic}, {Vanzella}, {Vaughan}, {Vignali}, {Vink}, {Vito},
  {Volonteri}, {Wang}, {Webb}, {Willingale}, {Wilms}, {Wise}, {Worrall},
  {Young}, {Zampieri}, {In't Zand}, {Zane}, {Zezas}, {Zhang}, \&
  {Zhuravleva}}]{athena_white_paper}
{Nandra}, K., {Barret}, D., {Barcons}, X., {et~al.} 2013, arXiv e-prints,
  arXiv:1306.2307.
\newblock \doarXiv{1306.2307}

\bibitem[{{Navarro} {et~al.}(1997){Navarro}, {Frenk}, \& {White}}]{NFW_1997}
{Navarro}, J.~F., {Frenk}, C.~S., \& {White}, S. D.~M. 1997, \apj, 490, 493,
  \dodoi{10.1086/304888}

\bibitem[{{Nelson} {et~al.}(2023){Nelson}, {Byrohl}, {Ogorzalek}, {Markevitch},
  {Khabibullin}, {Churazov}, {Zhuravleva}, {Bogdan}, {Chakraborty},
  {Kilbourne}, {Kraft}, {Pillepich}, {Sarkar}, {Schellenberger}, {Su},
  {Truong}, {Vladutescu-Zopp}, \& {Wijers}}]{nelson_byrohl_etal_2023}
{Nelson}, D., {Byrohl}, C., {Ogorzalek}, A., {et~al.} 2023, \mnras, 522, 3665,
  \dodoi{10.1093/mnras/stad1195}

\bibitem[{{Nicastro}(2016)}]{nicastro_etal_2016}
{Nicastro}, F. 2016, in XMM-Newton: The Next Decade, 27.
\newblock \doarXiv{1611.03722}

\bibitem[{{Nomoto} {et~al.}(2006){Nomoto}, {Tominaga}, {Umeda}, {Kobayashi}, \&
  {Maeda}}]{nomoto_tominaga_etal_2006}
{Nomoto}, K., {Tominaga}, N., {Umeda}, H., {Kobayashi}, C., \& {Maeda}, K.
  2006, \nphysa, 777, 424, \dodoi{10.1016/j.nuclphysa.2006.05.008}

\bibitem[{{Pandya} {et~al.}(2020){Pandya}, {Somerville},
  {Angl{\'e}s-Alc{\'a}zar}, {Hayward}, {Bryan}, {Fielding}, {Forbes},
  {Burkhart}, {Genel}, {Hernquist}, {Kim}, {Tonnesen}, \&
  {Starkenburg}}]{pandya_somerville_etal_2020}
{Pandya}, V., {Somerville}, R.~S., {Angl{\'e}s-Alc{\'a}zar}, D., {et~al.} 2020,
  \apj, 905, 4, \dodoi{10.3847/1538-4357/abc3c1}

\bibitem[{P\'erez \& Granger(2007)}]{ipython}
P\'erez, F., \& Granger, B.~E. 2007, Computing in Science \& Engineering, 9,
  21, \dodoi{10.1109/MCSE.2007.53}

\bibitem[{{Perna} \& {Loeb}(1998)}]{perna_loeb_1998}
{Perna}, R., \& {Loeb}, A. 1998, \apjl, 503, L135, \dodoi{10.1086/311544}

\bibitem[{{Ploeckinger} \&
  {Schaye}(2020{\natexlab{a}})}]{ploeckinger_schaye_2020_tableref}
{Ploeckinger}, S., \& {Schaye}, J. 2020{\natexlab{a}}, {Rates and fractions},
  V1,  Harvard Dataverse, \dodoi{10.7910/DVN/GR3L5N}

\bibitem[{{Ploeckinger} \&
  {Schaye}(2020{\natexlab{b}})}]{ploeckinger_schaye_2020}
---. 2020{\natexlab{b}}, \mnras, 497, 4857, \dodoi{10.1093/mnras/staa2172}

\bibitem[{{Power} {et~al.}(2003){Power}, {Navarro}, {Jenkins}, {Frenk},
  {White}, {Springel}, {Stadel}, \& {Quinn}}]{power_navarro_etal_2003}
{Power}, C., {Navarro}, J.~F., {Jenkins}, A., {et~al.} 2003, \mnras, 338, 14,
  \dodoi{10.1046/j.1365-8711.2003.05925.x}

\bibitem[{{Prochaska} {et~al.}(2019){Prochaska}, {Burchett}, {Tripp}, {Werk},
  {Willmer}, {Howk}, {Lange}, {Tejos}, {Meiring}, {Tumlinson}, {Lehner},
  {Ford}, \& {Dav{\'e}}}]{prochaska_burchett_etal_2019}
{Prochaska}, J.~X., {Burchett}, J.~N., {Tripp}, T.~M., {et~al.} 2019, \apjs,
  243, 24, \dodoi{10.3847/1538-4365/ab2b9a}

\bibitem[{{Qu} \& {Bregman}(2018)}]{qu_bregman_2018}
{Qu}, Z., \& {Bregman}, J.~N. 2018, \apj, 856, 5,
  \dodoi{10.3847/1538-4357/aaafd4}

\bibitem[{{Qu} {et~al.}(2024){Qu}, {Chen}, {Johnson}, {Rudie}, {Zahedy},
  {DePalma}, {Schaye}, {Boettcher}, {Cantalupo}, {Chen}, {Faucher-Gigu{\`e}re},
  {Li}, {Mulchaey}, {Petitjean}, \& {Rafelski}}]{qu_chen_etal_2024_preprint}
{Qu}, Z., {Chen}, H.-W., {Johnson}, S.~D., {et~al.} 2024, arXiv e-prints,
  arXiv:2402.08016, \dodoi{10.48550/arXiv.2402.08016}

\bibitem[{{Ravi}(2019)}]{ravi_2019}
{Ravi}, V. 2019, \apj, 872, 88, \dodoi{10.3847/1538-4357/aafb30}

\bibitem[{{Rodr{\'\i}guez-Puebla} {et~al.}(2016){Rodr{\'\i}guez-Puebla},
  {Behroozi}, {Primack}, {Klypin}, {Lee}, \&
  {Hellinger}}]{rodriguez-puebla_behroozi_etal_2016}
{Rodr{\'\i}guez-Puebla}, A., {Behroozi}, P., {Primack}, J., {et~al.} 2016,
  \mnras, 462, 893, \dodoi{10.1093/mnras/stw1705}

\bibitem[{{Schaye} {et~al.}(2015){Schaye}, {Crain}, {Bower}, {Furlong},
  {Schaller}, {Theuns}, {Dalla Vecchia}, {Frenk}, {McCarthy}, {Helly},
  {Jenkins}, {Rosas-Guevara}, {White}, {Baes}, {Booth}, {Camps}, {Navarro},
  {Qu}, {Rahmati}, {Sawala}, {Thomas}, \& {Trayford}}]{eagle_paper}
{Schaye}, J., {Crain}, R.~A., {Bower}, R.~G., {et~al.} 2015, \mnras, 446, 521,
  \dodoi{10.1093/mnras/stu2058}

\bibitem[{{Smith} {et~al.}(2016){Smith}, {Abraham}, {Allured}, {Bautz},
  {Bookbinder}, {Bregman}, {Brenneman}, {Brickhouse}, {Burrows}, {Burwitz},
  {Carvalho}, {Cheimets}, {Costantini}, {Dawson}, {DeRoo}, {Falcone}, {Foster},
  {Grant}, {Heilmann}, {Hertz}, {Hine}, {Huenemoerder}, {Kaastra}, {Madsen},
  {McEntaffer}, {Miller}, {Miller}, {Morse}, {Mushotzky}, {Nandra}, {Nowak},
  {Paerels}, {Petre}, {Plice}, {Poppenhaeger}, {Ptak}, {Reid}, {Sanders},
  {Schattenburg}, {Schulz}, {Smale}, {Temi}, {Valencic}, {Walker},
  {Willingale}, {Wilms}, \& {Wolk}}]{smith_abraham_etal_2016_arcus}
{Smith}, R.~K., {Abraham}, M.~H., {Allured}, R., {et~al.} 2016, in \procspie,
  Vol. 9905, Space Telescopes and Instrumentation 2016: Ultraviolet to Gamma
  Ray (Bellingham, Washington USA: Society of Photo-Optical Instrumentation
  Engineers (SPIE)), 99054M, \dodoi{10.1117/12.2231778}

\bibitem[{{Stern} {et~al.}(2018){Stern}, {Faucher-Gigu{\`e}re}, {Hennawi},
  {Hafen}, {Johnson}, \& {Fielding}}]{stern_faucher-giguere_etal_2018}
{Stern}, J., {Faucher-Gigu{\`e}re}, C.-A., {Hennawi}, J.~F., {et~al.} 2018,
  \apj, 865, 91, \dodoi{10.3847/1538-4357/aac884}

\bibitem[{{Stern} {et~al.}(2019){Stern}, {Fielding}, {Faucher-Gigu{\`e}re}, \&
  {Quataert}}]{stern_fielding_etal_2019}
{Stern}, J., {Fielding}, D., {Faucher-Gigu{\`e}re}, C.-A., \& {Quataert}, E.
  2019, \mnras, 488, 2549, \dodoi{10.1093/mnras/stz1859}

\bibitem[{{Strawn} {et~al.}(2024){Strawn}, {Roca-F{\`a}brega}, {Primack},
  {Kim}, {Genina}, {Hausammann}, {Kim}, {Lupi}, {Nagamine}, {Powell}, {Revaz},
  {Shimizu}, {Vel{\'a}zquez}, {Abel}, {Ceverino}, {Dong}, {Jung}, {Quinn},
  {Shin}, {Barrow}, {Dekel}, {Oh}, {Mandelker}, {Teyssier}, {Hummels}, {Maji},
  {Man}, {Mayerhofer}, \& {The Agora
  Collaboration}}]{strawn_roca-fabrega_etal_2024}
{Strawn}, C., {Roca-F{\`a}brega}, S., {Primack}, J.~R., {et~al.} 2024, \apj,
  962, 29, \dodoi{10.3847/1538-4357/ad12cb}

\bibitem[{Su {et~al.}(2019)Su, Hopkins, Hayward, Ma, Faucher-Giguère, Kereš,
  Orr, Chan, \& Robles}]{su_hopkins_etal_2019}
Su, K.-Y., Hopkins, P.~F., Hayward, C.~C., {et~al.} 2019, Monthly Notices of
  the Royal Astronomical Society, 487, 4393, \dodoi{10.1093/mnras/stz1494}

\bibitem[{{Su} {et~al.}(2021){Su}, {Hopkins}, {Bryan}, {Somerville}, {Hayward},
  {Angl{\'e}s-Alc{\'a}zar}, {Faucher-Gigu{\`e}re}, {Wellons}, {Stern},
  {Terrazas}, {Chan}, {Orr}, {Hummels}, {Feldmann}, \&
  {Kere{\v{s}}}}]{su_hopkins_etal_2021}
{Su}, K.-Y., {Hopkins}, P.~F., {Bryan}, G.~L., {et~al.} 2021, \mnras, 507, 175,
  \dodoi{10.1093/mnras/stab2021}

\bibitem[{{Sukhbold} {et~al.}(2016){Sukhbold}, {Ertl}, {Woosley}, {Brown}, \&
  {Janka}}]{sukhbold_ertl_etal_2016}
{Sukhbold}, T., {Ertl}, T., {Woosley}, S.~E., {Brown}, J.~M., \& {Janka}, H.~T.
  2016, \apj, 821, 38, \dodoi{10.3847/0004-637X/821/1/38}

\bibitem[{{The EAGLE team}(2017)}]{eagle-team_2017}
{The EAGLE team}. 2017, arXiv e-prints, arXiv:1706.09899.
\newblock \doarXiv{1706.09899}

\bibitem[{{The~Lynx~Team}(2018)}]{lynx_2018_08}
{The~Lynx~Team}. 2018, arXiv e-prints.
\newblock \doarXiv{1809.09642}

\bibitem[{{The pandas development team}(2020)}]{reback_2020_pandas}
{The pandas development team}. 2020, pandas-dev/pandas: Pandas, latest,
  Zenodo, \dodoi{10.5281/zenodo.3509134}

\bibitem[{{Torrey} {et~al.}(2020){Torrey}, {Hopkins}, {Faucher-Gigu{\`e}re},
  {Angl{\'e}s-Alc{\'a}zar}, {Quataert}, {Ma}, {Feldmann}, {Keres}, \&
  {Murray}}]{torrey_hopkins_etal_2020}
{Torrey}, P., {Hopkins}, P.~F., {Faucher-Gigu{\`e}re}, C.-A., {et~al.} 2020,
  \mnras, 497, 5292, \dodoi{10.1093/mnras/staa2222}

\bibitem[{{Tripp} {et~al.}(2011){Tripp}, {Meiring}, {Prochaska}, {Willmer},
  {Howk}, {Werk}, {Jenkins}, {Bowen}, {Lehner}, {Sembach}, {Thom}, \&
  {Tumlinson}}]{tripp_meiring_etal_2011}
{Tripp}, T.~M., {Meiring}, J.~D., {Prochaska}, J.~X., {et~al.} 2011, Science,
  334, 952, \dodoi{10.1126/science.1209850}

\bibitem[{{Tsai} \& {Mathews}(1995)}]{tsai_mathews_1995}
{Tsai}, J.~C., \& {Mathews}, W.~G. 1995, \apj, 448, 84, \dodoi{10.1086/175943}

\bibitem[{{Tumlinson} {et~al.}(2017){Tumlinson}, {Peeples}, \&
  {Werk}}]{tumlinson_peeples_werk_2017_cgmreview}
{Tumlinson}, J., {Peeples}, M.~S., \& {Werk}, J.~K. 2017, \araa, 55, 389,
  \dodoi{10.1146/annurev-astro-091916-055240}

\bibitem[{{Upton Sanderbeck} {et~al.}(2018){Upton Sanderbeck}, {McQuinn},
  {D'Aloisio}, \& {Werk}}]{upton-sanderbeck_mcquinn_etal_2018}
{Upton Sanderbeck}, P.~R., {McQuinn}, M., {D'Aloisio}, A., \& {Werk}, J.~K.
  2018, \apj, 869, 159, \dodoi{10.3847/1538-4357/aaeff2}

\bibitem[{{van de Voort} {et~al.}(2021){van de Voort}, {Bieri}, {Pakmor},
  {G{\'o}mez}, {Grand}, \& {Marinacci}}]{van-de-voort_bieri_etal_2021}
{van de Voort}, F., {Bieri}, R., {Pakmor}, R., {et~al.} 2021, \mnras, 501,
  4888, \dodoi{10.1093/mnras/staa3938}

\bibitem[{{Verner} {et~al.}(1994){Verner}, {Tytler}, \&
  {Barthel}}]{verner_tytler_barthel_1994}
{Verner}, D.~A., {Tytler}, D., \& {Barthel}, P.~D. 1994, \apj, 430, 186,
  \dodoi{10.1086/174392}

\bibitem[{Virtanen {et~al.}(2020)Virtanen, Gommers, Oliphant, Haberland, Reddy,
  Cournapeau, Burovski, Peterson, Weckesser, Bright, {van der Walt}, Brett,
  Wilson, Millman, Mayorov, Nelson, Jones, Kern, Larson, Carey, Polat, Feng,
  Moore, {VanderPlas}, Laxalde, Perktold, Cimrman, Henriksen, Quintero, Harris,
  Archibald, Ribeiro, Pedregosa, {van Mulbregt}, \& {SciPy 1.0
  Contributors}}]{scipy_2020}
Virtanen, P., Gommers, R., Oliphant, T.~E., {et~al.} 2020, Nature Methods, 17,
  261, \dodoi{10.1038/s41592-019-0686-2}

\bibitem[{{Voit}(2019)}]{voit_2019}
{Voit}, G.~M. 2019, \apj, 880, 139, \dodoi{10.3847/1538-4357/ab2bfd}

\bibitem[{{Wellons} {et~al.}(2023){Wellons}, {Faucher-Gigu{\`e}re}, {Hopkins},
  {Quataert}, {Angl{\'e}s-Alc{\'a}zar}, {Feldmann}, {Hayward}, {Kere{\v{s}}},
  {Su}, \& {Wetzel}}]{wellons_faucher-giguere_etal_2023}
{Wellons}, S., {Faucher-Gigu{\`e}re}, C.-A., {Hopkins}, P.~F., {et~al.} 2023,
  \mnras, 520, 5394, \dodoi{10.1093/mnras/stad511}

\bibitem[{Wendland(1995)}]{wendland_1995}
Wendland, H. 1995, Advances in Computational Mathematics, 4, 389,
  \dodoi{10.1007/BF02123482}

\bibitem[{{Werk} {et~al.}(2014){Werk}, {Prochaska}, {Tumlinson}, {Peeples},
  {Tripp}, {Fox}, {Lehner}, {Thom}, {O'Meara}, {Ford}, {Bordoloi}, {Katz},
  {Tejos}, {Oppenheimer}, {Dav{\'e}}, \&
  {Weinberg}}]{werk_proschaska_etal_2014}
{Werk}, J.~K., {Prochaska}, J.~X., {Tumlinson}, J., {et~al.} 2014, \apj, 792,
  8, \dodoi{10.1088/0004-637X/792/1/8}

\bibitem[{{W}es {M}c{K}inney(2010)}]{mckinney_2010_pandas}
{W}es {M}c{K}inney. 2010, in {P}roceedings of the 9th {P}ython in {S}cience
  {C}onference, ed. {S}t\'efan van~der {W}alt \& {J}arrod {M}illman, 56 -- 61,
  \dodoi{10.25080/Majora-92bf1922-00a}

\bibitem[{{Wetzel} {et~al.}(2023){Wetzel}, {Hayward}, {Sanderson}, {Ma},
  {Angl{\'e}s-Alc{\'a}zar}, {Feldmann}, {Chan}, {El-Badry}, {Wheeler},
  {Garrison-Kimmel}, {Nikakhtar}, {Panithanpaisal}, {Arora}, {Gurvich},
  {Samuel}, {Sameie}, {Pandya}, {Hafen}, {Hummels}, {Loebman},
  {Boylan-Kolchin}, {Bullock}, {Faucher-Gigu{\`e}re}, {Kere{\v{s}}},
  {Quataert}, \& {Hopkins}}]{wetzel_hayward_etal_2023}
{Wetzel}, A., {Hayward}, C.~C., {Sanderson}, R.~E., {et~al.} 2023, \apjs, 265,
  44, \dodoi{10.3847/1538-4365/acb99a}

\bibitem[{{Wiersma} {et~al.}(2009){Wiersma}, {Schaye}, \&
  {Smith}}]{wiersema_etal_2009_tables}
{Wiersma}, R.~P.~C., {Schaye}, J., \& {Smith}, B.~D. 2009, \mnras, 393, 99,
  \dodoi{10.1111/j.1365-2966.2008.14191.x}

\bibitem[{{Wijers} \& {Schaye}(2022)}]{wijers_schaye_2022}
{Wijers}, N.~A., \& {Schaye}, J. 2022, \mnras, 514, 5214,
  \dodoi{10.1093/mnras/stac1580}

\bibitem[{{Wijers} {et~al.}(2020){Wijers}, {Schaye}, \&
  {Oppenheimer}}]{wijers_schaye_oppenheimer_2020}
{Wijers}, N.~A., {Schaye}, J., \& {Oppenheimer}, B.~D. 2020, \mnras, 498, 574,
  \dodoi{10.1093/mnras/staa2456}

\bibitem[{{Yao} \& {Wang}(2005)}]{yao_wang_2005}
{Yao}, Y., \& {Wang}, Q.~D. 2005, \apj, 624, 751, \dodoi{10.1086/429537}

\bibitem[{{Zhang} {et~al.}(2024){Zhang}, {Comparat}, {Ponti}, {Meloni},
  {Nandra}, {Haberl}, {Locatelli}, {Zhang}, {Sanders}, {Zheng}, {Liu},
  {Popesso}, {Liu}, {Truong}, {Pillepich}, {Predehl}, \&
  {Salvato}}]{zhang_comparat_etal_2024_preprint}
{Zhang}, Y., {Comparat}, J., {Ponti}, G., {et~al.} 2024, arXiv e-prints,
  arXiv:2401.17308, \dodoi{10.48550/arXiv.2401.17308}

\bibitem[{{Zhu} \& {Springel}(2024)}]{zhu_springel_2024_preprint}
{Zhu}, B., \& {Springel}, V. 2024, arXiv e-prints, arXiv:2404.13837,
  \dodoi{10.48550/arXiv.2404.13837}

\bibitem[{{Zhu} {et~al.}(2021){Zhu}, {Xu}, {Hu}, {Shan}, {Zhu}, {Fan}, {Zhao},
  {Gu}, \& {Wu}}]{zhu_xu_2021}
{Zhu}, Z., {Xu}, H., {Hu}, D., {et~al.} 2021, \apj, 908, 17,
  \dodoi{10.3847/1538-4357/abd327}

\end{thebibliography}



\appendix

\section{Estimating halo masses from stellar masses}
\label{app:smhm}

Here, we explore some effects of different ways of calculating the halo mass from a galaxy stellar mass. The effects are largest at central galaxy stellar masses of $\Mstellar \gtrsim 10^{10.5} \Msun$. This is because the halo mass increases more strongly with stellar mass above this mass than below it, and because the scatter in halo masses at fixed stellar mass is higher above this mass than below it. 
Using the median halo mass at a given stellar mass, and not the median stellar mass at a given halo mass, to estimate halo masses is important. 
Furthermore, when calculating the best-estimate halo masses and the halo mass uncertainty, it is important to account for the fact that there are a range of possible halo masses for a given galaxy stellar mass.

First, we outline the way we calculate halo mass probability distributions throughout this work. 
To calculate halo masses for the \citet[CASBaH]{burchett_tripp_etal_2018} halos, we take their reported log stellar masses and uncertainties, and assume the probability distribution for the true log stellar mass is a gaussian distribution with the best-estimate stellar mass as the mean and the uncertainty as the variance.
We then use the distribution of halo masses at a given central galaxy stellar mass in the UniverseMachine catalog to translate these to probability distributions for parent halo masses. The probability for a given halo mass bin is 
\begin{equation}
\begin{aligned}
P(\mathrm{M}_{\mathrm{h}} & \in [\mathrm{M}_{\mathrm{h}, j}, \mathrm{M}_{\mathrm{h}, j + 1}])
=  \\
\sum_{i} & \left\{ 
P(\mathrm{M}_{\star} \in [\mathrm{M}_{\star, i}, \mathrm{M}_{\star, i + 1}]) \right.\\
& \cdot
\left. P(\mathrm{M}_{\mathrm{h}}\in [\mathrm{M}_{\mathrm{h}, j}, \mathrm{M}_{\mathrm{h}, j + 1}] \,|\, \mathrm{M}_{\star} \in [\mathrm{M}_{\star, i}, \mathrm{M}_{\star, i + 1}]) \right\},
\end{aligned}
\label{eq:mstomhmat}
\end{equation}
where $\mathrm{M}_{\mathrm{h}, j}$ and $\mathrm{M}_{\star, i}$ are edges of bins in halo and stellar mass, respectively. For the probabilities $P(\mathrm{M}_{\mathrm{h}} | \Mstellar)$, we simply use a normalized histogram of the UniverseMachine stellar and halo masses.
We note that there are further (systematic) uncertainties associated with the measurement of stellar masses, which we do not account for here.
As the relation depends on redshift, we use the halo catalog for the redshift closest to the measured galaxy redshift. 

We calculate the \citet[CUBS]{qu_chen_etal_2024_preprint} halo masses in the same way as the \citet{burchett_tripp_etal_2018} masses. However, the estimated probability distribution for the true stellar mass of a galaxy is a bit more complicated.
\citet{qu_chen_etal_2024_preprint} report different lower and upper error ranges for the stellar masses of some galaxies. The errors they report are $2\sigma$. We therefore approximate the true stellar mass probability distribution for a measured galaxy as two half lognormal distributions (each with a total probability of 0.5), centered at the measured value and with a variance equal to the upper/lower uncertainty. We estimate the $1\sigma$ upper/lower uncertainty $\sigma_{\mathrm{tot}}$ by combining the reported statistical uncertainty and the systematic uncertainty $\sigma_{\mathrm{sys}}$ of $0.2 \dex$ of the stellar mass measurements. We estimate the $1\sigma$ statistical uncertainty as half the reported $2\sigma$ statistical uncertainties $\sigma_{2, \mathrm{stat}}$. This means the variance of each half-lognormal distribution is estimated as $\sigma_{\mathrm{tot}} = \sqrt(\sigma_{2, \mathrm{stat}}^2 \,/\, 4 + \sigma_{\mathrm{sys}}^2)$.

\begin{figure*}
\centering
\includegraphics[width=0.9\textwidth]{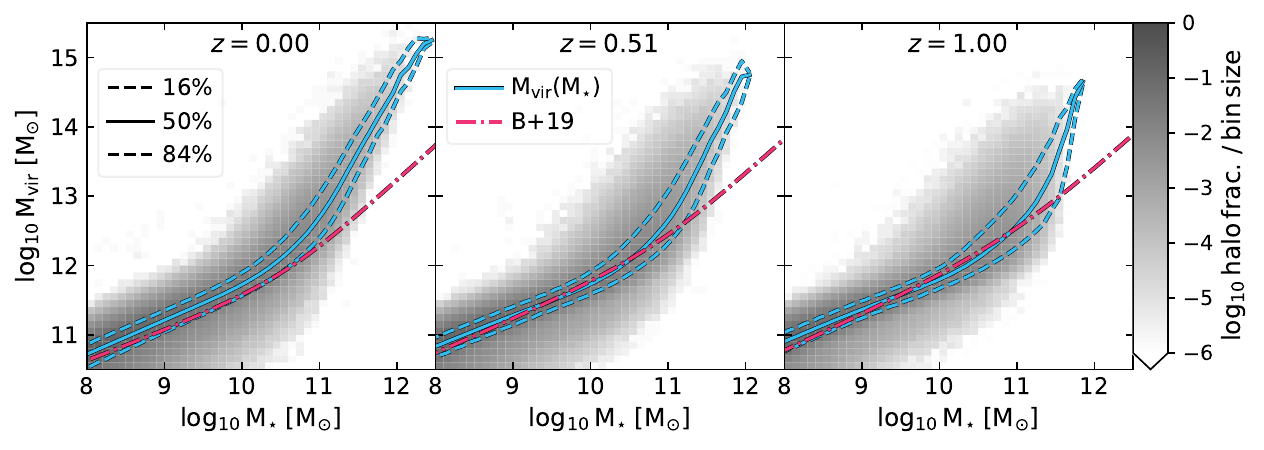}
\caption{A histogram of the joint (central galaxy) stellar and halo mass distribution (grayscale), taken from the UniverseMachine abundance-matching fits applied to the SMDPL simulations. The cyan lines show different percentiles of $\Mvir(\Mstellar)$, the halo mass as a function of stellar mass, for the UniverseMachine galaxies. The red curve shows a modified version of the \citet{moster_naab_white_2013} relation, introduced by \citet{burchett_tripp_etal_2016} and used by \citet{burchett_tripp_etal_2018} to calculate the virial radii in their tab.~1. We convert the \citet{burchett_tripp_etal_2018} $\mathrm{M}_{\mathrm{200c}}$ halo masses to the \citet{bryan_norman_1998} definition assuming their mass profiles are NFW \citep{NFW_1997}, with concentrations following \citet{dutton_maccio_2014}. Different relations give different halo masses for a measured stellar mass, and the scatter in halo mass at fixed stellar mass can be large. The differences and scatter are largest at $\Mstellar \gtrsim 10^{10.5} \Msun$.} 
\label{fig:msmh}
\end{figure*}

In Fig.~\ref{fig:msmh}, we show the distribution of the UniverseMachine stellar and halo masses for central galaxies at redshifts~0, 0.5, and~1. The colored lines show relations between stellar mass and halo mass. At galaxy stellar masses $\Mstellar \gtrsim 10^{10.5} \Msun$, the \citet{burchett_tripp_etal_2016} relation used by \citet{burchett_tripp_etal_2018} differs considerably from the median halo mass at a given stellar mass. The difference is 
$\lesssim 0.35 \dex$ at $\Mstellar \approx 10^{11} \Msun$ at these redshifts, but reaches $\gtrsim 0.6 \dex$ at $\Mstellar \approx 10^{11.5} \Msun$.
The scatter in halo mass at fixed stellar mass is also considerable, especially at $\Mstellar \gtrsim 10^{10.5} \Msun$: it is  $\approx 0.3$--$0.4 \dex$ at  $\Mstellar \approx 10^{11} \Msun$.

\begin{figure}
\centering
\includegraphics[width=\columnwidth]{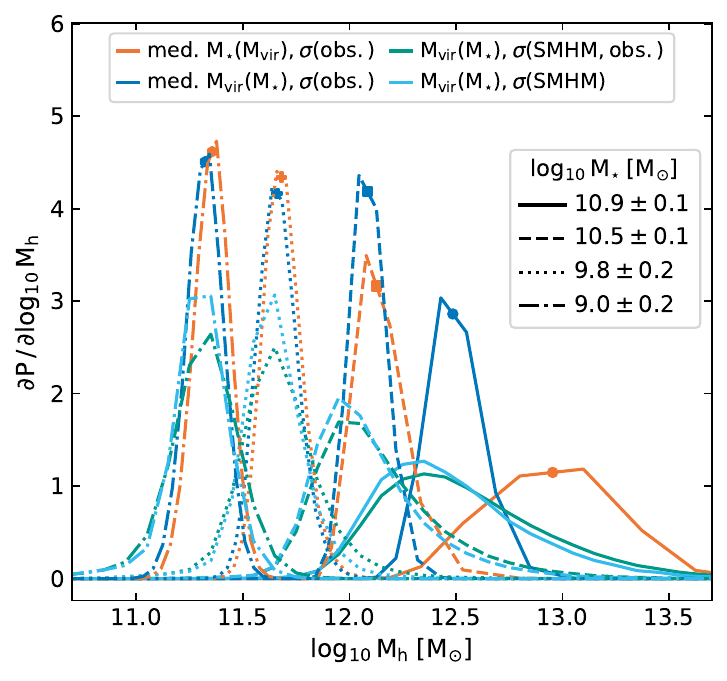}
\caption{Probability densities for host halo masses of galaxies with different stellar masses (different line styles) from \citet{burchett_tripp_etal_2018}, assuming a redshift $z=0.74$. The curves of different colors represent different methods for calculating the halo mass at a given stellar mass. The teal curves account for uncertainty in the measured stellar mass and scatter in the stellar-mass-halo-mass relation. The cyan curves instead account only for the scatter in the stellar-mass-halo-mass relation, and the blue curves only account for uncertainty in the measured stellar mass. The orange curves are obtained in the same way as the blue curves, except that we calculate the halo mass at a given stellar mass by mathematically inverting the median stellar mass as a function of halo mass.}
\label{fig:mherr}
\end{figure}

In Fig.~\ref{fig:mherr} we calculate the halo mass probability density functions for a few representative galaxy stellar masses and uncertainties from  \citet{burchett_tripp_etal_2018}. Different line styles are for different galaxies, different colors are for different relations. 
For the teal curves, we use the method outlined above, which we use throughout this work, except that we assume a single redshift $z=0.74$ for this example.

For the cyan curve, 
we ignored errors in the stellar mass measurements. Instead, the probability distribution is just the normalized halo mass histogram at the best-estimate stellar mass. The blue curve effectively does the opposite: we include uncertainties in the true stellar mass, but ignore scatter in the SMHM relation. Here, we still start with a probability for different stellar mass bins. However, we simply convert the stellar mass bin edges to halo mass bin edges using the median halo mass at a given stellar mass. The probabilities remain the same, and we simply divide by the resulting halo mass bin sizes to obtain the probability density. Markers on the blue curves show the halo mass corresponding to the best-estimate stellar mass.

Comparing the blue and cyan curves to the teal, we see that the uncertainty in the halo masses largely comes from the scatter in the SMHM relation at all these masses. However, the SMHM scatter is a more dominant error source at high stellar and halo masses. At masses $\Mstellar \gtrsim 10^{10.5} \Msun$, obtaining the most likely halo masses also requires accounting for the SMHM scatter, although at lower masses, the median relation suffices for this.

The orange curves show a method of calculating the halo masses which we strongly caution against. The approach is similar to that of the blue curve, using a one-to-one relation between halo mass and stellar mass, and simply propagating the uncertainties in the stellar mass through that relation to obtain the uncertainties in the halo mass. However, here, we use the median \emph{stellar} mass at a given \emph{halo} mass to define the relation. We calculate median log stellar masses over a range of log halo masses, and simply linearly interpolate between the resulting points to obtain the halo mass at a given stellar mass. 

At low stellar masses ($\lesssim 10^{10.5} \Msun$), this provides reasonable best-estimate halo masses. The lack of inclusion of SMHM scatter does lead to an underestimate of the halo mass uncertainty, but the distributions are very similar to those obtained using the more appropriate one-to-one relation, median halo mass at a given stellar mass. However, at higher stellar masses 
($\gtrsim 10^{10.5} \Msun$), the median stellar mass at a given halo mass yields considerably higher halo mass estimates than the median halo mass at a given stellar mass. At these masses, it is important to both use the correct relation, and to account for SMHM scatter.

\citet{burchett_tripp_etal_2018} calculated halo masses for their galaxies as well. They use the method of
\citet{burchett_tripp_etal_2016}, who used the stellar-mass-halo-mass relation of \citet{moster_naab_white_2013} as a starting point for their relation. \citet{moster_naab_white_2013} fit a function for the mean stellar mass at a given halo mass, not the mean or median halo mass at a given stellar mass. \citet{burchett_tripp_etal_2016} argued that at high galaxy masses, a flatter slope than \citet{moster_naab_white_2013} found was more appropriate for their isolated galaxies.

We argue that the differences between the \citet{burchett_tripp_etal_2016} stellar-mass-halo-mass relation and the median UniverseMachine halo mass at a given stellar mass (Fig.~\ref{fig:msmh}) are most likely explained by the difference between the median halo mass at a given stellar mass, and a mathematically inverted median stellar mass as a function of halo mass. 
We note that the flatter \citet{burchett_tripp_etal_2016} slope at high stellar masses reasonably matches the median halo mass at a given stellar mass up to almost $\Mstellar \approx 10^{11} \Msun$, which is above the $\Mstellar \approx 10^{10.5} \Msun$ where the $\Mstellar(\Mvir)$ and $\Mvir(\Mstellar)$ median relations diverge.

\section{m12 FIRE-3 NoBH halos with higher halo masses}
\label{app:m12plus}

In the main body of the paper, we analysed samples of m12 halos with similar halo masses for the four different physics models we explored. However, these halo masses correspond to different central galaxy stellar masses in the different FIRE models. In particular, the FIRE-2 NoBH model produces higher central galaxy stellar masses in its m12 halos than the FIRE-3 models.

\begin{table*}
\caption{As Tab.~\ref{tab:sims}, but for an additional set of m12 FIRE-3 NoBH halos, some of which have higher halo and stellar masses than the FIRE-3 NoBH sample we analyse in the main body of this work.}
\label{tab:m12plus}
\begin{tabular}{l l l l l l l l l}
\hline
         &       &       & \multicolumn{3}{c}{$z=1.0$}   & \multicolumn{3}{c}{$z=0.5$}   \\
ICs      & model         & resolution    & $\mathrm{M}_{\mathrm{vir}}$   & $\mathrm{M}_{\star}$          & $\mathrm{R}_{\mathrm{vir}}$   & $\mathrm{M}_{\mathrm{vir}}$   & $\mathrm{M}_{\star}$  & $\mathrm{R}_{\mathrm{vir}}$    \\
         &       & $[\mathrm{M}_{\odot}]$        & $[\mathrm{M}_{\odot}]$        & $[\mathrm{M}_{\odot}]$        & [pkpc]        & $[\mathrm{M}_{\odot}]$        & $[\mathrm{M}_{\odot}]$        & [pkpc]         \\
\hline
m12a     & FIRE-3 NoBH       & 6e4   & 8.3e11        & 1.1e10        & 146   & 1.2e12        & 2.3e10        & 211   \\
m12d     & FIRE-3 NoBH       & 6e4   & 5.2e11        & 5.1e9         & 124   & 7.0e11        & 9.7e9         & 175   \\
m12e     & FIRE-3 NoBH       & 6e4   & 1.0e12        & 2.8e9         & 157   & 1.4e12        & 5.6e9         & 221   \\
m12g     & FIRE-3 NoBH       & 6e4   & 1.7e12        & 1.7e10        & 186   & 1.9e12        & 4.5e10        & 245   \\
m12j     & FIRE-3 NoBH       & 6e4   & 2.8e11        & 1.7e9         & 102   & 7.1e11        & 4.8e9         & 176   \\
m12j     & FIRE-3 NoBH       & 7e3   & 3.1e11        & 1.9e9         & 105   & 7.1e11        & 9.1e9         & 176   \\
m12k     & FIRE-3 NoBH       & 6e4   & 1.2e12        & 1.2e10        & 166   & 1.9e12        & 2.5e10        & 245   \\
m12n     & FIRE-3 NoBH       & 6e4   & 7.8e11        & 2.1e9         & 143   & 8.9e11        & 5.8e9         & 190   \\
m12n     & FIRE-3 NoBH       & 7e3   & 8.3e11        & 1.3e10        & 146   & 9.3e11        & 1.5e10        & 193   \\
m12u     & FIRE-3 NoBH       & 3e4   & 5.5e11        & 1.7e9         & 127   & 5.2e11        & 4.3e9         & 158   \\
m12x     & FIRE-3 NoBH       & 3e4   & 2.7e11        & 1.6e9         & 100   & 5.0e11        & 2.4e9         & 157   \\
m12x     & FIRE-3 NoBH       & 4e3   & 2.6e11        & 1.1e9         & 98    & 4.1e11        & 1.4e9         & 147   \\
\hline
\end{tabular}
\end{table*}

After the simulations in the main body of this paper were run, additional m12 galaxies were simulated with the FIRE-3 NoBH model. Some of these have higher halo masses than our main m12 FIRE-3 NoBH sample, and central galaxy stellar masses comparable to those of our FIRE-2 NoBH halo sample. Three of the halos have resolutions matching those of our other m12 FIRE-3 NoBH halos, the other nine were run at lower resolution. These halos are listed in Tab.~\ref{tab:m12plus}. 

\begin{figure}
\centering
\includegraphics[width=\columnwidth]{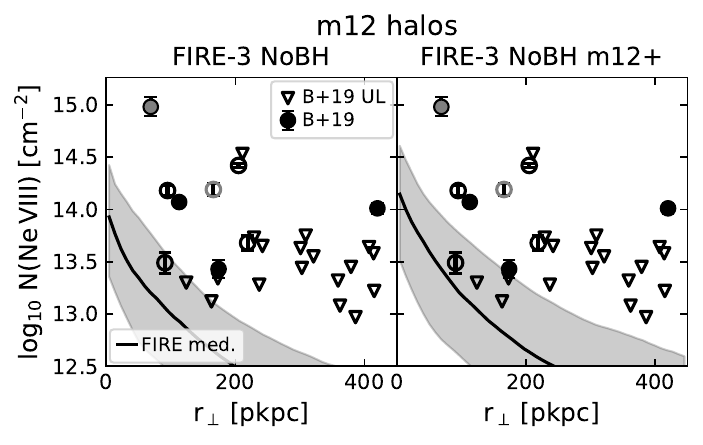} 
\includegraphics[width=\columnwidth]{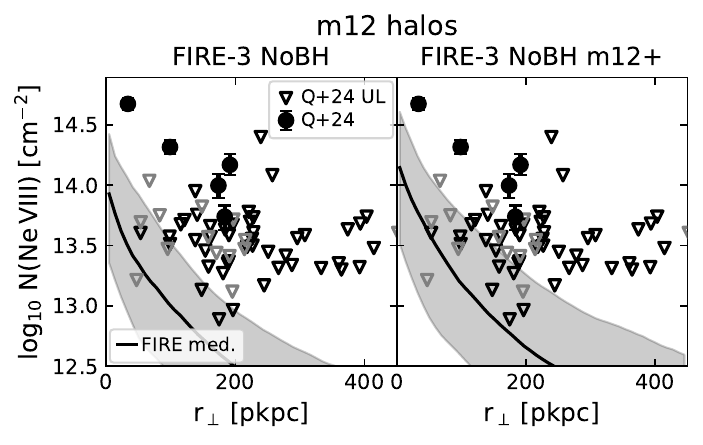} 
\caption{A comparison of the m12 FIRE-3 NoBH halos to the \citet[CASBaH, B+19]{burchett_tripp_etal_2018} data (top panels) and the \citet[CUBS, Q+24]{qu_chen_etal_2024_preprint} data (bottom panels). Like Figs.~\ref{fig:ne8_obscomp_b19} and~\ref{fig:ne8_obscomp_q23}, the solid lines show the median for the FIRE models across m12 halos and redshifts~0.5, 0.6, 0.7, 0.8, 0.9, and~1.0. Here, we compare the FIRE-3 NoBH sample from the main body of the paper to the FIRE-3 NoBH m12+ sample listed in Tab.~\ref{tab:m12plus}. Including m12 halos with higher halo and stellar masses in the FIRE-3 NoBH sample decreases the difference with the original FIRE-2 NoBH sample, but the FIRE-3 NoBH model still predicts lower \ion{Ne}{8} column densities than the FIRE-2 NoBH model.}
\label{fig:m12plus_obscomp}
\end{figure}

Here, we test to what extent the difference in central galaxy stellar masses explains the column density difference between the FIRE-2 NoBH and FIRE-3 NoBH m12 halos.
In Fig.~\ref{fig:m12plus_obscomp}, we compare to good and plausible halo-mass-matched data points in the same way as in \S\ref{sec:obscomp}, but we base the halo mass selection range on the range of halo masses in the combined m12 FIRE-3 NoBH sample.

 We find that the lower stellar masses in the FIRE-3 NoBH m12 halos might play a part in their lower column densities relative to the FIRE-2 NoBH halos. However, as the stellar yield differences would suggest, the stellar masses do not fully explain the differences (Fig.~\ref{fig:m12plus_obscomp}).

\section{The power-law CGM model}
\label{app:plmodel}

For our simple power law CGM model, we assume a spherically symmetrical gas distribution within a dark-matter-dominated potential well. We assume power-law circular velocity ($v_{\mathrm{c}}$) and entropy ($K$) profiles
\begin{align}
&v_{\mathrm{c}} \propto r^{m}, \label{eq:mdef2} \\
&K \propto r^{l}, \label{eq:ldef2}
\end{align}
where $r$ is the distance to the halo center, and $m$ and $l$ are the exponents of the circular velocity and entropy profiles, respectively.
The entropy is defined as $K = k \mathrm{T}\mathrm{n}^{-\frac{2}{3}}$, where $k$ is the Boltzmann constant, and $\mathrm{n}$ is the gas particle number density.  

Our model is based on the analytical cooling flow model of 
\citet{stern_fielding_etal_2019}, but generalized to arbitrary entropy profile slopes.
We assume a steady-state inflow within the halo. If the only relevant forces are gravity and gas pressure, Newton's second law applied to a gas element gives
\begin{equation}
\rho V \frac{\mathrm{d^{2}} r}{\mathrm{d} t^2} = -V \frac{\mathrm{d} P}{\mathrm{d} r}  - \rho V \frac{G M(<r)}{r^2},
\end{equation}
where $\rho$ is density, $V$ is volume, $t$ is time, $P$ is pressure, $G$ is Newton's constant, and $M(<r)$ is the enclosed mass within radius $r$. 
Using $v_{\mathrm{c}} = \sqrt{G M(<r) \,/\, r}$ and the fact that velocity depends only on radius (and not explicitly on time) in this model,
we obtain
\begin{equation}
\frac{1}{2}\frac{\mathrm{d} v^2}{\mathrm{d} r} = - \frac{1}{\rho} \frac{\mathrm{d} P}{\mathrm{d} r}  - \frac{v_{\mathrm{c}}^2}{r}.
\end{equation}
We multiply both sides of this equation by $r \, /\, c_{\mathrm{s}}^2$, where $c_{\mathrm{s}} = \sqrt{\gamma P \,/\, \rho}$ is the adiabatic sound speed and $\gamma$ is the adiabatic index. This gives
\begin{equation}
\mathcal{M}^2 \frac{\mathrm{d} \ln v}{\mathrm{d} \ln  r} = 
 - \frac{1}{\gamma} \frac{\mathrm{d} \ln P}{\mathrm{d} \ln  r}
 - \frac{v_{\mathrm{c}}^2}{c_{\mathrm{s}}^2}, 
\label{eq:s19_16}
\end{equation}
where $\mathcal{M} = v \,/\, c_{\mathrm{s}}$ is the Mach number. 
From here, we make another assumption: that the inflows are subsonic, i.e., $\mathcal{M} \ll 1$.
Setting the left-hand term of equation~\ref{eq:s19_16} to zero, we then obtain 
\begin{equation}
- \frac{1}{\gamma} \frac{\mathrm{d} \ln P}{\mathrm{d} \ln  r} = \frac{v_{\mathrm{c}}^2}{c_{\mathrm{s}}^2}.
\label{eq:Tfromcs}
\end{equation}
We further assume that the CGM gas is a monatomic ideal gas, meaning $\gamma = \frac{5}{3}$. 

Since we are assuming the thermodynamical quantities follow power laws, $\mathrm{d} \ln P \, /\, \mathrm{d} \ln  r$ is a constant, meaning $v_{\mathrm{c}}^2 \,/\, c_{\mathrm{s}}^2$ must be as well. Since $c_{\mathrm{s}}^2 = \gamma k T \,/\, \mu \mathrm{m}_{\mathrm{H}}$, where $T$ is the temperature and $\mu$ is the mean molecular mass in units of the hydrogen atom mass $\mathrm{m}_{\mathrm{H}}$,
\begin{equation}
T \propto c_{\mathrm{s}}^2 \propto v_{\mathrm{c}}^2 \propto r^{2m}.
\end{equation}
We have assumed here that the mean molecular mass $\mu$ is constant. This is reasonable for the warm/hot CGM, where hydrogen and helium are fully ionized, and electron contributions from metals are small.

We can then also solve for the density slope using the entropy, and again assuming $\rho \, / \, n = \mu \mathrm{m}_{\mathrm{H}}$ is constant. Since $K = k T n^{-2/3}$, we get
\begin{equation}
l = \frac{\mathrm{d} \ln K}{\mathrm{d} \ln r} = \frac{\mathrm{d} \ln T}{\mathrm{d} \ln r} - \frac{2}{3} \frac{\mathrm{d} \ln \rho}{\mathrm{d} \ln r} = 2 m -  \frac{2}{3} \frac{\mathrm{d} \ln \rho}{\mathrm{d} \ln r},
\end{equation}
which means
\begin{equation}
\rho \propto r^{- 3l/2 + 3 m}.
\end{equation}

Finally, this also gives the pressure slope, which is simply the sum of the temperature and density slopes. We use this to solve eq.~\ref{eq:Tfromcs} for the sound speed
\begin{equation}
\frac{v_{\mathrm{c}}^2}{c_{\mathrm{s}}^2} = - 3 m  + \frac{9}{10} l.
\end{equation}
This gives the normalization of the temperature profile. Using 
\begin{equation}
T_{\mathrm{vir}} = \frac{\mu \mathrm{m}_{\mathrm{H}} v_{\mathrm{c}}(R_{\mathrm{vir}})}{2 k},    
\end{equation}
where $T_{\mathrm{vir}}$ is the virial temperature and $R_{\mathrm{vir}}$ is the virial radius,
we get
\begin{equation}
T(R_{\mathrm{vir}}) = \frac{6}{5} \frac{1}{\frac{9}{10}l - 3m} T_{\mathrm{vir}}.
\end{equation}

If we chose the entropy slope to be $l = 1 + 4 m \,/\, 3$, we recover the power-law cooling-flow solution of \citet{stern_fielding_etal_2019}.

We set the normalization of the density profile 
using the parameter $\mathrm{f}_\mathrm{CGM} = \mathrm{M}_{\mathrm{CGM}} \,/\, (\Omega_{\mathrm{b}} \Mvir \,/\, \Omega_{\mathrm{m}})$, where 
$\mathrm{M}_{\mathrm{CGM}}$ is the mass of the CGM gas 0.1--$1 \Rvir$ from the halo center. (Specifically, it is the mass of warm/hot phase we are modelling here.) $\Omega_{\mathrm{b}}$ and $\Omega_{\mathrm{m}}$ are the cosmic mean baryon and matter densities, respectively, normalized by the critical density. We convert the mass $\mathrm{M}_{\mathrm{CGM}}$ to a hydrogen number density normalization assuming a hydrogen mass fraction of 0.752.


\end{document}